\documentclass[preprint,11pt]{elsarticle}

\usepackage{amssymb}
\usepackage{amsmath}
\usepackage{amsthm}
\usepackage{amsfonts}

\usepackage{lineno}

\usepackage[T1]{fontenc}
\usepackage[utf8]{inputenc}
\usepackage{graphicx}
\usepackage{moreverb}
\usepackage{mathrsfs}
\usepackage[top=3.5cm, bottom=3.5cm, left=2.5cm, right=2.5cm]{geometry}
\usepackage{fancyhdr}
\usepackage{float}
\usepackage{lastpage}
\usepackage{sectsty}
\usepackage{fixltx2e}
\usepackage{gensymb}
\usepackage{xfrac}
\usepackage{lmodern}
\usepackage{color}
\usepackage{caption}
\usepackage{subcaption}
\usepackage{epstopdf}
\usepackage{listings}
\usepackage{todonotes}
\usepackage{animate}
\usepackage{booktabs}
\usepackage{bm}
\usepackage[ pagebackref=true]{hyperref}

\usepackage{placeins}

\theoremstyle{plain}
\newtheorem{theoremm}{Theorem}[section]

\theoremstyle{definition}

\theoremstyle{remark}

\let\oldsqrt\sqrt
\def\sqrt{\mathpalette\DHLhksqrt}
\def\DHLhksqrt#1#2{%
\setbox0=\hbox{$#1\oldsqrt{#2\,}$}\dimen0=\ht0
\advance\dimen0-0.2\ht0
\setbox2=\hbox{\vrule height\ht0 depth -\dimen0}%
{\box0\lower0.4pt\box2}}

\newcommand{\R}{\mathbb{R}}
\newcommand{\E}{\mathbb{E}}

\newcommand{\x}{\mathbf{x}}
\newcommand{\colvec}[2]{\begin{bmatrix}#1 \\ #2\end{bmatrix}}

\newcommand{\rd}{\mathrm{d}}

\newcommand{\tr}{\mathrm{tr}}

\usepackage{listings}
\usepackage[ruled, lined, linesnumbered, commentsnumbered, longend]{algorithm2e}

\SetKwInput{KwInput}{Input}
\SetKwInput{KwOutput}{Output}

\newcommand{\norm}[1]{\left|\left| #1 \right| \right|}

\newcommand{\f}{\vt{f}} 

\newcommand{\vt}[1]{\bm{#1}}
\newcommand{\Rey}{\mathrm{Re}}

\newcommand{\dd}[2]{\frac{\partial #1}{\partial #2}}

\newcommand{\q}{\vt{u}} 
\newcommand{\qbar}{\bar{\q}} 
\newcommand{\pphys}{\mu} 
\newcommand{\pcal}{\theta} 

\newcommand{\Qbar}{\bar{\vt{U}}} 
\newcommand{\Qhbar}{\Qbar_{h}} 

\newcommand{\xSI}{\vt{x}} 
\newcommand{\xsde}{\vt{X}} 
\newcommand{\bW}{\bm{W}} 
\newcommand{\bz}{\bm{z}} 
\newcommand{\bI}{\bm{I}} 
\newcommand{\bR}{\bm{R}} 
\newcommand{\drift}{\bm{b}} 
\newcommand{\score}{s} 
\newcommand{\velocity}{v} 
\newcommand{\diffusion}{g} 
\newcommand{\normalvar}{\vt{Z}} 

\newcommand{\QOI}{Q} 

\newcommand{\qrec}{\vt{v}} 
\newcommand{\qred}{\bar{\qrec}} 

\newcommand{\proj}{\Pi} 

\newcommand{\Fh}{\vt{F}_{h}} 
\newcommand{\qh}{\vt{u}_{h}} 
\newcommand{\Gh}{\vt{G}_{h,\pcal}} 
\newcommand{\qhred}{\qred_{h}} 
\newcommand{\qhbar}{\qbar_{h}} 

\newcommand{\A}{A}

\makeatletter
\newcommand{\NoSubscriptUppercase}[1]{%
    \begingroup
    \expandafter\NoSubscriptUppercase@split#1\relax
    \endgroup
}

\def\NoSubscriptUppercase@split#1{%
    \@ifnextchar^%
        {\NoSubscriptUppercase@withsup{#1}}%
        {\@ifnextchar_%
            {\NoSubscriptUppercase@withsub{#1}}%
            {\text{\uppercase{$#1$}}}%
        }%
}

\def\NoSubscriptUppercase@withsup#1^#2{%
    \text{\uppercase{$#1$}}^{#2}%
    \@ifnextchar_%
        {\NoSubscriptUppercase@addsub}%
        {}%
}

\def\NoSubscriptUppercase@withsub#1_#2{%
    \text{\uppercase{$#1$}}_{#2}%
    \@ifnextchar^%
        {\NoSubscriptUppercase@addsup}%
        {}%
}

\def\NoSubscriptUppercase@addsup^#1{^{#1}}
\def\NoSubscriptUppercase@addsub_#1{_{#1}}

\makeatother

\newcommand{\prob}[1]{p(#1)}

\newcommand{\condprob}[2]{p(#1 \left |#2 \right. )}

\journal{Computers and Fluids}

\begin{document}

\begin{frontmatter}



\title{Physics-aware generative models for turbulent fluid flows through energy-consistent stochastic interpolants}

        
\author[cwi]{Nikolaj T. M{\"u}cke} 
\author[cwi,tue]{Benjamin Sanderse} 

\address[cwi]{
    Scientific Computing, Centrum Wiskunde \& Informatica, Science Park 123, Amsterdam, 1098 XG, The Netherlands
}
\address[tue]{
    Centre for Analysis, Scientific Computing and Applications, Eindhoven University of Technology, PO Box 513, Eindhoven, 5600 MB, The Netherlands
}

\begin{abstract}
Generative models have demonstrated remarkable success in domains such as text, image, and video synthesis. In this work, we explore the application of generative models to fluid dynamics, specifically for turbulence simulation, where classical numerical solvers are computationally expensive. We propose a novel stochastic generative model based on stochastic interpolants, which enables probabilistic forecasting while incorporating physical constraints such as energy stability and divergence-freeness. Unlike conventional stochastic generative models, which are often agnostic to underlying physical laws, our approach embeds energy consistency by making the parameters of the stochastic interpolant learnable coefficients. We evaluate our method on a benchmark turbulence problem -- Kolmogorov flow -- demonstrating superior accuracy and stability over state-of-the-art alternatives such as autoregressive conditional diffusion models (ACDMs) and PDE-Refiner. Furthermore, we achieve stable results for significantly longer roll-outs than standard stochastic interpolants. Our results highlight the potential of physics-aware generative models in accelerating and enhancing turbulence simulations while preserving fundamental conservation properties.
\end{abstract}



\begin{keyword}



Stochastic interpolants \sep generative models \sep stochastic differential equations \sep fluid dynamics \sep energy conservation \sep turbulence
\end{keyword}

\end{frontmatter}


\section{Introduction}\label{section:introduction}
Recently developed AI tools like ChatGPT \cite{ray_chatgpt_2023}, Sora \cite{liu_sora_2024}, and DALL-E \cite{ramesh_hierarchical_2022} mark a ground-breaking era in generative modelling across text, video, and image domains. A natural question is whether generative models can also be used to support or replace simulation codes that have been classically used to model complex problems arising in domains like physics, chemistry, or biology. In particular, our interest lies in generative models for fluid dynamics problems. Fluid dynamics simulation codes have been developed in the last decades based on known physical principles (e.g.\ conservation laws), but with the main limitation that they are computationally typically very expensive to run, and also very difficult to maintain. Pre-trained generative models hold the promise that they can significantly accelerate physics simulations while at the same time providing a notion of uncertainty in the outcome, and having the ability to model unresolved scale processes \cite{chen2023b}. 

A first step in this direction are the ``foundation models'' that have been developed for weather prediction \cite{bodnar2024,pathak2022,nguyen2023b,lam2023}. These foundation models \cite{bommasani2022} are very large neural networks (sometimes having more than 1 billion parameters \cite{bodnar2024}), that are pre-trained on vast (re-analysis) datasets, and can be finetuned in space and time to specific tasks. These models have shown the potential to accelerate classical numerical weather predictions to the point that a 5-day forecast can be made in less than a minute \cite{bodnar2024}. The foundation model approach has also been applied to computational chemistry \cite{batatia2024}, biology \cite{rosen2023}, and fluid mechanics \cite{herde2024}. In \cite{herde2024}, a foundation model was pre-trained on certain PDEs (compressible Euler and incompressible Navier-Stokes) and gave accurate results on PDEs that were not in the pre-training set.

These foundation models are pre-trained, but unlike their counterparts in text, video and image generation, they are still usually deterministic in nature and mainly focus on forecasting the mean of the possible trajectories \cite{bodnar2024,lam2023,herde2024,price2024}. This can lead to blurred forecast states \cite{lam2023,keisler2022}. A few stochastic machine learning approaches have recently been introduced to address this issue \cite{li2024e,price2024,kochkov2024}. More in general, stochasticity naturally arises when developing reduced models of fluid dynamics problems, in which the state is typically split into large (resolved) and small (unresolved) scales, and the purpose is to infer models for only the large scales. An example is through the Mori-Zwanzig formalism, in which the effect of the initial condition of the unresolved scales is typically modelled by a noise term. Similarly, one can argue that the evolution of the large scales is \textit{not} deterministic, even when the evolution of the full model is deterministic. This manifests itself in non-uniqueness of the problem (different small scale realizations can have the same effect on the large scales), which warrants the use of a stochastic approach \cite{pope2004}. We note here that, outside the realm of machine learning, stochastic approaches have been long in use to model unresolved processes, typically known as `stochastic parameterizations' \cite{chen2023b,berner2017}. The stochastic approach that we are considering is in line with those approaches, but benefits from the advanced approximation capabilities of neural networks.

Previous research has explored the approximation of SDEs using neural networks. In \cite{li_scalable_2020} neural networks are trained with gradients computed via adjoint methods. While such methods alleviate some of the computational burden compared with naively backpropagating through a solver, it is still prohibitively expensive for very high-dimensional problems. Specifically, systems of up to 50 dimensions are studied in \cite{li_scalable_2020}. In \cite{boral2023} similar approaches are utilized for learning a neural SDE as a closure term for LES simulations. To minimize the computational restrictions of high-dimensional systems, the discrete state dimension is reduced via an autoencoder and the SDE is trained in the latent space. While this approach is indeed promising, the use of autoencoders makes it difficult to infuse physics knowledge into the model due to the non-physical nature of the latent space. Furthermore, this approach relies on an LES solver, as the neural network only approximates a closure term. It is unclear how this approach would fare as a full non-intrusive surrogate model. In this work, as an alternative to the expensive adjoint-based training methods, we will draw inspiration from the field of generative models. 

The current state-of-the-art in stochastic generative models for fluid flows relies mainly on denoising diffusion probabilistic models (DDPMs) \cite{kohl2024,price2024} -- similar to what is used in machine learning models for image generation \cite{ho2020,song_score-based_2021}. DDPMs are based on sampling from a Gaussian distribution and transforming it through a sequence of steps into a realistic image. In the context of time-dependent problems, one deals with a sequence of time steps, and a realistic `image' (e.g.\ a flow field) needs to be created \textit{at each time step}. This significantly complicates the process compared to image generation. An example of how this can be achieved is given in \cite{price2024}: sample from a Gaussian at each time step, denoise, and condition on the state at the previous time step. A similar approach is used in \cite{kohl2024}. Denoising at each time step is rather involved and not in line with the actual physical process. 
In general, a common issue with existing stochastic generative models is that they are agnostic of the underlying physical processes that are being modelled. For example, conservation of mass and energy are fundamental physical principles in fluid flows, but are not obeyed by existing generative models. This can lead to issues with the stability of predictions, especially when considering long prediction horizons.



In this work, we develop a pre-trained, stochastic generative model that (to the authors' knowledge) incorporates for the first time a sense of physical awareness. Our approach is built upon two key components: stochastic interpolants and energy stability. Stochastic interpolants, introduced in \cite{albergo2023a, albergo2023stochastic}, possess the crucial ability to bridge two arbitrary probability distributions over a finite time interval. This property enables their use in time-dependent simulations for time stepping, as demonstrated in \cite{chen2024probabilistic}, without requiring the sampling and transformation of Gaussian distributions \cite{kohl2024,price2024}. In \cite{chen2024probabilistic} only modest roll-outs of two time steps are considered. In this work we aim to achieve significantly longer roll-outs of several hundreds of steps. 

\begin{figure}[t]
    \centering
    \includegraphics[width=1.0\linewidth]{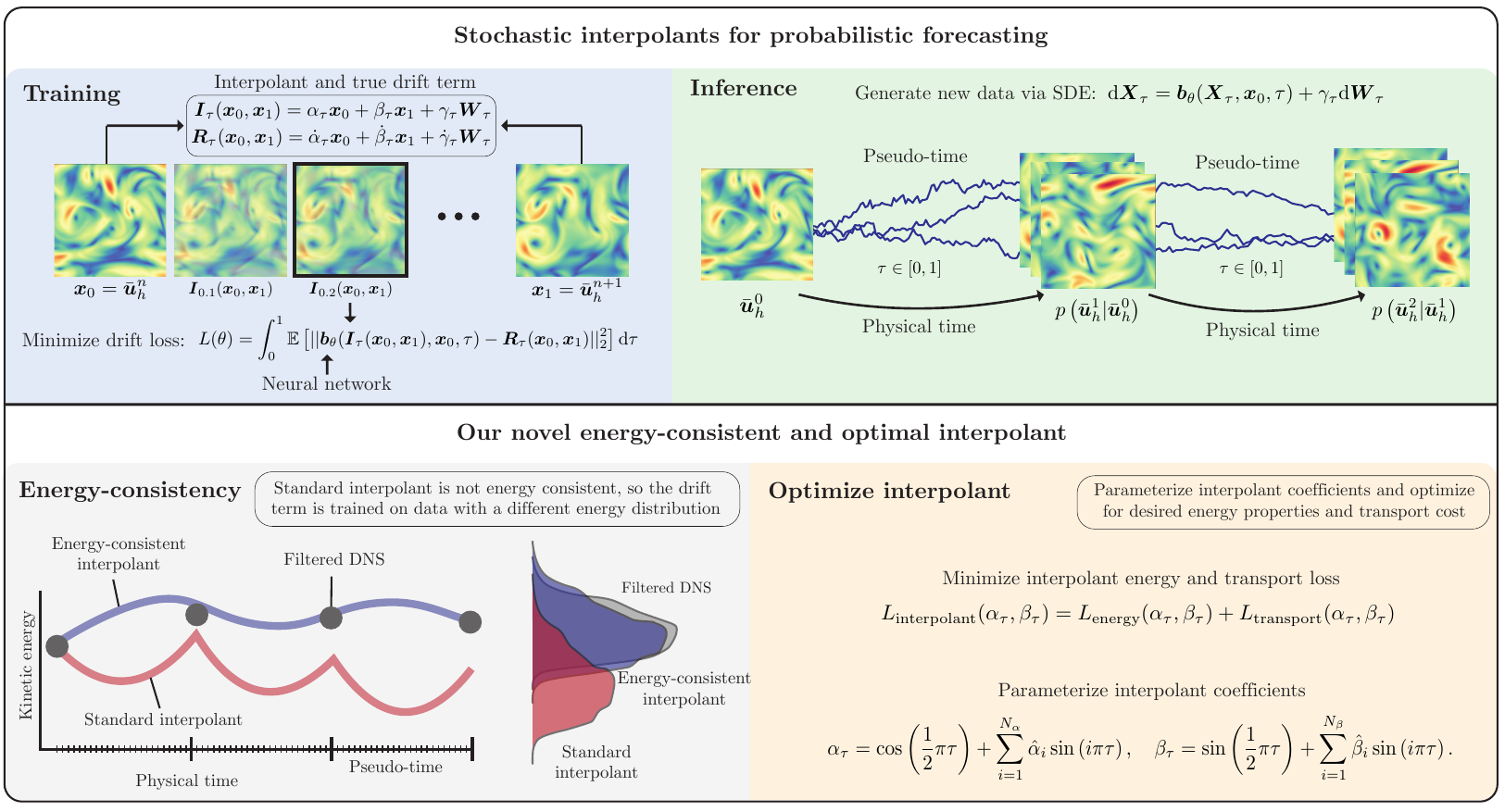}
    \caption{Visualization of the various steps and components in the energy-consistent stochastic interpolant framework. At the top, the existing framework presented in \cite{chen_probabilistic_2024} is visualized. The training is performed by sampling two consecutive physical states and interpolating between them via the stochastic interpolant in pseudo-time. The interpolant is used to train a drift term in an SDE that will be solved in pseudo-time during inference to generate new states conditioned on the initial state. Choosing the interpolant without including physics knowledge can result in inconsistencies in energy with respect to the physical energy, as visualized in the lower left part of the figure. In this paper, we propose to optimize the interpolant for energy-consistency by minimizing a loss over the interpolant coefficients as shown in the lower right part of the figure. }
    \label{fig:stochastic_interpolant}
\end{figure}

The main novelty of our work lies in the second component: energy stability through the imposition of a carefully designed stochastic interpolant (and, consequently, the drift term). Energy stability plays a dual role: it not only improves numerical stability \cite{vangastelen2024,sanderse2020}, but also promotes physical correctness in the generative process. By designing an interpolant that respects energy conservation, we ensure that the drift term is trained on physically consistent data, leading to generated samples that adhere to this constraint. See Figure \ref{fig:stochastic_interpolant} for a visual representation of our energy-consistent stochastic interpolant framework. Our work can be seen as a new approach to achieve stability with neural SDEs, which is known to be difficult to achieve with neural ODEs, requiring either specialized equation forms \cite{vangastelen2024}, specialized filters \cite{agdestein2025}, clipping \cite{park2021}, noise addition \cite{kurz2021}, or online learning \cite{rasp2020} (for an overview, see \cite{sanderse2024a}).

This paper is structured as follows. In Section \ref{section:preliminaries}, we introduce the problem setting and provide an overview of the stochastic interpolant framework for probabilistic forecasting. Section \ref{section:methodology} details our proposed modifications to the stochastic interpolant framework, emphasizing energy consistency, divergence-freeness, and efficient sampling. Implementation details, including the neural network architecture, training procedures, and inference strategies, are discussed in Section \ref{section:implementation}. In Section \ref{section:results}, we evaluate our methodology on a Kolmogorov flow test case and compare its performance with state-of-the-art alternatives. Finally, in Section \ref{section:conclusion}, we summarize our findings and outline potential directions for future research.

\section{Preliminaries}\label{section:preliminaries}

We consider probability spaces, $(\Omega, \mathcal{F}, P)$, where $\Omega$ is the sample space, $\mathcal{F}$ is a $\sigma$ algebra of events, and $P$ is a probability measure on $\mathcal{F}$. Stochastic variables are denoted with capital letters, $X(\omega)$, where $\omega\in\Omega$. For simplicity, we omit the explicit dependence on $\omega$ and simply write $X$ for stochastic variables and $X_t:= X(\Omega, t)$ for time-dependent stochastic variables. We assume the existence of a probability density function, $p$, associated with the probability measure, $P$, and will refer to the density when referring to the underlying distribution. Furthermore, to simplify notation we identify the densities through the arguments rather than by using subscripts. Thus, we will write $X\sim \prob{\x}$ when $X$ is distributed according to $P(X)$ and $X\sim \condprob{x}{y}$ when $X$ is distributed according to $P(X|Y)$.

\subsection{Problem setting}
In this work we are interested in stochastic approaches to approximate the solution of (deterministic) PDEs, and in particular the incompressible Navier-Stokes equations:
\begin{equation}\label{eq:NS}
        \dd{\q}{t} + \nabla \cdot (\q \otimes \q) = -\nabla p + \frac{1}{\Rey} \nabla^2 \q  + \f(\q), 
 \end{equation}
supplemented with the divergence-free constraint
\begin{equation}\label{eq:divfree}
    \nabla \cdot \q = 0.
\end{equation} 
Here $\q(\vt{x},t)$ is the velocity field, $p(\vt{x},t)$ is the (scaled) pressure, and $\Rey$ is the Reynolds number. Discretization in space and time leads to the fully discrete equation
\begin{align} \label{eq:high_fidelity_PDE}
    \qh^{n+1} = \Fh\left(\qh^{n} ; \pphys\right).
\end{align}
Here $\qh^{n}\in \R^{D} \approx \q(\vt{x},t^{n})$ and $\Fh$ is the discrete right-hand side incorporating both the spatial and temporal discretization, and $\pphys$ represents physical parameters such as the Reynolds number and parameterized initial and boundary conditions. We assume that the discretized model can resolve all relevant scales of motion present in \eqref{eq:NS}, at the cost of computation time. Therefore, we refer to Eq.\ \eqref{eq:high_fidelity_PDE} as the high-fidelity model, also known as Direct Numerical Simulation (DNS).

The high-fidelity model is often computationally prohibitive to solve, and a common approach is to reduce the range of scales present in \eqref{eq:high_fidelity_PDE} by a model reduction step. For the case of the Navier-Stokes equations, a common model reduction step that we will also employ in this work is \textit{filtering}. Filtering aims to remove the smallest scales of the flow through a low-pass filter $\A$, $\qhbar:=\A \qh$, such that $\qhbar \in \mathbb{R}^{d}$ can be represented and simulated on a much coarser grid than $\qh$ ($d \ll D$) \cite{sagaut2006}. However, $\qhbar$ does not satisfy the Navier-Stokes equations because the filter does not commute with the PDE operations.

A main ongoing challenge is thus to find a parameterized evolution equation that approximates the (exact) large scales $\qhbar$ \cite{sanderse2024a}, i.e.\ an equation of the form
\begin{align} \label{eq:reduced_PDE}
    \qhred^{n+1} = \Gh\left(\qhred^{n} ; \pphys\right),
\end{align}
such that $\qhred \approx \qhbar$ and $\theta$ are learnable parameters. Many existing approaches formulate the parametric model $\Gh$ in terms of $\Fh$ with an additional correction term, also known as a closure model \cite{sanderse2024a}. Usually, these models are deterministic, which corresponds to the fact that equation \eqref{eq:NS} is deterministic. However, there are good reasons to model the reduced dynamics instead with a stochastic approach as highlighted in the introduction \cite{pope2004,pope2000,ahmed2021}.
SDEs are therefore a natural fit for turbulence, and neural SDEs are in particular interesting, given their combination of stochasticity, time continuity, and the expressive capability of neural networks. So far, neural SDEs seem under-explored for turbulent flows, most likely due to the costs of training and simulating stochastic systems. We address this issue through recent developments in training SDE-based generative models, and in particular by using so-called stochastic interpolants (SIs). For some very recent related contributions that use DDPMs, see \cite{kohl_benchmarking_2024,dong2024a,molinaro2025}.

In our stochastic approach we replace the deterministic state, $\qhbar^{n}$, with a stochastic variable, $\Qhbar^{n}$. Due to the stochastic nature, the time evolution of the state can be  rewritten in terms of sampling from a conditional distribution:
\begin{align}\label{eq:sampling_conditional}
    \Qhbar^{n+1} \sim \condprob{\qhbar^{n+1}}{\qhbar^{n}}.
\end{align}
We typically assume that the initial density, $p(\qhbar^{0})$, is known or that the initial condition is simply given $\Qhbar^{0}=\qhbar^{0}$. The conditional distribution in \eqref{eq:sampling_conditional} is generally not available and is  approximated via ensembles. 
The task is thus to find a stochastic version of \eqref{eq:reduced_PDE} that produces ensemble members that are distributed according to $\condprob{\qhbar^{n+1}}{\qhbar^{n}}$.


\subsection{Stochastic interpolants for probabilistic forecasting} \label{subsection:stochastic_interpolants}
In this section, we present the principles of the generative model for probabilistic forecasting trained via the stochastic interpolant (SI) framework. The aim is to generate samples from the conditional distribution, $\condprob{\qhbar^{n+1}}{\qhbar^{n}}$, via a stochastic differential equation (SDE) that transforms samples from a base distribution to samples from the target distribution. In the standard version of the SI framework, the SDE is not trained to mimic the physical evolution of the state. It should rather be interpreted as a means of transforming samples from one distribution to another. Hence, the SDE is not solved with respect to physical time, $t$, discussed in the previous section, but it is solved in pseudo-time, $\tau$, introduced with the sole purpose of facilitating the transformation. The pseudo-time interval can therefore be chosen freely, but is typically chosen to be $[0,1]$ for simplicity.

SIs are a type of generative models introduced in \cite{albergo2023a} and expanded upon in \cite{albergo2023, chen2024c}. The general idea is similar to SDE-based denoising diffusion models where a normal distribution is being transformed into the target distribution. In the SI framework, however, one can transform samples from an arbitrary distribution into samples from another arbitrary distribution. This property makes it a suitable choice for physics-based modeling. In this section, we summarize the findings from \cite{chen2024c}, which focuses on utilizing SIs for probabilistic forecasting.

Consider the densities $\prob{\xSI_0}$ and $\condprob{\xSI_1}{\xSI_0}$, with $\xSI_0, \xSI_1 \in \R^d$. Given a sample $\xSI_0$, the aim of the  generative model is to sample from the conditional distribution:
\begin{align}
\condprob{\xSI_1}{\xSI_0} = \frac{\prob{\xSI_0, \xSI_1}}{\prob{\xSI_0}} > 0.  
\end{align}
With the SI framework, such samples are generated by solving an SDE with initial condition $\xSI_0$ on the pseudo-time interval $\tau\in[0,1]$:
\begin{align}\label{eq:SI_SDE}
    \rd \xsde_{\tau} = \drift_\theta(\xsde_{\tau}, \xSI_0, \tau) \rd \tau + \gamma_\tau \rd \bW_\tau, \quad \tau \in [0,1], \quad \xsde_0 = \xSI_0.
\end{align}
The drift term, $\drift_\theta$, is modeled as a neural network with weights, $\theta$. The SI framework provides a way of training it such that solutions of the SDE at $\tau=1$, are distributed according to the conditional distribution, $\xsde_{\tau=1} \sim \condprob{\xSI_1}{\xSI_0}$. The diffusion term $\gamma_{\tau}$\footnote{not to be confused with the diffusion term $\nu \nabla^2 \q$ in the Navier-Stokes equations} and Wiener process $\bW_\tau$ form the source of stochasticity. In this work $\gamma_{\tau}=0.1(1-\tau)$ is chosen. As mentioned earlier, the SDE is solved in pseudo-time, $\tau$, in contrast to the physical time, $t$.

The key element in training the drift term is the so-called stochastic interpolant:
\begin{align} \label{eq:stochastic_interpolant}
    \bI_{\tau}(\xSI_0, \xSI_1) = \alpha_\tau \xSI_0 + \beta_\tau \xSI_1 + \gamma_\tau \bW_\tau = \xSI_\tau.
\end{align}
The Wiener process, $\bW_\tau$, can be sampled with $\bW_\tau = \sqrt{\tau}\bz$, where $\bz\sim N(0,1)$. There is freedom in the choice of the $\tau$-dependent functions $\alpha_\tau$, $\beta_\tau$, and $\gamma_\tau$, but they need to satisfy the temporal boundary conditions, $\alpha_0 = \beta_1 = 1$ and $\alpha_1 = \beta_0 = \gamma_1 = 0$. This ensures that $\bI_{\tau=0}(\xSI_0, \xSI_1)\sim \prob{\xSI_0}$ and $\bI_{\tau=1}(\xSI_0, \xSI_1)\sim \condprob{\xSI_1}{\xSI_0}$. The dynamics of the interpolant in Eq. \eqref{eq:stochastic_interpolant} given an initial condition $\xSI_0$ can be written as
\begin{align}\label{eq:true_SI_SDE}
    \rd \bI_{\tau}(\xSI_0, \xSI_1) = \underbrace{\left(\dot{\alpha}_\tau \xSI_0 + \dot{\beta}_\tau \xSI_1 + \dot{\gamma}_\tau \bW_\tau \right)}_{:=\bR_{\tau}(\xSI_0, \xSI_1)} \rd \tau + \gamma_\tau \rd \bW_\tau, \quad \tau \in [0,1], \quad \xsde_0 = \xSI_0,
\end{align}
where $\dot{(\:)}$ denotes differentiation with respect to $\tau$. Hence, solving the SDE from $0$ to $\tau$ can be seen as a mapping of a samples, $\xSI_0$, to a sample from $\xsde_\tau\sim p(\xSI_\tau | \xSI_0)$. In particular, solving Eq. \eqref{eq:true_SI_SDE} from 0 to 1 maps a sample, $\xSI_0$, to a sample from $\xsde_1\sim p(\xSI_1 | \xSI_0)$. Solving the SDE in Eq. \eqref{eq:true_SI_SDE} requires that both $\xSI_0$ and $\xSI_1$ are available. However, the goal is to generate samples, $\xSI_1$, when only having access to $\xSI_0$. Therefore, we train $\drift_\theta$, which only takes in $\xSI_0$, $\tau$, and the intermediate state, $\xsde_{\tau}$, to match $\bR_{\tau}$. With proper training the solution of Eq. \eqref{eq:SI_SDE} approximates the solution of Eq. \eqref{eq:true_SI_SDE}, $\xsde_{\tau}\approx \bI_{\tau}(\xSI_0, \xSI_1)$ for all $\tau$, including the endpoint of interest, $\xsde_{1}\approx \bI_{1}(\xSI_0, \xSI_1) = \xSI_1$. This enables generation of samples from $p(\xSI_1 | \xSI_0)$ by using $\xSI_0$ alone via Eq. \eqref{eq:SI_SDE}. 

Given choices of $\alpha_\tau$, $\beta_\tau$, $\gamma_\tau$, and training data $(\xSI_0, \xSI_1)\sim p(\xSI_0, \xSI_1)$, the interpolant can be evaluated and the drift term is trained by minimizing the following loss function with respect to the model weights, $\theta$ \cite{chen2024c}:
\begin{align} \label{eq:SI_drift_loss}
    L(\theta) = \int_0^1 \E_{(\xSI_0, \xSI_1, \bW_\tau)}\left[ \norm{ \drift_{\theta}(\bI_\tau(\xSI_0, \xSI_1), \xSI_0, \tau) -  \bR_\tau(\xSI_0, \xSI_1) }^2_2 \right] \rd \tau.
\end{align}
$\norm{\cdot}_{2}$ denotes the standard $l^2$-norm. During training, the integral in \eqref{eq:SI_drift_loss} is approximated by sampling $\tau$ uniformly. The expected value is approximated using the samples $(\xSI_0, \xSI_1) \sim \prob{\xSI_0, \xSI_1}$ from a training set and samples from the Wiener process are sampled via  $\bW_\tau = \sqrt{\tau}\bz$, $\bz\sim N(0,1)$. In practice this is minimized by sampling mini-batches of $(\xSI_0, \xSI_1)$ and $\tau$. We found that sampling a single $\tau$-value for each pair $(\xSI_0, \xSI_1)$ did not hurt the training when compared to sampling several $\tau$-values for each state pair. It is important to note that this setup enables training a model to approximate a drift term via a simple mean squared error-type loss without ever solving any SDEs during the training stage. See Alg.\ \ref{alg:SI_training} for pseudo-code of the training stage. Note that Alg.\ \ref{alg:SI_training} is a simplified algorithm. In practice, there are additional considerations to be taken into account, and we refer to Section \ref{section:implementation:training_considerations} for details. 

An important property of the SI framework is that we learn the drift term of a \textit{continuous} SDE that generates new samples. This means that one can choose an SDE solver and number of pseudo-time steps after training. Such choices determine the computational time and the quality of the samples. This trade-off between accuracy and computation time can be made based on the application at hand.  

After training, new samples, $\xsde_1 \sim \condprob{\xSI_0}{\xSI_1}$, can be generated by solving \eqref{eq:SI_SDE} with the trained drift term and an appropriate SDE solver. In particular, by choosing $\xSI_0=\qhbar^n$ and $\xSI_1=\qhbar^{n+1}$ the model learns to sample from the distribution of interest, $\condprob{\qhbar^{n+1}}{\qhbar^{n}}$ -- namely the distribution of the next physical state given the current state. Then, by setting $\xSI_0=\qhbar^{n+1}$ and solving \eqref{eq:SI_SDE} again, we can obtain samples from $\condprob{\qhbar^{n+2}}{ \qhbar^{n+1}}$. Thus, we can obtain arbitrarily long trajectories in the \textit{physical} space by repeatedly solving the trained SDE, provided the solution is stable. Typically, at each physical time step we solve the SDE in pseudo-time several times in order to get an ensemble representation of the distribution. A visualization of this process is given at the top of Figure \ref{fig:stochastic_interpolant}.

\begin{figure*}[t!]
    \centering
    \begin{subfigure}[b]{0.55\textwidth}
        \centering
        \includegraphics[width=0.9\linewidth]{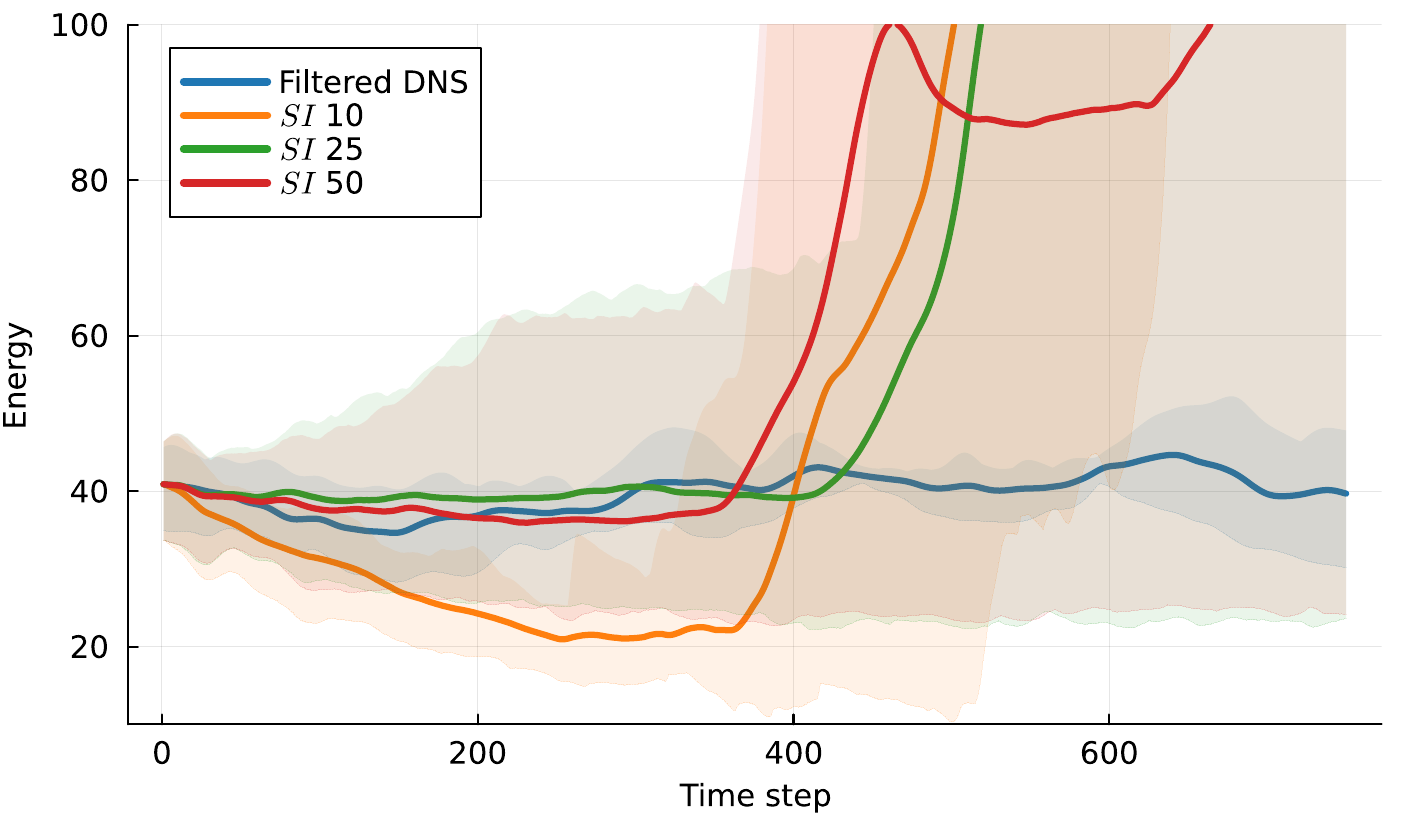}
        \caption{Time evolution of kinetic energy}
        \label{fig:kolmogorov_energy_evolution}
    \end{subfigure}
    \begin{subfigure}[b]{0.40\textwidth}
        \centering
        \includegraphics[width=1.0\linewidth]{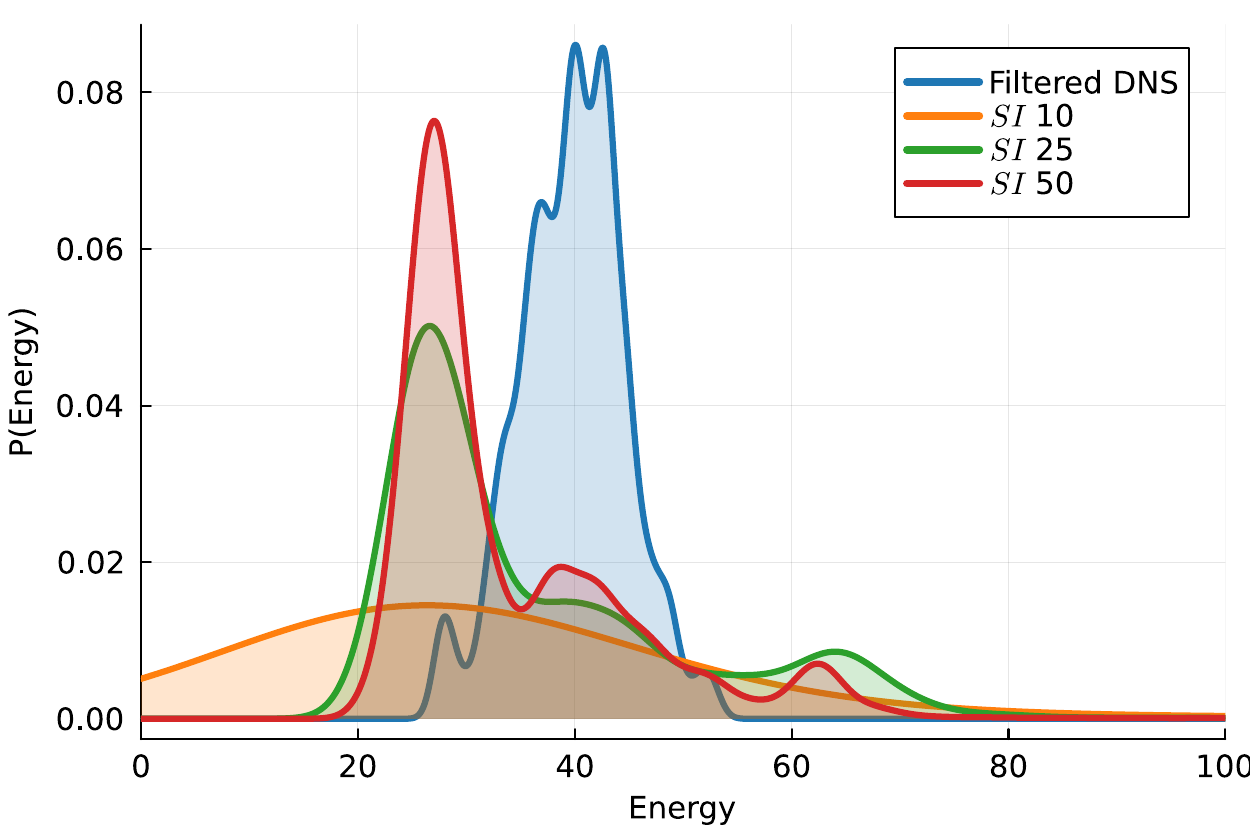}
        \caption{Kinetic energy distributions.}
        \label{fig:kolmogorov_energy_distributions}
    \end{subfigure}
    \caption{Energy results when simulating Kolmogorov flow with the stochastic interpolants as presented in Section \ref{subsection:stochastic_interpolants}.}
    \label{fig:_vanilla_si_kolmogorov_energy}
\end{figure*}

The SI framework for probabilistic time stepping does not necessarily generate physically plausible trajectories. As an example, we visualize the kinetic energy of the state as a function of physical time, as well as the distribution of the energy for a Kolmogorov flow approximated with the SI framework in Figure \ref{fig:_vanilla_si_kolmogorov_energy}. In the Kolmogorov flow the energy should remain constant in expectation. However, we see that after 300-400 physical time steps, the energy starts to grow. This is reminiscent of instabilities that have been reported when using neural networks and neural ODEs to represent turbulence \cite{vangastelen2024, agdestein2025, beck_deep_2019}

Furthermore, the distributions clearly do not match. Solving the SDE with more time steps improves this slightly, but not enough to a satisfactory degree. In the following section, we outline improvements to the SI setup that mitigates these issues. For more details on the Kolmogorov flow, see Section \ref{section:results}.

\begin{algorithm}[!t]
\caption{Training drift term in stochastic interpolant framework\label{alg:SI_training}}
\small
\KwInput{$\alpha_{\tau}$, $\beta_{\tau}$, $\gamma_{\tau}$, untrained $\drift_{\theta}$, training data, $N_{epochs}$, Optimizer}
\For{$i=1:N_{epochs}$}{
    \For{$(\xSI_0, \xSI_1)$ in training data}{
        Sample pseudo time, $\tau \sim U[0,1]$; \\
        Sample Wiener process, $\bW_\tau = \sqrt{\tau}\bz$, $\bz\sim N(0,1)$; \\
        Evaluate interpolant, $\bI_{\tau}(\xSI_0, \xSI_1) = \alpha_\tau \xSI_0 + \beta_\tau \xSI_1 + \gamma_\tau \bW_\tau$; \\
        Evaluate $\bR_{\tau}(\xSI_0, \xSI_1) = \dot{\alpha}_\tau \xSI_0 + \dot{\beta}_\tau \xSI_1 + \dot{\gamma}_\tau \bW_\tau$; \\
        Compute drift loss, $L(\theta)$, via Eq.\ \eqref{eq:SI_drift_loss}; \\
        Update drift model weights: $\theta \leftarrow \text{Optimizer}(\theta, L(\theta))$; \\
    }
}
\KwOutput{Trained $\drift_{\theta}$}
\end{algorithm}

\section{A new energy-consistent stochastic interpolant}\label{section:methodology}

\subsection{Physics consistency in generative models}
With proper training and neural network architecture, it is theoretically possible to generate samples that resemble the true conditional distribution. However, in practice this is not an easy task because errors can accumulate during long roll-outs, leading to inaccurate or unstable simulations, similar to what has been observed for neural ODEs \cite{melchers2023}. This is largely due to the time-dependent nature of the problem, which gives an additional complexity that is not present in image generation. While this is theoretically alleviated with bigger neural networks and more training, it is in practice difficult to achieve the long-term stability with such approaches alone.  

As a consequence, with current SIs only short roll-outs have been considered. For example, \cite{chen2024c} considered roll-outs of two physical steps. In this paper, we aim to generate significantly longer trajectories by promoting stability through physical consistency of the SI. 

To achieve physical consistency in the generation procedure there are several steps that can be modified in the SI framework. Firstly, the interpolant in Eq.\ \eqref{eq:stochastic_interpolant} can be adjusted to have desired properties. Secondly, the loss function in Eq.\ \eqref{eq:SI_drift_loss} can be extended. Thirdly, the neural network architecture used to approximate the drift term can be designed with physics consistency in mind. Lastly, the generative SDE in Eq.\ \eqref{eq:SI_SDE} can be modified, to impose desired properties after training.

In general, physical consistency can either be imposed as \textit{soft} or \textit{hard} constraints. When imposing soft constraints, one typically adds additional terms to the loss function during training. Although this does not guarantee exact adherence to the constraints, it is often easier to implement. Imposing hard constraints, on the other hand, ensures adherence to the constraint. This is typically done via the architecture \cite{brantner_volume-preserving_2024} or additional computations in shape of projections \cite{agdestein2025}, constraint optimization \cite{dener_training_2020}, or by modifying the equations of interest \cite{vangastelen2023}. The choice between hard constraints or soft constraints depends on the problem at hand. Additional computations sometimes associated with hard constraints can lead to high computational costs. A specific architecture might make the training easier and the predictions more stable, but it could also impose limitations as one does not allow the neural network to potentially find an optimal representation. In addition, in a stochastic setting, there are additional considerations to be taken into account. The system under consideration might only conserve certain properties in \textit{distribution}, while other properties are conserved for all \textit{realizations}.

For the specific case of the incompressible Navier-Stokes equations \eqref{eq:NS}-\eqref{eq:divfree}, arguably the most important physical structures are the divergence-freeness of the flow field and the conservation of kinetic energy (in the absence of boundaries and viscosity) \cite{foias2001navier,sanderse2020,vangastelen2023}. This is further detailed in \ref{section:NS_energy}. Our aim is therefore to ensure that the generated trajectories are \textit{energy-consistent} and \textit{divergence-free}. Energy-consistency will be promoted in distribution via a soft constraint which determines the parameterization of the SI in Section \ref{subsection:energy_consistent_interpolant}. Divergence-freeness will be enforced via a hard constraint while time-stepping the SDE in Section \ref{subsection:divergence_consistency}. 

\subsection{Energy-consistent interpolant} \label{subsection:energy_consistent_interpolant}
The interpolant defines the stochastic paths between $\qhbar^{n}$ and $\qhbar^{n+1}$. The drift term is trained directly on the interpolated states, such that the generating SDE approximates these paths. Therefore, the specific choice of interpolant plays a major role in the training and by extension also the generation. When generating data without any immediate need for physics-consistency, it typically does not matter if these paths adhere to any physical laws. However, when physics-consistency is of importance, the choice of interpolant should adhere to the desired properties so the model is trained on physics-consistent data. 

As mentioned above, some properties should only be enforced on average. Energy-consistency is one such property where individual trajectories should not necessarily adhere to the constraint, but an ensemble of realizations should be energy-consistent (see \ref{section:NS_energy}). Therefore, we choose to impose energy-consistency in a \textit{soft} manner. In this section, we show how one can optimize the choice of $\alpha_\tau$ and $\beta_\tau$ to achieve such properties. 

The following theorem forms the basis of the optimization:
\begin{theoremm}[Energy Evolution of the Interpolant] \label{theorem:energy_interpolant}
For a stochastic interpolant defined as in Eq. \eqref{eq:stochastic_interpolant} and energy defined by $\frac{1}{2}\norm{\bI_\tau}_2^2$, the time evolution of the interpolant energy is given by:
\begin{align}
        \rd \left[\frac{1}{2}\norm{\bI_\tau(\xSI_0, \xSI_1)}_2^2\right] = (\bI_\tau(\xSI_0, \xSI_1) \cdot \bR_\tau(\xSI_0, \xSI_1) + \frac{d}{2} \gamma_\tau^2) \rd \tau + \gamma_\tau \bI_\tau(\xSI_0, \xSI_1) \cdot \rd \bW_\tau.
\end{align}
Furthermore, the expected time evolution is given by:
\begin{align} \label{eq:expected_time_evolution_of_energy}
\begin{split}
    &\E_{(\xSI_0, \xSI_1, \bW)} \left[\rd \left(\frac{1}{2}\norm{\bI_\tau(\xSI_0, \xSI_1)}_2^2 \right) \right]  = \E_{(\xSI_0, \xSI_1)}\left[H_\tau(\xSI_0, \xSI_1; \alpha_\tau, \beta_\tau, \gamma_\tau)\right]\rd\tau,
\end{split}
\end{align}
where
\begin{align}
    H_\tau(\xSI_0, \xSI_1; \alpha_\tau, \beta_\tau, \gamma_\tau)=\dot{\alpha} _{\tau}\alpha_{\tau}\norm{\xSI_0}_2^2 + \dot{\beta}_{\tau} \beta_{\tau}\norm{\xSI_1}_2^2 + (\dot{\beta}_{\tau} \alpha_{\tau} + \dot{\alpha}_{\tau} \beta_{\tau}) \langle \xSI_0, \xSI_1 \rangle + \dot{\gamma}_{\tau} \gamma_{\tau} \tau d + \frac{d}{2} \gamma_\tau^2.
\end{align}
\end{theoremm}
See \ref{appendix:proof_of_theorem} for the proof of the theorem. Note that the energy depends on the grid size, $h$. As $h$ is assumed constant, it does not change any of the results above nor the derivations below besides a scalar multiplication. See \ref{section:appendix_metrics_and_qoi} for more details.  

With Eq. \eqref{eq:expected_time_evolution_of_energy}, we can control the expected rate of change of the kinetic energy of the SI through the choice of its parameters $\alpha_\tau$ and $\beta_\tau$. Since $\gamma_\tau$ controls the noise levels, we choose to only consider $\alpha_\tau$ and $\beta_\tau$ for this purpose. 

The key idea is that in many physical systems we have a-priori knowledge about the rate of change of kinetic energy between $\xSI_0$ and $\xSI_1$ -- see \ref{section:NS_energy} for the Navier-Stokes equations. This knowledge can be used to determine $\alpha_\tau$ and $\beta_\tau$. Denoting the known rate of change by $k_\tau(\xSI_0, \xSI_1)$, the loss function quantifying the expected discrepancy between the desired rate of change and the actual rate of change induced by the interpolant reads:
\begin{align}\label{eq:L_energy}
\begin{split}
    L_{\text{energy}}(\alpha_\tau, \beta_\tau) 
    &=
    \int_{0}^{1} \E_{(\xSI_0, \xSI_1)} \left[\rd \left[\frac{1}{2}\norm{\bI_\tau(\xSI_0, \xSI_1)}_2^2 \right] - k_{\tau}(\xSI_0, \xSI_1)\right] \rd \tau \\
    &=
    \int_{0}^{1} \int_{\R^d}\int_{\R^d} \left[H_\tau(\xSI_0, \xSI_1 ; \alpha_\tau, \beta_\tau, \gamma_\tau) - k_\tau(\xSI_0, \xSI_1)\right]\prob{\xSI_0, \xSI_1} \rd\xSI_0 \rd \xSI_1 \rd \tau \\
    &\approx 
    \frac{1}{N_\tau N_s}
    \sum_{i=1}^{N_\tau} \sum_{j=1}^{N_s} H_\tau(\xSI_{0,j}, \xSI_{1,j} ; \alpha_{\tau_i}, \beta_{\tau_i}, \gamma_{\tau_i}) - k_\tau(\xSI_{0,j}, \xSI_{1,j}).
\end{split}
\end{align}
In this expression, $\tau_i$, $i=1,\dots, N_\tau$ are pseudo-time samples between 0 and 1, and $\xSI_{0,j}$ and $\xSI_{1,j}$, $j=1,\ldots N_s$ are samples from the training set. In our simulations we are interested in the case where energy is conserved on average. This corresponds to choosing $k_\tau(\xSI_0, \xSI_1) = 0$, optimizing the interpolant such that the change in energy is zero in expectation. 

Minimizing $L_{\text{energy}}(\alpha_\tau, \beta_\tau)$ with respect to $\alpha_\tau$ and $\beta_\tau$ requires parameterizing $\alpha_\tau$ and $\beta_\tau$. In order to do so, we write $\alpha_{\tau}$ and $\beta_{\tau}$ as Fourier series and optimize with respect to the coefficients, $\hat{\alpha}_i$ and $\hat{\beta}_i$:
\begin{align} \label{eq:interpolant_parameterization}
   \alpha_{\tau} = \cos\left( \frac{1}{2} \pi \tau \right) + \sum_{i=1}^{N_{\alpha}} \hat{\alpha}_i \sin\left(i \pi \tau \right), \quad \beta_{\tau} = \sin\left( \frac{1}{2} \pi \tau \right) + \sum_{i=1}^{N_\beta} \hat{\beta}_i \sin\left(i \pi \tau \right).
\end{align}
Note that these expressions satisfy the temporal boundary conditions by construction. While Fourier series have been chosen for this study, they are not the only option. Other basis functions, such as Legendre or Chebyshev polynomials are also a potential option. Further investigation of these approaches will be pursued in future work. Choosing $\alpha_\tau$ and $\beta_\tau$ to minimize \eqref{eq:L_energy} is effectively a \textit{soft} constraint on the training data: on average, the resulting trajectories will conserve energy, but each individual trajectory can still have locally increasing or decreasing energy. This is in line with the properties of the incompressible Navier-Stokes equations with body force (see \ref{section:NS_energy}).

The minimization of $L_{\text{energy}}$ with respect to $\hat{\alpha}_\tau$ and $\hat{\beta}_i$ will be performed in a pretraining step, i.e., before the training of the drift term as described in Section \ref{subsection:stochastic_interpolants}. The optimization problem to be solved is of dimension $N_{\alpha} + N_{\beta}$. Our experiments indicate that $N_{\alpha} = N_{\beta} < 10$ is sufficient. Hence, higher-order optimization methods, such as Newton methods, can be used without computational issues.  

\subsubsection{Minimizing path complexity}
If one only minimizes the energy discrepancy, there is a chance that the resulting interpolant can be complex and include high-frequency oscillations. Therefore, we also want to promote low \textit{complexity} in the trajectories. In this section we describe further improvements to be made to the interpolant like an additional loss term for fitting the $\alpha_\tau$ and $\beta_\tau$. The aim is to reduce the complexity of the paths defined by the interpolant connecting $\xSI_0$ and $\xSI_1$. In this context, ``complexity'' refers to factors that make the paths difficult to learn and integrate. Hence, reducing complexity must simplify the training stage, and reduce the number of necessary pseudo-time steps when generating new data. To this end, we use the transport cost as a metric for complexity. In \cite{albergo_building_2023} the connection between the SI framework for normalizing flows and the optimal transport problem in the framework of \cite{benamou_computational_2000} is established. The transport cost is defined by \cite{benamou_computational_2000}:
\begin{align} \label{eq:transport_loss}
\begin{split}
    C_{\text{transport}}(\bR_\tau) &= \int_0^1 \E_{(\xSI_0, \xSI_1)}\left[\norm{\bR_\tau(\xSI_1, \xSI_0)}^2_2\right] \rd\tau \\
    &= \int_0^1 \int_{\R^d} \int_{\R^d}\norm{\bR_\tau(\xSI_1, \xSI_0)}^2_2 \prob{\xSI_1, \xSI_0} \rd\xSI_0\rd\xSI_1 \rd\tau.
\end{split}
\end{align}
The transport cost measures the cost of transforming one distribution to another. Minimizing the transport cost minimizes the traveled distance between $\prob{\xSI_0}$ and $\condprob{\xSI_1}{\xSI_0}$. 

In \cite{albergo_building_2023}, it is briefly discussed how one can minimize the transport cost while training the drift term by solving a max-min problem. However, max-min problems are typically difficult to handle due to the saddle point structure of the optimum. In this paper, we take a slightly different approach and perform this optimization \textit{before} training the drift term. By decoupling the training of drift term and the identification of the energy-consistent interpolant we avoid solving a max-min problem which is generally more difficult to deal with. However, we now have to solve two optimization problems where the outcome of the first problem is used in the second. This does add some complexity in the implementation. 

To minimize the transport cost, we use the same approach as discussed in Section \ref{subsection:energy_consistent_interpolant}. Using the parameterization from Eq.\  \eqref{eq:interpolant_parameterization} we simultaneously minimize the transport cost as well as $L_{\text{energy}}$. This gives the full loss term for the interpolant:
\begin{align} \label{eq:interpolant_loss}
    L_{\text{interpolant}}(\alpha_\tau, \beta_\tau) = L_{\text{energy}}(\alpha_\tau, \beta_\tau) + L_{\text{transport}}(\alpha_\tau, \beta_\tau),
\end{align}
where
\begin{align}
\begin{split}
    L_{\text{transport}}(\alpha_{\tau}, \beta_{\tau}) 
    &= \int_0^1 \int_{\R^d}\int_{\R^d} \norm{\bR_{\tau}(\xSI_{0}, \xSI_{1})}^2_2 \prob{\xSI_0,\xSI_1} \rd\xSI_0 \rd\xSI_1 \rd\tau \\
    &= \int_0^1 \int_{\R^d}\int_{\R^d} \norm{\dot{\alpha}_{\tau} \xSI_{0} + \dot{\beta}_{\tau} \xSI_{1} + \dot{\gamma}_{\tau} \bW_{\tau} }^2_2 \prob{\xSI_0,\xSI_1} \rd\xSI_0 \rd\xSI_1 \rd\tau \\
    &\approx \frac{1}{N_\tau N_s}\sum_{i=1}^{N_\tau} \sum_{j=1}^{N_s} \norm{\dot{\alpha}_{\tau_i} \xSI_{0, j} + \dot{\beta}_{\tau_i} \xSI_{1, j} + \dot{\gamma}_{\tau_i} \bW_{\tau_i} }^2_2,
\end{split}
\end{align}
with $(\xSI_{0, j}, \xSI_{1, j}) \sim \prob{\xSI_0,\xSI_1}$ for $j=1,\ldots,N_s$ being training samples. Note that $L_{\text{transport}}$ is simply $C_{\text{transport}}$ considered as a function of $\alpha_\tau$ and $\beta_\tau$ instead of $\bR_\tau$. See Algorithm \ref{alg:interpolant_optimization} for the pseudo-code of the interpolant optimization procedure.

\begin{algorithm}[!t]
\caption{Optimize interpolant}
 \label{alg:interpolant_optimization}
\small
\KwInput{training data, $N_\alpha$, $N_\beta$, $N_{epochs}$, Optimizer}
\For{$i=1:N_{epochs}$}{
    Evaluate $\alpha_\tau$ and $\beta_\tau$ via Eq. \eqref{eq:interpolant_parameterization}; \\
    Compute interpolant loss, $L_{\text{interpolant}}(\alpha, \beta)$, via Eq.\eqref{eq:interpolant_loss}; \\
    Update $\alpha_\tau$ and $\beta_\tau$: $(\alpha_\tau, \beta_\tau) \leftarrow \text{Optimizer}(\alpha_\tau, \beta_\tau, L(\alpha_\tau, \beta_\tau))$; \\

}
\KwOutput{Trained $\alpha, \beta$}
\end{algorithm}

\begin{figure*}[t!]
    \centering
    \begin{subfigure}[b]{0.75\textwidth}
        \centering
        \includegraphics[width=1.0\linewidth]{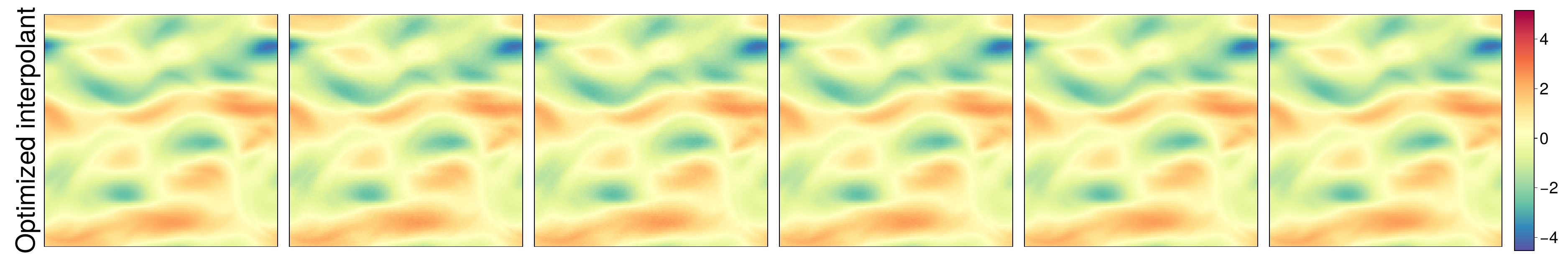}
    \end{subfigure}
    \begin{subfigure}[b]{0.75\textwidth}
        \centering
        \includegraphics[width=1.0\linewidth]{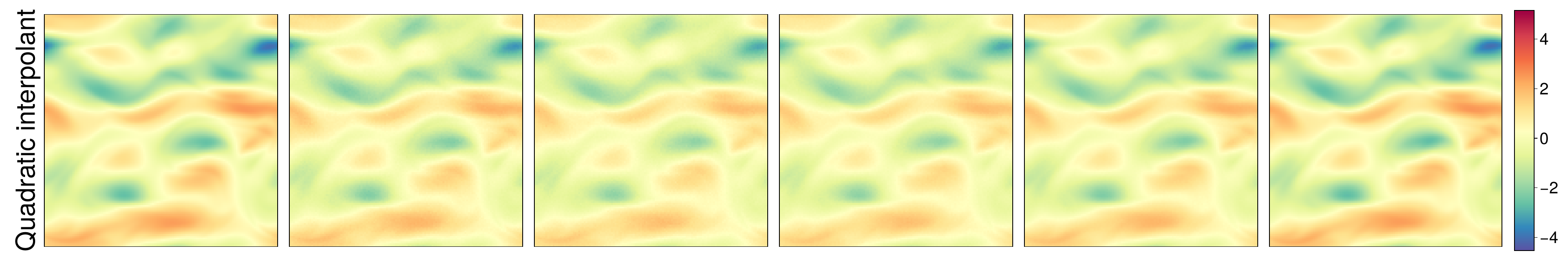}
        \caption{Transport and energy optimized (top row) and quadratic (bottom row) interpolant, $\bI_\tau(\xSI_0, \xSI_1)$, of the $x$-velocity for pseudo-time step from left to right: $\tau=0.0, 0.2, 0.4, 0.6, 0.8, 1.0$.}
        \label{fig:interpolant_visualization}
    \end{subfigure}
    \begin{subfigure}[b]{0.76\textwidth}
        \centering
        \includegraphics[width=1.0\linewidth]{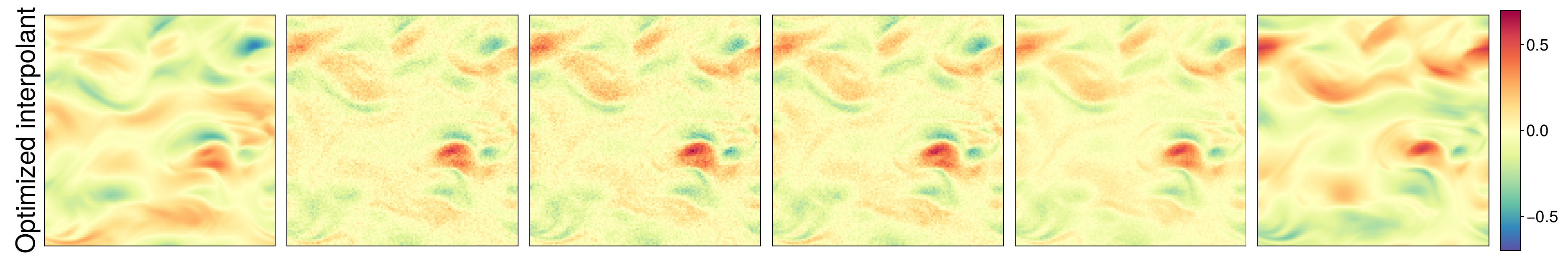}
    \end{subfigure}
    \begin{subfigure}[b]{0.75\textwidth}
        \centering
        \includegraphics[width=1.0\linewidth]{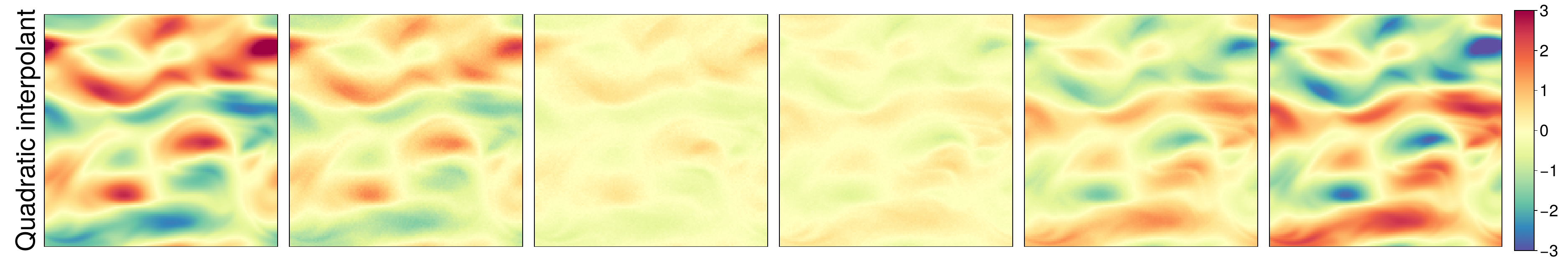}
        \caption{Transport and energy optimized (top row) and quadratic (bottom row) interpolant drift, $\bR_\tau(\xSI_0, \xSI_1)$, of the $x$-velocity for pseudo-time step from left to right: $\tau=0.0, 0.2, 0.4, 0.6, 0.8, 1.0$. Note the different limits in the colorbar.}
        \label{fig:interpolant_drift_visualization}
    \end{subfigure}
    \begin{subfigure}[b]{0.32\textwidth}
        \centering
        \includegraphics[width=0.9\linewidth]{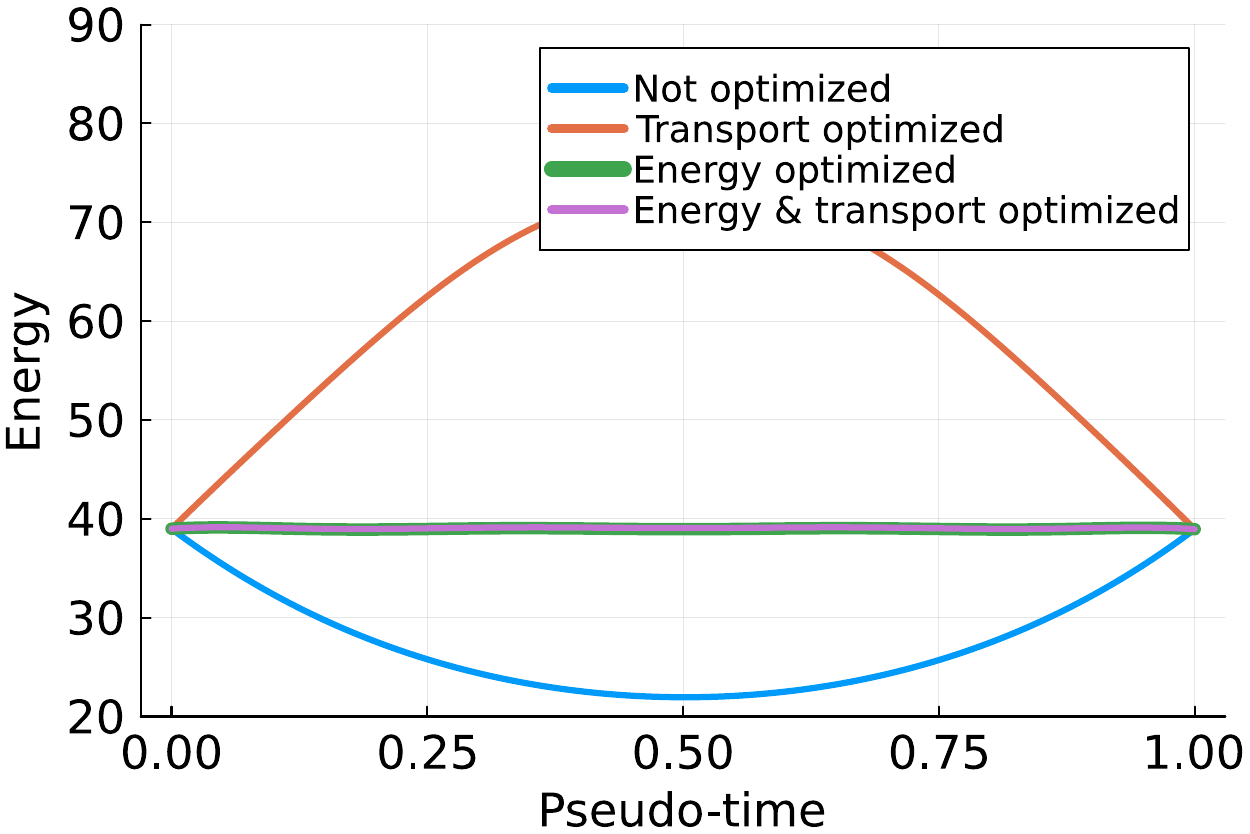}
        \caption{Energy evolution of the interpolant.}
        \label{fig:interpolant_energy}
    \end{subfigure}
    \begin{subfigure}[b]{0.32\textwidth}
        \centering
        \includegraphics[width=0.9\linewidth]{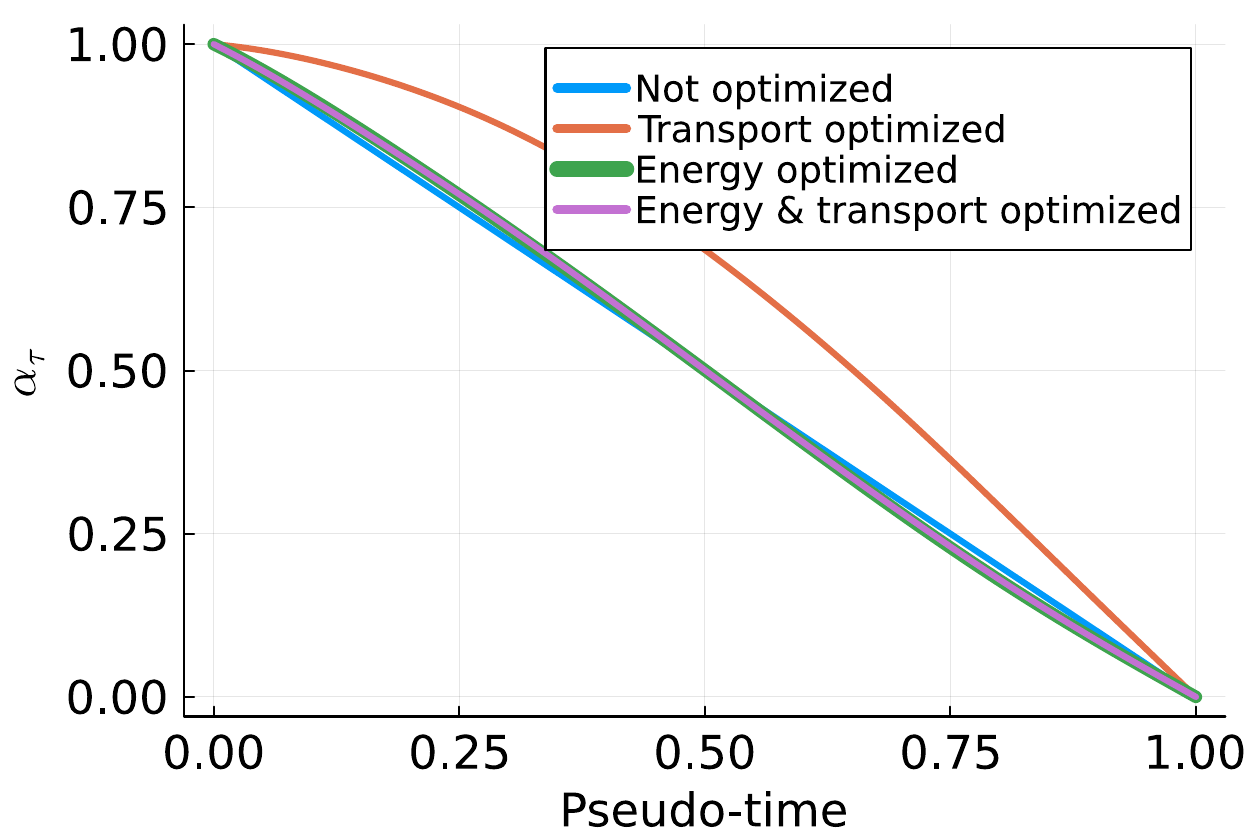}
        \caption{$\alpha_\tau$.}
        \label{fig:alpha}
    \end{subfigure}
    \begin{subfigure}[b]{0.32\textwidth}
        \centering
        \includegraphics[width=0.9\linewidth]{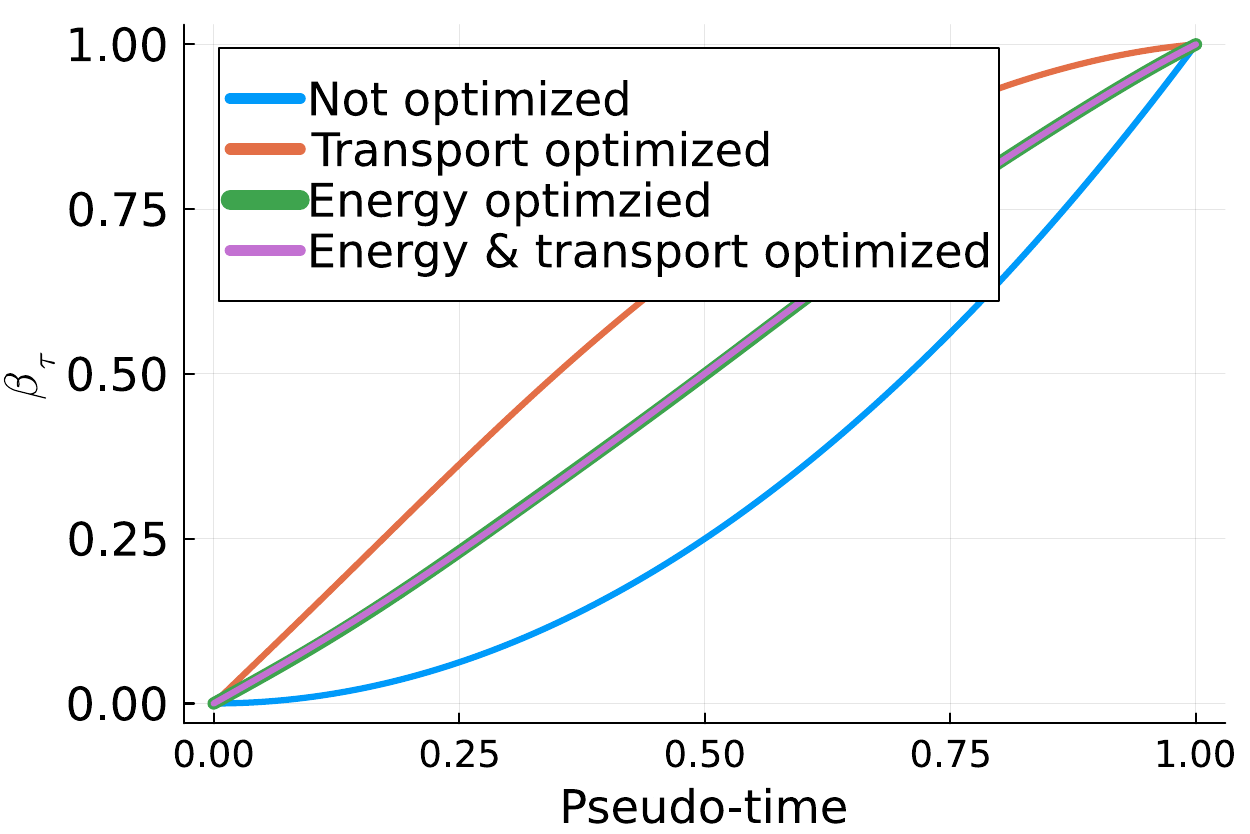}
        \caption{$\beta_\tau$.}
        \label{fig:beta}
    \end{subfigure}
    \caption{Comparison of the optimized interpolant and the interpolant proposed in \cite{chen_probabilistic_2024} with $\alpha_\tau = 1-\tau$ and $\beta_\tau = \tau^2$.}
    \label{fig:interpolant_comparison}
\end{figure*}

In Figure \ref{fig:interpolant_comparison}, we visually compare the interpolant proposed in \cite{chen_probabilistic_2024} with interpolants optimized with respect to the transport and energy loss individually and together. We compare with the choices $\alpha_\tau = 1-\tau$ and $\beta_\tau = \tau^2$, as they are stated in \cite{chen_probabilistic_2024} to yield the best results. This comparison is performed using samples from Kolmogorov flow trajectories where the energy is constant on average. Hence, we set $k_\tau(\xSI_0, \xSI_1) = 0$. In Figure \ref{fig:interpolant_visualization} it is apparent that the optimized interpolant and the interpolant from \cite{chen_probabilistic_2024} are visually similar. However, the drift terms, $\bR_\tau(\xSI_0, \xSI_1)$, vary quite a lot as shown in Figure \ref{fig:interpolant_drift_visualization}. Furthermore, in Figure \ref{fig:interpolant_energy} we see that despite the similarities in the interpolant realizations, the energy of the two interpolants differ significantly. This is primarily due to the differences in the $\beta_\tau$ term. Additionally, we see that minimizing only the transport loss results in poor energy conservation, while minimizing the  energy loss, whether the transport loss is included or not, results in an interpolant with good energy conservation. We also see that the resulting $\alpha_\tau$ and $\beta_\tau$ terms, in Figure \ref{fig:alpha} and \ref{fig:beta} respectively, are visually very similar when minimizing the energy loss, even without including the transport loss. This suggests that the energy loss is dominating in this particular case. However, for other problems this might not necessarily be the case. More details about the Kolmogorov flow test case can be found in Section \ref{section:results}.

\subsection{Divergence-consistency} \label{subsection:divergence_consistency}
For incompressible fluid flows the velocity field is divergence-free. However, a generative model does generally not adhere to this property, despite being trained on divergence-free data. This is especially the case when dealing with long roll-outs due to accumulation of errors.

In contrast to energy-consistency, divergence-freeness in the incompressible Navier-Stokes equations is an algebraic constraint and not an evolution equation. In other words, every realization arising from solving the SDE in Eq. \eqref{eq:SI_SDE} should be divergence-free. For this reason we impose divergence-consistency as a hard constraint on every single realization. 

Ensuring that the generated velocity fields are divergence-free can be done in several ways. For example, it can be done on the neural network architecture level \cite{vangastelen2023}. Another approach is to learn an SDE for the stream function $\vt{\psi}$ instead of the velocity, such that $\q = \nabla \times \vt{\psi}$, and its divergence is by construction zero. As an alternative, we will make use of \textit{projections} and the face-averaging filter as presented in \cite{agdestein2025}, which are designed to keep the filtered velocity field divergence-free. The projection operator $\proj:\R^d \rightarrow \R^d$,  projects any field onto its divergence-free part. In \cite{agdestein2025} the projection is performed at every stage of the Runge-Kutta method by solving a Poisson equation. We use the same projection in this work - for details on $\Pi$, see equations \eqref{eqn:poisson}-\eqref{eqn:velocity_update} in \ref{sec:NS_filtered}.

Ideally, all interpolated states should be divergence-free both during training and inference. However, this would require projecting the state after every single pseudo-time step when solving the generating SDE. As the projection requires solving a Poisson equation, it adds significant computational time. Therefore, we propose to project the state once during each physical time step, i.e.\ after the SDE has been solved in pseudo-time. Hence, the generation of samples is given by:
\begin{subequations}
\begin{align}
    &\rd \xsde_{\tau} = \drift(\xsde_{\tau}, \qhbar^{n}, \tau) \rd \tau + \gamma_\tau \rd \bW_\tau, \quad \xsde_0 = \qhbar^{n}, \quad \tau=0\ldots1, \\
    &\qhbar^{n+1} = \proj \xsde_1.
\end{align}
\end{subequations}

\begin{algorithm}[!t]
\caption{Trajectory generation with drift term learned via the stochastic interpolant}\label{alg:generation}
\small
\KwInput{Initial condition $\qhbar^{0}$, $N_{t}$, $N_{\tau}$, $\drift_{\theta}$, $\gamma_\tau$, $\Pi$, SDE integrator}
\For{$n_t=0:N_{t}$}{
    $\xsde_{0} = \qhbar^{n_t}$; \\
    \For{$n_j=0:N_{\tau}$}{
        $\xsde_{n_\tau + 1} =$ SDE step$(\xsde_{n_\tau}, \drift_{\theta}, \gamma_\tau)$ ; \\
  }  
  $\qhbar^{n_t+1} = \Pi \xsde_{1}$
}
\KwOutput{$\left\{\qhbar^{n}\right\}_{n=1}^{N_t}$ }
\end{algorithm}

\section{Implementation}\label{section:implementation}
Besides the general framework presented in the previous sections, there are still several decisions regarding implementation to make. Here, we briefly discuss such considerations. 

All implementations are done in Julia. The source code can be found on GitHub \footnote{\url{https://github.com/nmucke/StochasticInterpolants.jl.git}}. All trainings are performed on a single Nvidia RTX 3090 GPU. 

\subsection{Neural network architecture}
The SI framework does not require a specific family of models to parameterize $\drift_{\theta}$. However, due to the universality of neural networks, they are generally chosen for approximating the drift term. This choice of family of models naturally leads one to ask what architecture to use. In this regard, previous work on SIs for image generation have taken inspiration from work on DDPMs \cite{chen_probabilistic_2024}. We do the same and take inspiration from work using DPPMs for fluid dynamics \cite{lippe2023, kohl_benchmarking_2024}. 

\begin{figure}
    \centering
    \includegraphics[width=1.0\linewidth]{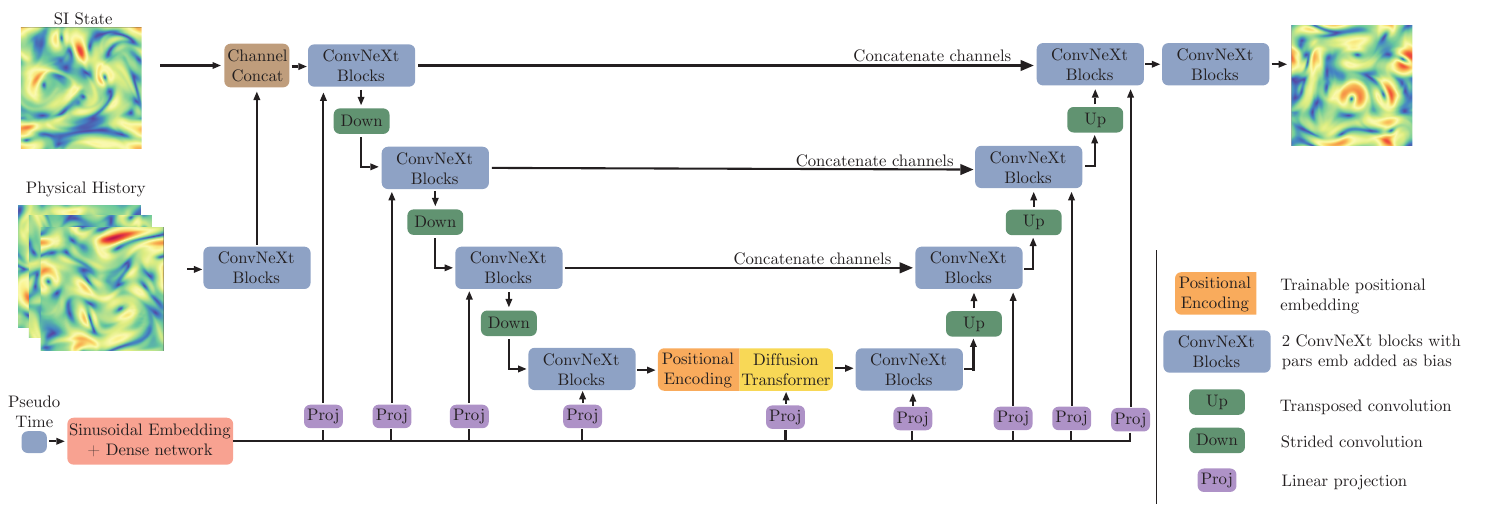}
    \caption{Drift term neural network architecture. }
    \label{fig:drift_term_unet}
\end{figure}

We make use of a UNet architecture, originally introduced in \cite{ronneberger_u-net_2015}, with ConvNeXt layers instead of normal residual layers \cite{liu_convnet_2022}. A sketch of the architecture is shown in Figure \ref{fig:drift_term_unet}. The pseudo-time, $\tau$, is first embedded and then passed to the ConvNeXt layers as a bias term. It is embedded via a sinusoidal positional embedding followed by a shallow neural network and then passed onto the convolutional layers. 

The downsampling is performed via strided convolutions and transposed convolutions are used for the upsampling. It is generally known that up- and downsampling via convolutional layers rather than pooling and interpolating gives better and less smoothed results \cite{springenberg_striving_2015, zeiler_deconvolutional_2010}.  

In the bottleneck we use a diffusion transformer \cite{peebles_scalable_2023} to compute attention globally while also incorporating embedded time. We add a trainable positional encoding to the tensor before passing it to the diffusion transformer.  The transformer layer ensures a global receptive field and the specific choice of diffusion transformer is made due to the efficient incorporation of parametric dependence. 

Instead of only inputting the current physical state as conditions for the model, we make use of several previous states. While this has not been included in discussions and derivations in previous sections, it does not change the approach, as this only requires changing the conditioning in the drift term. It has been shown that providing a history as conditioning to the model and not just the current state results in higher accuracy \cite{mucke_deep_2024, mucke_reduced_2021, geneva_transformers_2022}. The drift term and corresponding SDE in Eq.\ \eqref{eq:SI_SDE} change to:
\begin{align}
    \rd \xsde_{\tau} = \drift_\theta(\xsde_{\tau}, \qhbar^{n-l:n}, \tau) \rd \tau + \gamma_\tau \rd \bW_\tau, \quad \xsde_0 = \qhbar^{n}, \quad \tau=0 \ldots 1,
\end{align}
where $\qhbar^{n-l:n} = (\qhbar^{n-l}, \ldots, \qhbar^{n})$ and $l$ is the length of the history to be included.  

Throughout the network we use the GELU activation function \cite{hendrycks_gaussian_2023}. In the ConvNeXt layers and diffusion transformer, we use layer normalization. Lastly, we do not make use of dropout as experiments did not show improved performance. 

See Figure \ref{fig:drift_term_unet} for a visualization of the UNet architecture. 

\subsection{Training considerations} \label{section:implementation:training_considerations}
In this section we briefly describe the specific implementation of the training of the SI drift term and the interpolant coefficients.

The training states are standardized per channel. Each channel corresponds to a physical field, e.g. velocity in $x$-direction and $y$-direction. Each field is standardized to have zero mean and unit standard deviation. For the optimization, we use the AdamW optimizer \cite{loshchilov_decoupled_2019} with a cosine annealing learning rate scheduler with a warmup \cite{loshchilov_sgdr_2017}. The weights are regularized with L2-regularization. To further prevent overfitting, we make use of early stopping based on a validation dataset. The validation loss is computed by time-stepping via the generating SDE for multiple physical time steps. This ensures that the trained model actually performs well for the task it is intended for.  

Optimizing the interpolant is performed using a batched Newton optimization algorithm with a backtracking line search for determining how much to update in the optimal search direction.

\subsection{Inference}
One of the key advantages of the SI method, and SDE-based generative models in general, is the flexibility it provides in the inference stage. As the drift term of a (continuous) SDE is learned, any numerical SDE solver can be used after training. Furthermore, the amount of pseudo-time steps used can also be chosen according to quality and time restrictions. In this work, we use the Heun SDE integrator \cite{thygesen_stochastic_2023}.

\section{Results}\label{section:results}

We present results for a Kolmogorov flow test case. The governing equations are the incompressible Navier-Stokes equations in two dimensions given by Eq.\ \eqref{eq:NS}. For an overview of the test case settings, see Table \ref{tab:test_cases}. 

We evaluate the proposed framework on a series of metrics and quantities of interest serving different purposes. We compute the mean squared error (MSE), the Pearson correlation, and LSiM \cite{kohl_learning_2020}, which measures how well the generated trajectories match the filtered DNS trajectories directly. Since the underlying dynamics are chaotic in nature, it is only to be expected that the generated trajectories match the filtered DNS trajectories in the short-term. Therefore, we also compute the kinetic energy, the energy spectrum, and the rate of change of the states, to assess if the generated trajectories have the same \textit{characteristics} in terms of energy and rate of change as the filtered DNS trajectories. Such metrics are more suitable for assessing long-term behavior. Furthermore, we compare the distributions of some of these quantities to ensure that the statistics match. See \ref{section:appendix_metrics_and_qoi} for details on the metrics and quantities of interest. 

We compare our proposed methodology with three other approaches. Namely, the PDE-refiner \cite{lippe2023} (referred to as Refiner from hereon), the Autoregressive Conditional Diffusion Model (ACDM) \cite{kohl_benchmarking_2024}, and the original version of the SI for probabilistic forecasting without the proposed improvements \cite{chen_probabilistic_2024}. These methods are generative models for probabilistic forecasting of states governed by PDEs. The Refiner and the ACDM utilize the DDPM framework. For the implementation of the ACDM and the Refiner, we use the code from the GitHub repository associated with \cite{kohl_benchmarking_2024} \footnote{\url{https://github.com/tum-pbs/autoreg-pde-diffusion}}. The specific models used are taken directly from that repository and are slightly modified to approximately match the amount of model weights that we use for the stochastic interpolant. Note that a key difference between the ACDM and the Refiner compared to the SI framework is that the ACDM and Refiner need to be trained for a specific amount of diffusion steps. Hence, the number of generation steps must be chosen before training, unlike for the SI where the number of SDE generation steps can be chosen freely after training. 

We generate 5 trajectories per test trajectory. That is, for a given initial condition, we generate 5 realizations with the generative models. All 5 realization are compared with the filtered DNS solution using the same initial condition. Since we are testing against 5 test trajectories, we generate a total of 25 realizations. 

\begin{table}[]
    \centering
    \begin{tabular}{lc}
        \toprule
        $\Rey$ & $10^3$\\
        Forcing & Yes \\
        \# Train trajectories & 45 \\
        \# Test trajectories & 5   \\
        \# Train time steps & 250   \\
        \# Test time steps & 750  \\
        \# High-fidelity DOFs & $2048^2 \cdot 2 = 8388608$  \\
        \# Reduced DOFs & $128^2 \cdot 2 = 32768$  \\ 
        High-fidelity step size & $5\cdot 10^{-4}$ \\
        Generative model step size & $100 \cdot 5\cdot 10^{-4}= 5\cdot 10^{-2}$ \\
        Boundary conditions & Periodic  \\
        \bottomrule
    \end{tabular}
    \caption{Overview of Kolmogorov test case and parameter settings.}
    \label{tab:test_cases}
\end{table}


\subsection{Kolmogorov flow}\label{sec:Kolmogorov}
In this test case we assess the models ability to perform accurate and stable long-horizon simulations with respect to the statistics of the fluid flow. 

Kolmogorov flow is a type of forced turbulent flow that obeys the Navier-Stokes equations. Specifically, we use the forcing:
\begin{align}
    \f(\q) =  \sin(4 y) \colvec{1}{0} - 0.1 \q.
\end{align}
We consider the domain, $\Omega = [0, 2\pi]^2$, and the time horizon $T=62.5$. The first term in the forcing injects energy into the system and the second is a dissipative term that depends on the velocity. In total, the two terms ensure that the flow converges towards a statistically stationary state. 

We simulate the Kolmogorov flow using a finite volume method on a staggered grid implemented in the IncompressibleNavierStokes.jl library \cite{agdestein_incompressiblenavierstokesjl_2024}\footnote{\url{https://github.com/agdestein/IncompressibleNavierStokes.jl}}. The high-fidelity simulations are performed on a $2048\times 2048$ grid and are downsampled to a $128\times 128$ grid using face-averaging. We refer to those trajectories as the filtered DNS solutions and consider those to be the ground truth. For more details, see \cite{agdestein2025}. Each trajectory is initiated with a random initial condition. To ensure that the flow is fully developed and has reached the stationary distribution, we discard the first $t=25$ seconds of the trajectories. Furthermore, for the training of the models we use every 100th state from the high-fidelity simulations in time. Hence, the models take significantly larger time steps than the high-fidelity simulations. We train on 250 time steps and predict up to 750 steps, starting from the same time step, which corresponds to training on the time interval $t\in[25,37.5]$ and predicting on the time interval $t\in[25,62.5]$.

We perform several tests using various settings of the models. We train three ACDM models with 10, 25, and 50 diffusion steps, and three Refiner models with 2, 4, and 8 diffusion steps. These choices were made based on the recommendations in the respective papers and for the purpose of comparison with the SI framework. For the SI framework, we train two models. One without the optimized interpolant, and one with the optimized interpolant. The non-optimized interpolant uses $\alpha_\tau = 1-\tau$ and $\beta_\tau = \tau^2$ as recommended in \cite{chen_probabilistic_2024}. For the optimized interpolants we found that $N_\alpha=N_\beta=5$ was sufficient to achieve the desired energy distribution properties of the interpolant. We will refer to the optimized SI model as SI$_{\text{opt}}$ and the non-optimized as SI. Furthermore, we test SI framework with divergence projection and without. As this is not imposed until after training, no additional models need to be trained. We refer to the models with divergence-free projection as SI$_{\text{div}}$ and SI$_{\text{opt,div}}$.

To simplify the presentation of results, we only show the best results from each model class (ACDM, Refiner, SI, SI$_{\text{opt,div}}$), and refer to \ref{section:appendix_additional_results} for additional results. 

\begin{figure*}[t!]
    \centering
    \begin{subfigure}[b]{0.32\textwidth}
        \centering
        \includegraphics[width=1.0\linewidth]{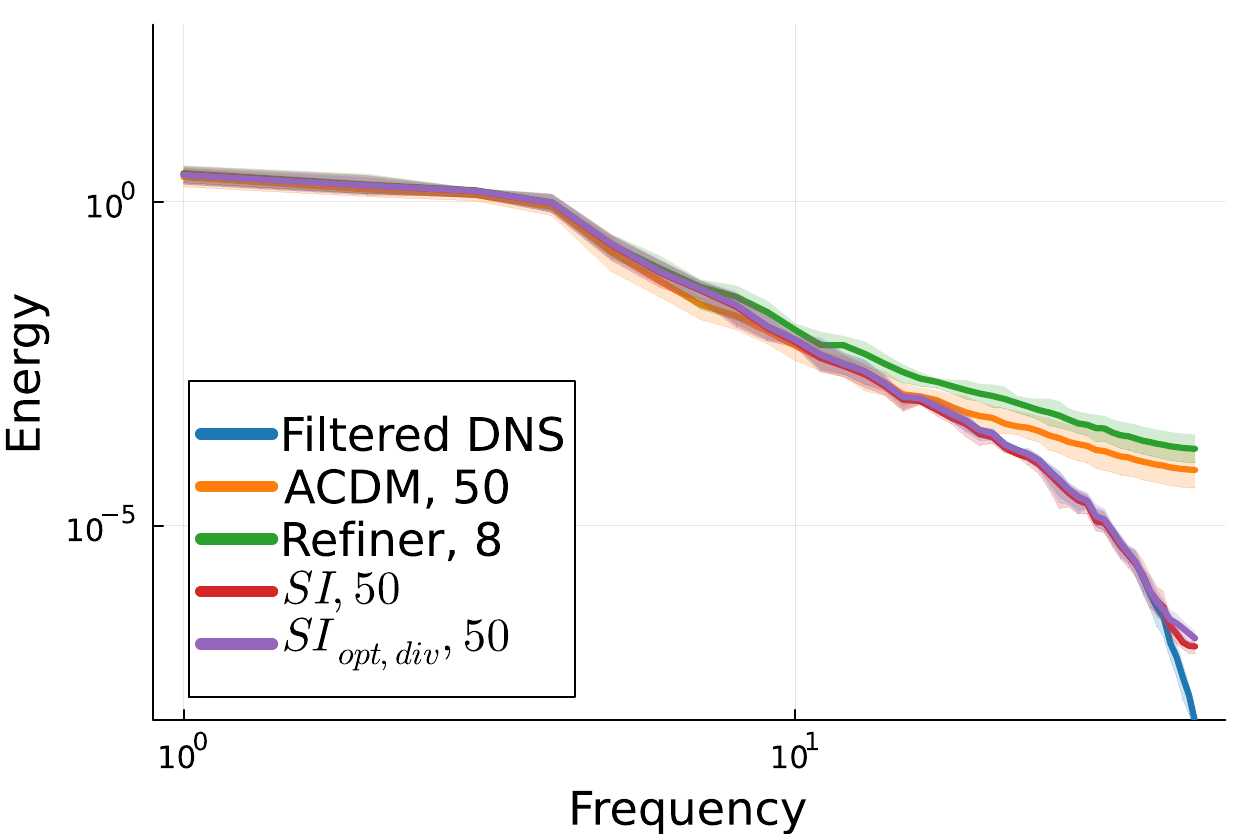}
        \caption{After $n=10$ time steps.}
    \end{subfigure}
    \begin{subfigure}[b]{0.32\textwidth}
        \centering
        \includegraphics[width=1.0\linewidth]{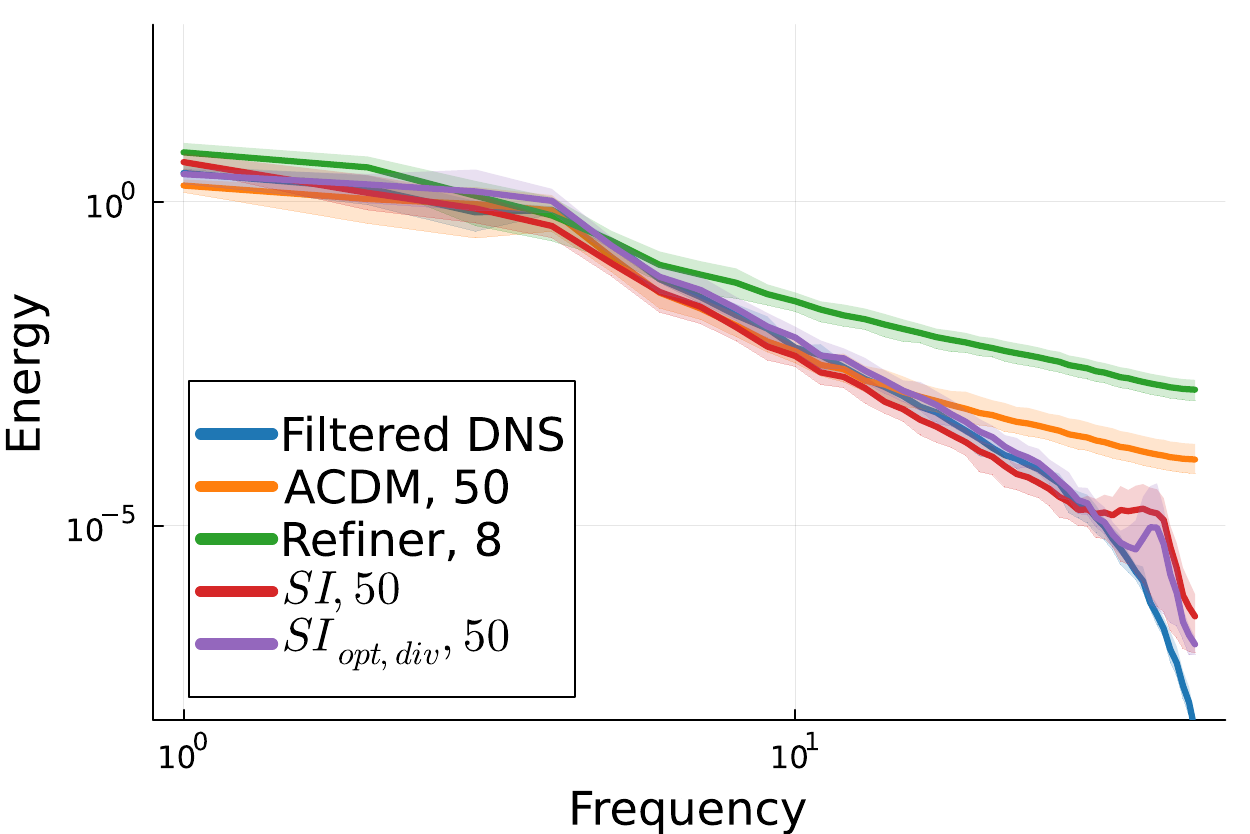}
        \caption{After $n=100$ time steps.}
    \end{subfigure}
    \begin{subfigure}[b]{0.32\textwidth}
        \centering
        \includegraphics[width=1.0\linewidth]{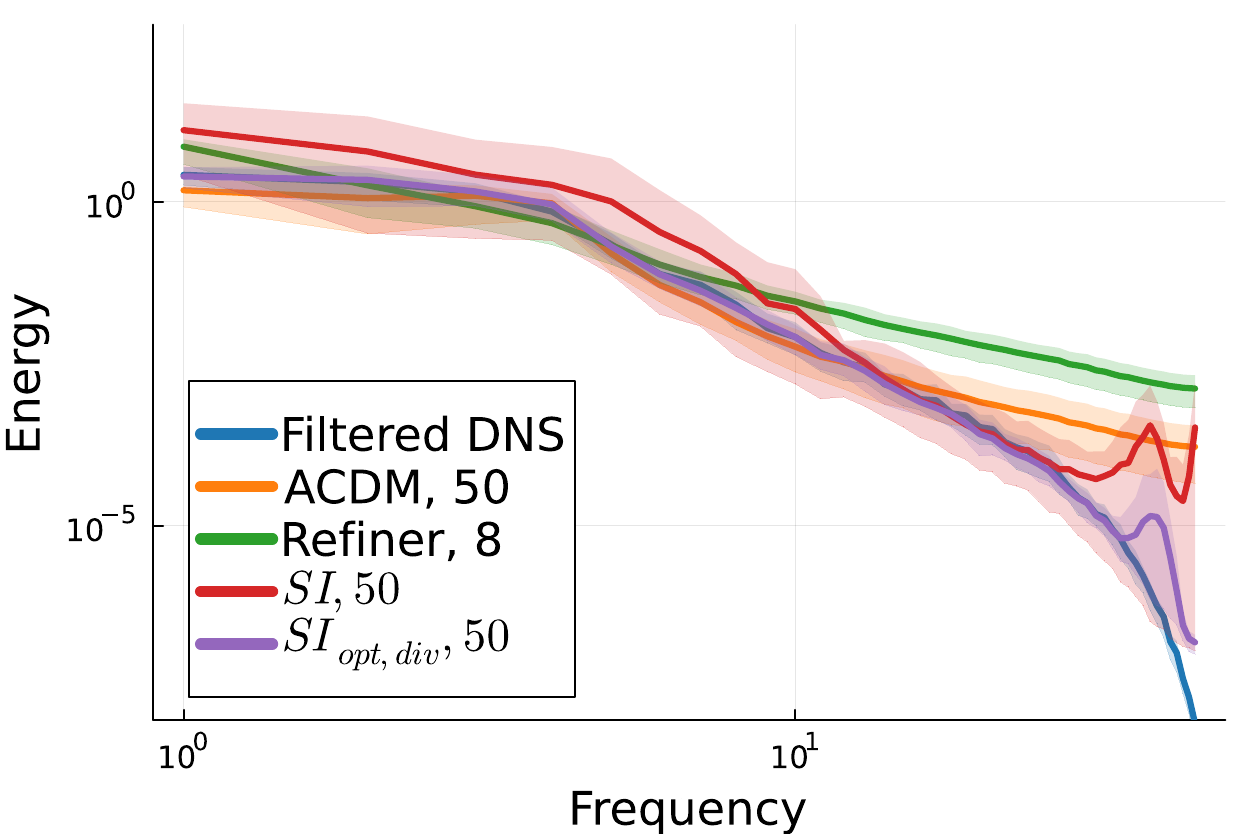}
        \caption{After $n=750$ time steps.}
    \end{subfigure}
    \caption{Energy spectra for various generative at three different physical time steps.}
    \label{fig:kolmogorov_spectra}
\end{figure*}

In Figure \ref{fig:kolmogorov_energy_evolution} we see that the alternative  methods either over- or undershoot the energy. Furthermore, in Figure \ref{fig:kolmogorov_energy_distributions} we clearly see that the distribution for the SI$_{\text{opt,div}}$ method matches the filtered DNS energy significantly better. In Figure \ref{fig:kolmogorov_spectra}, this is further emphasized as we see that the energy spectra are much better matched for both the low and high frequencies. Despite the better performance of the SI$_{\text{opt,div}}$ model, we still see a small bump in the high frequencies after a series of physical time steps. This suggests that high-frequency errors accumulate slightly with time. However, the difference between 100 and 750 time steps is small which supports the claim of stable long rollouts. In Figure \ref{fig:energy_spectra_all_SI_models} we also see that with more SDE pseudo-time steps this problem becomes smaller. 

\begin{figure*}[t!]
    \centering
    \begin{subfigure}[b]{0.55\textwidth}
        \centering
        \includegraphics[width=0.85\linewidth]{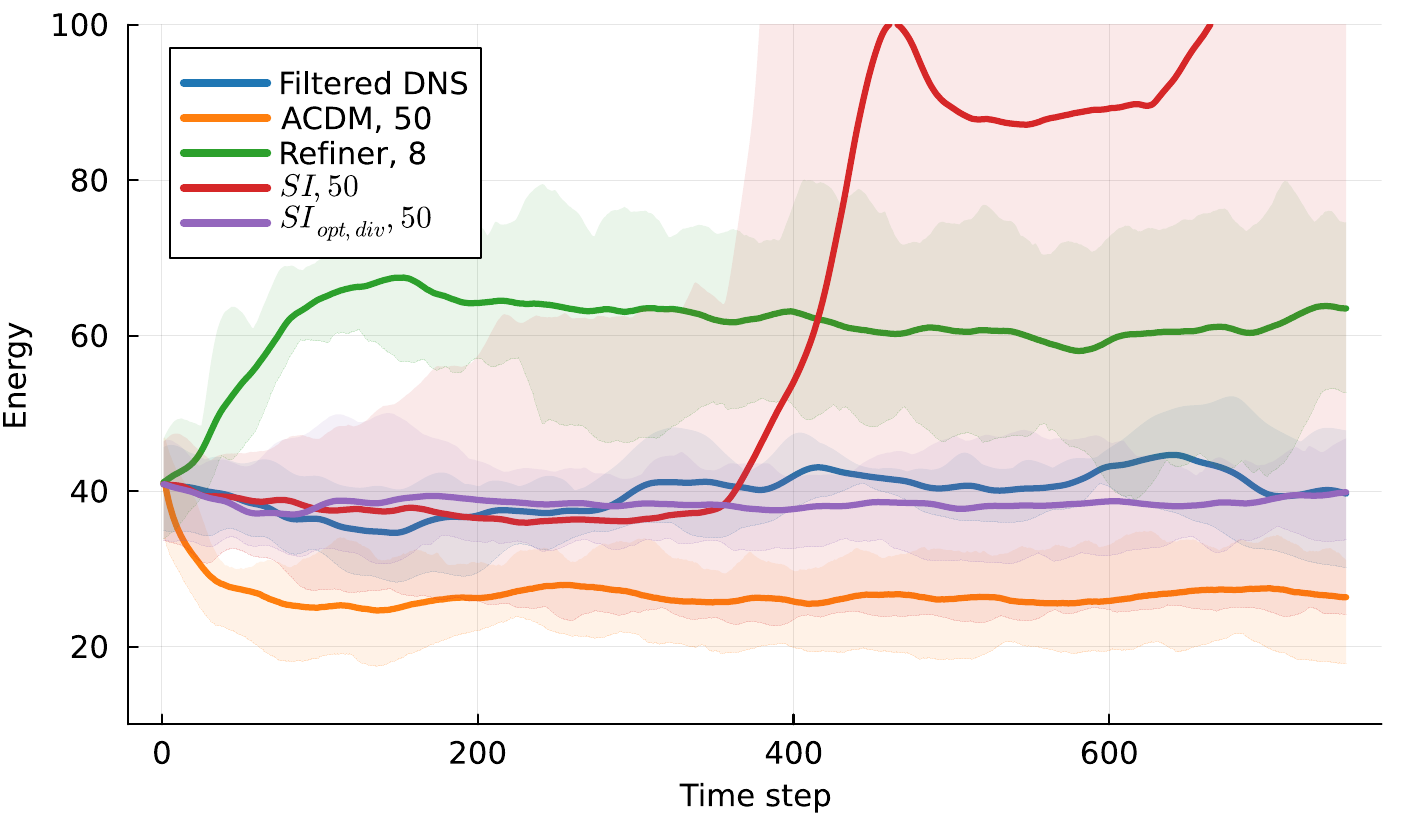}
        \caption{Time evolution of kinetic energy}
        \label{fig:kolmogorov_energy_evolution}
    \end{subfigure}
    \begin{subfigure}[b]{0.40\textwidth}
        \centering
        \includegraphics[width=1.0\linewidth]{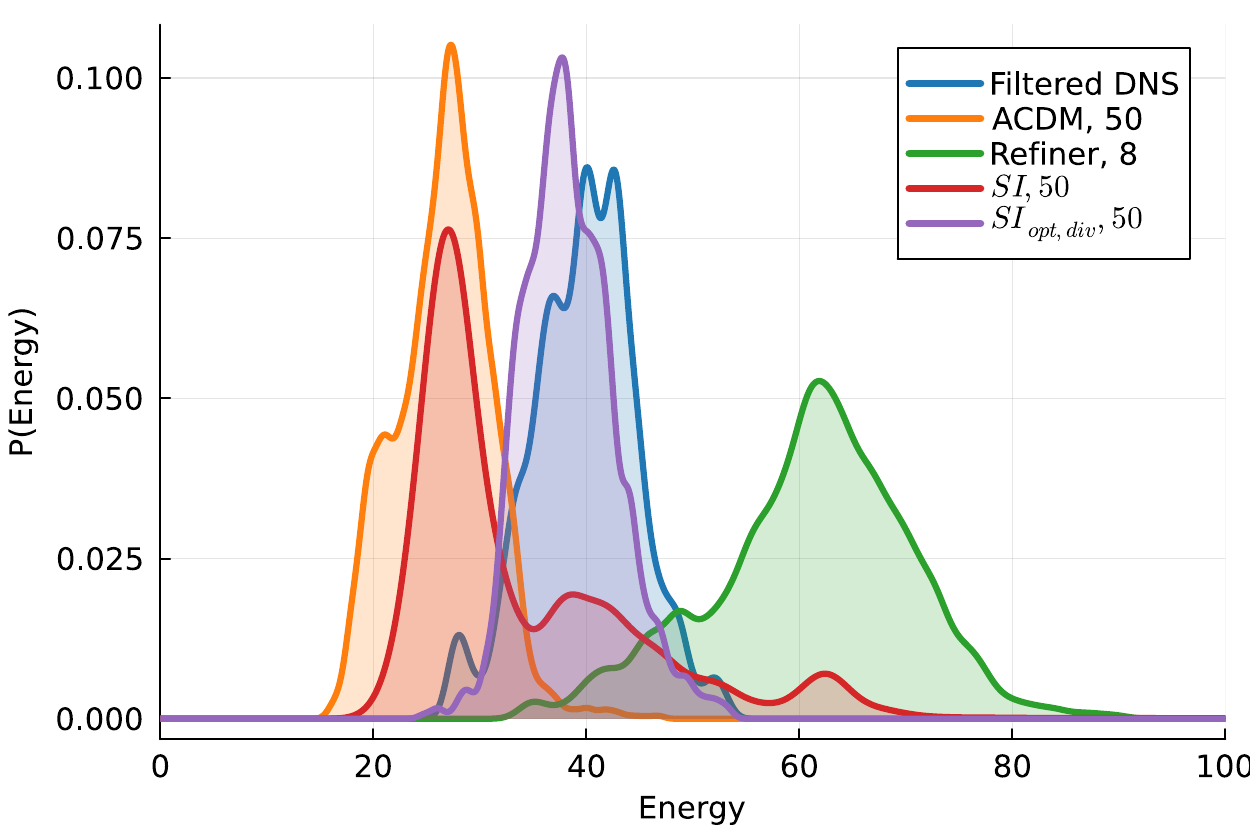}
        \caption{Kinetic energy distributions.}
        \label{fig:kolmogorov_energy_distributions}
    \end{subfigure}
    \caption{Kinetic energy results for various generative models.}
    \label{fig:kolmogorov_energy}
\end{figure*}

\begin{figure*}[t!]
    \centering
    \begin{subfigure}[b]{0.75\textwidth}
        \centering
        \includegraphics[width=1.0\linewidth]{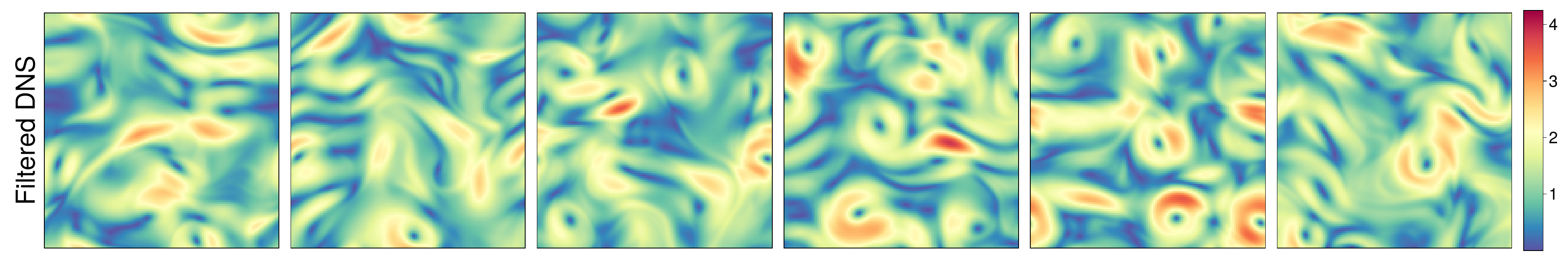}
    \end{subfigure}
    \begin{subfigure}[b]{0.75\textwidth}
        \centering
        \includegraphics[width=1.0\linewidth]{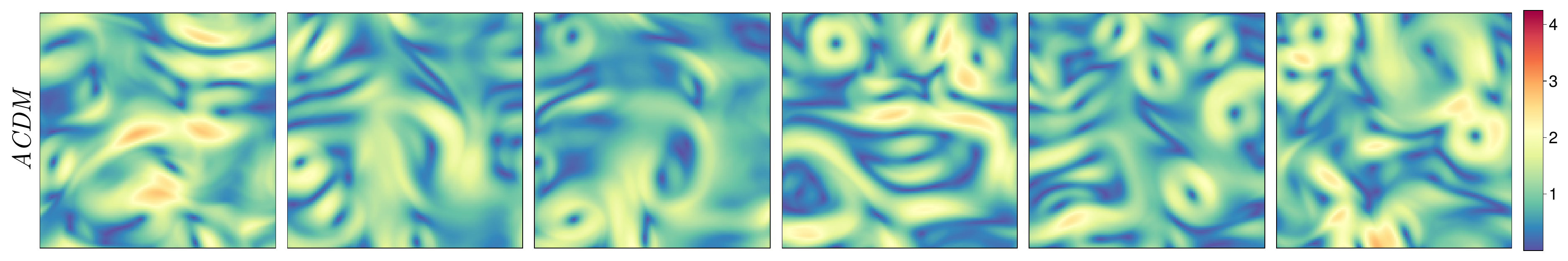}
    \end{subfigure}
    \begin{subfigure}[b]{0.75\textwidth}
        \centering
        \includegraphics[width=1.0\linewidth]{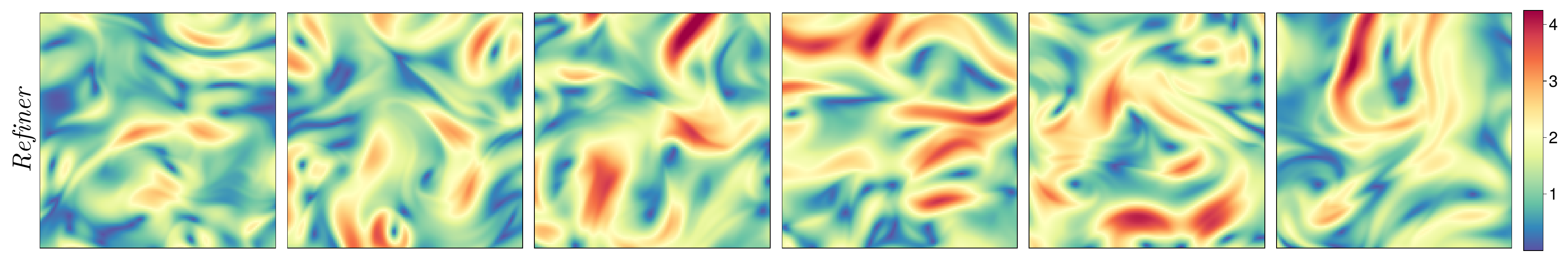}
    \end{subfigure}
    \begin{subfigure}[b]{0.75\textwidth}
        \centering
        \includegraphics[width=1.0\linewidth]{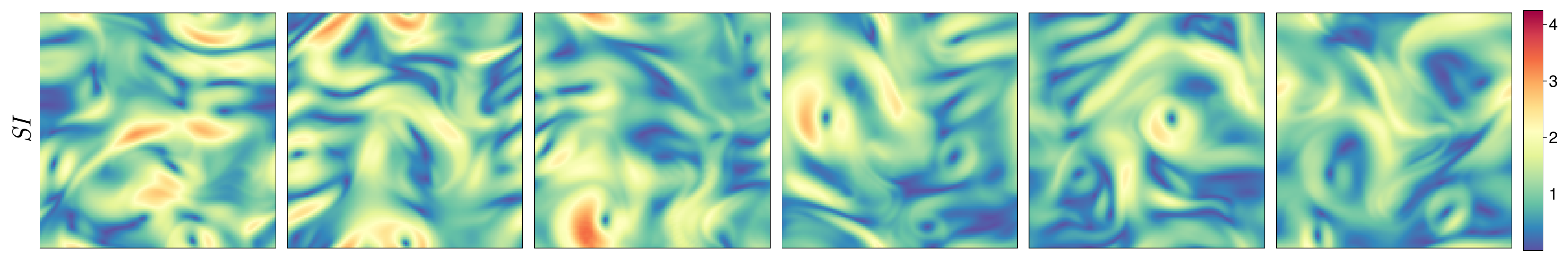}
    \end{subfigure}
    \begin{subfigure}[b]{0.75\textwidth}
        \centering
        \includegraphics[width=1.0\linewidth]{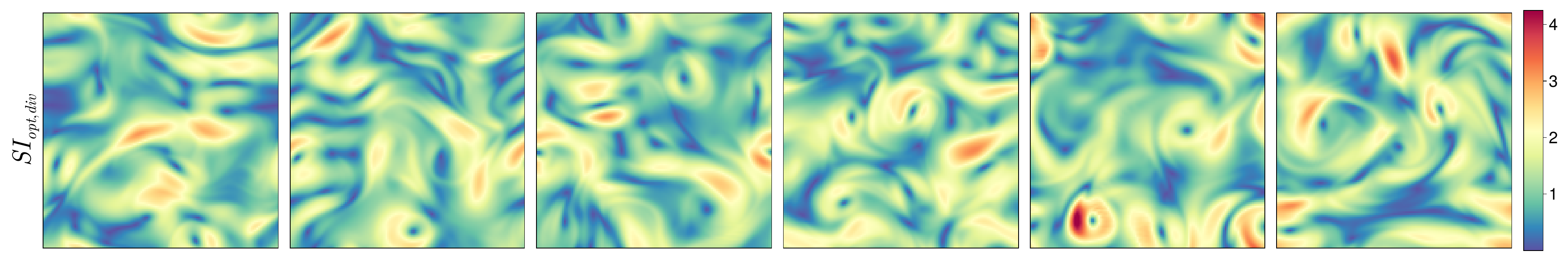}
    \end{subfigure}
    \caption{Velocity magnitude for the various models at different time steps. The same initial condition is used for all realizations. From left to right: $n=10$, $n=50$, $n=100$, $n=200$, $n=400$, $n=750$.}
    \label{fig:kolmogorov_velocity_magnitude}
\end{figure*}

In Figure \ref{fig:kolmogorov_velocity_magnitude} we see the velocity magnitude for various methods at 6 different physical time steps. Qualitatively, we see some differences between the various models. The ACDM and the SI seem generate slightly smoothed states. On the other hand, the Refiner results in states with significantly larger magnitudes. The SI$_{\text{opt,div}}$, however, generates trajectories that visually look physically plausible when comparing with the characteristics of the filtered DNS states. This is backed by the results in Figures \ref{fig:kolmogorov_energy} and \ref{fig:kolmogorov_spectra}. 

We summarize the results in Table \ref{tab:kolmogorov}. Here we see that SI$_{\text{opt,div}}$ model also performs quantitatively better than the alternatives. We see that the SI$_{\text{opt,div}}$ model approximates the energy distribution at least an order of magnitude better than all the alternatives. In particular, the SI$_{\text{opt,div}}$ model with only 10 SDE steps outperforms the rest. Furthermore, we see that the SI$_{\text{opt,div}}$ method achieves better LSiM accuracy for the first 50 time steps by an order of magnitude. For 750 steps, however, we see that all methods achieve similar accuracy. Since the system is chaotic, this is to be expected for long roll-outs. We see similar behaviour for the MSE. Lastly, we see that the SI$_{\text{opt,div}}$ method remains correlated with the filtered DNS solution for longer time than the other approaches. 

\begin{table}[]
    \centering
    \begin{tabular}{lcccccc}
    \toprule
        & & \multicolumn{2}{c}{LSiM $\downarrow$} & \multicolumn{2}{c}{MSE $\downarrow$} & \\
        & Energy W-1 $\downarrow$ & 50 steps & 750 steps & 50 steps & 750 steps & Corr > 0.8 time $\uparrow$ \\ \midrule
    ACDM, 10  & 24.0 $\cdot 10^7$  & 0.515 & 0.181 & 5.87 $\cdot 10^6$ & 6.1$\cdot 10^{6}$  & 0.02  \\
    ACDM, 25  &  16.871  & 0.183 & 0.154 &0.230 & \textbf{1.320} & 0.391  \\
    ACDM, 50  & 13.139  & 0.189  &0.149 &0.204 & \textbf{1.380} & 0.412  \\
    Refiner, 2   & 611.430  & 0.345 & 0.167 & 0.403& 17.3 &  0.33  \\
    Refiner, 4   & 8832.179  & 0.360 & 0.199 & 4.939 & 226.435 &  0.27  \\
    Refiner, 8   & 21.301  & 0.196 & 0.174  & 0.442 & 2.379  & 0.312  \\
    SI, 10  & 80.174  & 0.128 & 0.175 & 0.126 & 3.509 &  0.479  \\
    SI, 25  & 44.459  & 0.135 &0.168 & 0.136 & 2.591 &  0.477  \\
    SI, 50  & 37.866  & 0.132 &0.169 &0.132  & 2.438 &  0.477  \\
    SI$_{\text{opt,div}}$, 10  & 4.865 &  0.088 & 0.153 &0.043 & 1.561  & 0.703  \\
    SI$_{\text{opt,div}}$, 25  & \textbf{2.686} & \textbf{0.056} & 0.153 & \textbf{0.024} &  1.638  & \textbf{0.822}  \\
    SI$_{\text{opt,div}}$, 50  & \textbf{2.598} & \textbf{0.053} & 0.153 & \textbf{0.023}  & 1.65  & \textbf{0.841}  \\
    \bottomrule
    \end{tabular}
    \caption{Results for Kolmogorov. The arrow next to the metric denotes whether larger is better ($\uparrow$) or smaller is better ($\downarrow$). Note that we show results for MSE and LSiM averaged over 50 and 750 times steps. Since the Kolmogorov flow is highly chaotic, the long-term performance with respect to those metrics are not representative for the model performance alone. The number following the model is the amount of diffusion/SDE steps in the generation procedure. we highlight the best results in boldface. In cases where there is no significant difference between several results, we highlight all values that are approximately similar.}
    \label{tab:kolmogorov}
\end{table}

\FloatBarrier

\section{Conclusion}\label{section:conclusion}
In this work, we introduced a novel stochastic generative model for turbulence simulation, leveraging stochastic interpolants to enable probabilistic forecasting while maintaining physical consistency. Unlike conventional generative models, which often fail to incorporate physical constraints, our approach ensures energy-stable time stepping and divergence-free velocity fields, thereby improving both numerical stability and physical reliability. In particular, we have tuned the parameters of the stochastic interpolant in such a way that it is conserving kinetic energy, which is a crucial property in the incompressible Navier-Stokes equations. By training the interpolant on single time steps, we do not need unrolling over multiple time steps.

We demonstrate the effectiveness of our framework on Kolmogorov flow, where it outperforms state-of-the-art generative models, including autoregressive conditional diffusion models (ACDMs) and PDE-Refiners, in terms of energy conservation, spectral accuracy, and long-term stability. Our model not only achieves more accurate statistical properties but also allows for flexible inference, overcoming the rigid step-size constraints of diffusion-based methods.

Overall, our findings suggest that stochastic interpolants provide a promising foundation for physics-aware generative modeling in fluid dynamics. Two important limitations are: (i) our framework needs knowledge of $k_\tau$ (the average change of energy of the system), which in this article could be set to zero; (ii) we embed energy conservation as a soft constraint through parameterizing the interpolant, and not in a strong way (e.g.\ through parameterizing  the SDE or the NN). Future work will focus on applying the framework to more complex cases, including cases that do not necessarily reach a statistically stationary distribution. Furthermore, we will extend the framework to be able to handle to other relevant physical properties such as entropy, momentum, and symmetries.

\section*{Acknowledgment}
This research was funded by the National Growth Fund of the Netherlands and administered by the Netherlands Organisation for Scientific Research (NWO) under the AINed XS grant NGF.1609.242.037. The authors furthermore acknowledge the help and support of Syver Agdestein with the Julia package IncompressibleNavierStokes.jl.

\section*{CRediT authorship contribution statement}
\textbf{N. M{\"u}cke:} Conceptualization, methodology, software, formal analysis, writing - original draft. \textbf{B. Sanderse:} Writing -review \& editing, formal analysis, supervision, project administration.

\section*{Declaration of competing interest}
The authors declare that they have no competing financial or personal interests that have influenced the work presented in this paper.

\section*{Declaration of generative AI and AI-assisted technologies in the writing process} 
During the preparation of this work the authors used the Claude large language model inside the Cursor IDE to assist in writing code. After using this tool, the authors reviewed and edited the content as needed and takes full responsibility for the content of the published article.

\bibliographystyle{elsarticle-num}
\bibliography{references,library_BS, library_NTM}

\begin{thebibliography}{10}
\expandafter\ifx\csname url\endcsname\relax
  \def\url#1{\texttt{#1}}\fi
\expandafter\ifx\csname urlprefix\endcsname\relax\def\urlprefix{URL }\fi
\expandafter\ifx\csname href\endcsname\relax
  \def\href#1#2{#2} \def\path#1{#1}\fi

\bibitem{ray_chatgpt_2023}
P.~P. Ray, \href{https://www.sciencedirect.com/science/article/pii/S266734522300024X}{{ChatGPT}: {A} comprehensive review on background, applications, key challenges, bias, ethics, limitations and future scope}, Internet of Things and Cyber-Physical Systems 3 (2023) 121--154.
\newblock \href {https://doi.org/10.1016/j.iotcps.2023.04.003} {\path{doi:10.1016/j.iotcps.2023.04.003}}.
\newline\urlprefix\url{https://www.sciencedirect.com/science/article/pii/S266734522300024X}

\bibitem{liu_sora_2024}
Y.~Liu, K.~Zhang, Y.~Li, Z.~Yan, C.~Gao, R.~Chen, Z.~Yuan, Y.~Huang, H.~Sun, J.~Gao, L.~He, L.~Sun, \href{http://arxiv.org/abs/2402.17177}{Sora: {A} {Review} on {Background}, {Technology}, {Limitations}, and {Opportunities} of {Large} {Vision} {Models}}, arXiv:2402.17177 [cs] (Apr. 2024).
\newblock \href {https://doi.org/10.48550/arXiv.2402.17177} {\path{doi:10.48550/arXiv.2402.17177}}.
\newline\urlprefix\url{http://arxiv.org/abs/2402.17177}

\bibitem{ramesh_hierarchical_2022}
A.~Ramesh, P.~Dhariwal, A.~Nichol, C.~Chu, M.~Chen, \href{http://arxiv.org/abs/2204.06125}{Hierarchical {Text}-{Conditional} {Image} {Generation} with {CLIP} {Latents}}, arXiv:2204.06125 [cs] (Apr. 2022).
\newblock \href {https://doi.org/10.48550/arXiv.2204.06125} {\path{doi:10.48550/arXiv.2204.06125}}.
\newline\urlprefix\url{http://arxiv.org/abs/2204.06125}

\bibitem{chen2023b}
N.~Chen, Stochastic {{Methods}} for {{Modeling}} and {{Predicting Complex Dynamical Systems}}: {{Uncertainty Quantification}}, {{State Estimation}}, and {{Reduced-Order Models}}, Synthesis {{Lectures}} on {{Mathematics}} \& {{Statistics}}, Springer International Publishing, Cham, 2023.
\newblock \href {https://doi.org/10.1007/978-3-031-22249-8} {\path{doi:10.1007/978-3-031-22249-8}}.

\bibitem{bodnar2024}
C.~Bodnar, W.~P. Bruinsma, A.~Lucic, M.~Stanley, J.~Brandstetter, P.~Garvan, M.~Riechert, J.~Weyn, H.~Dong, A.~Vaughan, J.~K. Gupta, K.~Tambiratnam, A.~Archibald, E.~Heider, M.~Welling, R.~E. Turner, P.~Perdikaris, Aurora: {{A Foundation Model}} of the {{Atmosphere}} (May 2024).
\newblock \href {http://arxiv.org/abs/2405.13063} {\path{arXiv:2405.13063}}, \href {https://doi.org/10.48550/arXiv.2405.13063} {\path{doi:10.48550/arXiv.2405.13063}}.

\bibitem{pathak2022}
J.~Pathak, S.~Subramanian, P.~Harrington, S.~Raja, A.~Chattopadhyay, M.~Mardani, T.~Kurth, D.~Hall, Z.~Li, K.~Azizzadenesheli, P.~Hassanzadeh, K.~Kashinath, A.~Anandkumar, {{FourCastNet}}: {{A Global Data-driven High-resolution Weather Model}} using {{Adaptive Fourier Neural Operators}} (Feb. 2022).
\newblock \href {http://arxiv.org/abs/2202.11214} {\path{arXiv:2202.11214}}, \href {https://doi.org/10.48550/arXiv.2202.11214} {\path{doi:10.48550/arXiv.2202.11214}}.

\bibitem{nguyen2023b}
T.~Nguyen, J.~Brandstetter, A.~Kapoor, J.~K. Gupta, A.~Grover, {{ClimaX}}: {{A}} foundation model for weather and climate (Dec. 2023).
\newblock \href {http://arxiv.org/abs/2301.10343} {\path{arXiv:2301.10343}}, \href {https://doi.org/10.48550/arXiv.2301.10343} {\path{doi:10.48550/arXiv.2301.10343}}.

\bibitem{lam2023}
R.~Lam, A.~{Sanchez-Gonzalez}, M.~Willson, P.~Wirnsberger, M.~Fortunato, F.~Alet, S.~Ravuri, T.~Ewalds, Z.~{Eaton-Rosen}, W.~Hu, A.~Merose, S.~Hoyer, G.~Holland, O.~Vinyals, J.~Stott, A.~Pritzel, S.~Mohamed, P.~Battaglia, {{GraphCast}}: {{Learning}} skillful medium-range global weather forecasting (Aug. 2023).
\newblock \href {http://arxiv.org/abs/2212.12794} {\path{arXiv:2212.12794}}, \href {https://doi.org/10.48550/arXiv.2212.12794} {\path{doi:10.48550/arXiv.2212.12794}}.

\bibitem{bommasani2022}
R.~Bommasani, D.~A. Hudson, E.~Adeli, R.~Altman, S.~Arora, S.~von Arx, M.~S. Bernstein, J.~Bohg, A.~Bosselut, E.~Brunskill, E.~Brynjolfsson, S.~Buch, D.~Card, R.~Castellon, N.~Chatterji, A.~Chen, K.~Creel, J.~Q. Davis, D.~Demszky, C.~Donahue, M.~Doumbouya, E.~Durmus, S.~Ermon, J.~Etchemendy, K.~Ethayarajh, L.~{Fei-Fei}, C.~Finn, T.~Gale, L.~Gillespie, K.~Goel, N.~Goodman, S.~Grossman, N.~Guha, T.~Hashimoto, P.~Henderson, J.~Hewitt, D.~E. Ho, J.~Hong, K.~Hsu, J.~Huang, T.~Icard, S.~Jain, D.~Jurafsky, P.~Kalluri, S.~Karamcheti, G.~Keeling, F.~Khani, O.~Khattab, P.~W. Koh, M.~Krass, R.~Krishna, R.~Kuditipudi, A.~Kumar, F.~Ladhak, M.~Lee, T.~Lee, J.~Leskovec, I.~Levent, X.~L. Li, X.~Li, T.~Ma, A.~Malik, C.~D. Manning, S.~Mirchandani, E.~Mitchell, Z.~Munyikwa, S.~Nair, A.~Narayan, D.~Narayanan, B.~Newman, A.~Nie, J.~C. Niebles, H.~Nilforoshan, J.~Nyarko, G.~Ogut, L.~Orr, I.~Papadimitriou, J.~S. Park, C.~Piech, E.~Portelance, C.~Potts, A.~Raghunathan, R.~Reich, H.~Ren, F.~Rong, Y.~Roohani, C.~Ruiz, J.~Ryan,
  C.~R{\'e}, D.~Sadigh, S.~Sagawa, K.~Santhanam, A.~Shih, K.~Srinivasan, A.~Tamkin, R.~Taori, A.~W. Thomas, F.~Tram{\`e}r, R.~E. Wang, W.~Wang, B.~Wu, J.~Wu, Y.~Wu, S.~M. Xie, M.~Yasunaga, J.~You, M.~Zaharia, M.~Zhang, T.~Zhang, X.~Zhang, Y.~Zhang, L.~Zheng, K.~Zhou, P.~Liang, On the {{Opportunities}} and {{Risks}} of {{Foundation Models}} (Jul. 2022).
\newblock \href {http://arxiv.org/abs/2108.07258} {\path{arXiv:2108.07258}}, \href {https://doi.org/10.48550/arXiv.2108.07258} {\path{doi:10.48550/arXiv.2108.07258}}.

\bibitem{batatia2024}
I.~Batatia, P.~Benner, Y.~Chiang, A.~M. Elena, D.~P. Kov{\'a}cs, J.~Riebesell, X.~R. Advincula, M.~Asta, M.~Avaylon, W.~J. Baldwin, F.~Berger, N.~Bernstein, A.~Bhowmik, S.~M. Blau, V.~C{\u a}rare, J.~P. Darby, S.~De, F.~D. Pia, V.~L. Deringer, R.~Elijo{\v s}ius, Z.~{El-Machachi}, F.~Falcioni, E.~Fako, A.~C. Ferrari, A.~{Genreith-Schriever}, J.~George, R.~E.~A. Goodall, C.~P. Grey, P.~Grigorev, S.~Han, W.~Handley, H.~H. Heenen, K.~Hermansson, C.~Holm, J.~Jaafar, S.~Hofmann, K.~S. Jakob, H.~Jung, V.~Kapil, A.~D. Kaplan, N.~Karimitari, J.~R. Kermode, N.~Kroupa, J.~Kullgren, M.~C. Kuner, D.~Kuryla, G.~Liepuoniute, J.~T. Margraf, I.-B. Magd{\u a}u, A.~Michaelides, J.~H. Moore, A.~A. Naik, S.~P. Niblett, S.~W. Norwood, N.~O'Neill, C.~Ortner, K.~A. Persson, K.~Reuter, A.~S. Rosen, L.~L. Schaaf, C.~Schran, B.~X. Shi, E.~Sivonxay, T.~K. Stenczel, V.~Svahn, C.~Sutton, T.~D. Swinburne, J.~Tilly, C.~van~der Oord, E.~{Varga-Umbrich}, T.~Vegge, M.~Vondr{\'a}k, Y.~Wang, W.~C. Witt, F.~Zills, G.~Cs{\'a}nyi, A foundation
  model for atomistic materials chemistry (Mar. 2024).
\newblock \href {http://arxiv.org/abs/2401.00096} {\path{arXiv:2401.00096}}, \href {https://doi.org/10.48550/arXiv.2401.00096} {\path{doi:10.48550/arXiv.2401.00096}}.

\bibitem{rosen2023}
Y.~Rosen, Y.~Roohani, A.~Agarwal, L.~Samotor{\v c}an, T.~S. Consortium, S.~R. Quake, J.~Leskovec, Universal {{Cell Embeddings}}: {{A Foundation Model}} for {{Cell Biology}} (Nov. 2023).
\newblock \href {https://doi.org/10.1101/2023.11.28.568918} {\path{doi:10.1101/2023.11.28.568918}}.

\bibitem{herde2024}
M.~Herde, B.~Raoni{\'c}, T.~Rohner, R.~K{\"a}ppeli, R.~Molinaro, E.~{de B{\'e}zenac}, S.~Mishra, Poseidon: {{Efficient Foundation Models}} for {{PDEs}} (May 2024).
\newblock \href {http://arxiv.org/abs/2405.19101} {\path{arXiv:2405.19101}}, \href {https://doi.org/10.48550/arXiv.2405.19101} {\path{doi:10.48550/arXiv.2405.19101}}.

\bibitem{price2024}
I.~Price, A.~{Sanchez-Gonzalez}, F.~Alet, T.~R. Andersson, A.~{El-Kadi}, D.~Masters, T.~Ewalds, J.~Stott, S.~Mohamed, P.~Battaglia, R.~Lam, M.~Willson, {{GenCast}}: {{Diffusion-based}} ensemble forecasting for medium-range weather (May 2024).
\newblock \href {http://arxiv.org/abs/2312.15796} {\path{arXiv:2312.15796}}, \href {https://doi.org/10.48550/arXiv.2312.15796} {\path{doi:10.48550/arXiv.2312.15796}}.

\bibitem{keisler2022}
R.~Keisler, Forecasting {{Global Weather}} with {{Graph Neural Networks}} (Feb. 2022).
\newblock \href {http://arxiv.org/abs/2202.07575} {\path{arXiv:2202.07575}}, \href {https://doi.org/10.48550/arXiv.2202.07575} {\path{doi:10.48550/arXiv.2202.07575}}.

\bibitem{li2024e}
L.~Li, R.~Carver, I.~{Lopez-Gomez}, F.~Sha, J.~Anderson, Generative emulation of weather forecast ensembles with diffusion models, Science Advances 10~(13) (2024) eadk4489.
\newblock \href {https://doi.org/10.1126/sciadv.adk4489} {\path{doi:10.1126/sciadv.adk4489}}.

\bibitem{kochkov2024}
D.~Kochkov, J.~Yuval, I.~Langmore, P.~Norgaard, J.~Smith, G.~Mooers, M.~Kl{\"o}wer, J.~Lottes, S.~Rasp, P.~D{\"u}ben, S.~Hatfield, P.~Battaglia, A.~{Sanchez-Gonzalez}, M.~Willson, M.~P. Brenner, S.~Hoyer, Neural general circulation models for weather and climate, Nature 632~(8027) (2024) 1060--1066.
\newblock \href {https://doi.org/10.1038/s41586-024-07744-y} {\path{doi:10.1038/s41586-024-07744-y}}.

\bibitem{pope2004}
S.~B. Pope, Ten questions concerning the large-eddy simulation of turbulent flows, New Journal of Physics 6~(1) (2004) 35.
\newblock \href {https://doi.org/10.1088/1367-2630/6/1/035} {\path{doi:10.1088/1367-2630/6/1/035}}.

\bibitem{berner2017}
J.~Berner, U.~Achatz, L.~Batt{\'e}, L.~Bengtsson, A.~de~la C{\'a}mara, H.~M. Christensen, M.~Colangeli, D.~R.~B. Coleman, D.~Crommelin, S.~I. Dolaptchiev, C.~L.~E. Franzke, P.~Friederichs, P.~Imkeller, H.~J{\"a}rvinen, S.~Juricke, V.~Kitsios, F.~Lott, V.~Lucarini, S.~Mahajan, T.~N. Palmer, C.~Penland, M.~Sakradzija, J.-S. von Storch, A.~Weisheimer, M.~Weniger, P.~D. Williams, J.-I. Yano, Stochastic {{Parameterization}}: {{Toward}} a {{New View}} of {{Weather}} and {{Climate Models}}, Bulletin of the American Meteorological Society 98~(3) (2017) 565--588.
\newblock \href {https://doi.org/10.1175/BAMS-D-15-00268.1} {\path{doi:10.1175/BAMS-D-15-00268.1}}.

\bibitem{li_scalable_2020}
X.~Li, T.-K.~L. Wong, R.~T.~Q. Chen, D.~Duvenaud, \href{http://arxiv.org/abs/2001.01328}{Scalable {Gradients} for {Stochastic} {Differential} {Equations}}, arXiv:2001.01328 [cs] (Oct. 2020).
\newblock \href {https://doi.org/10.48550/arXiv.2001.01328} {\path{doi:10.48550/arXiv.2001.01328}}.
\newline\urlprefix\url{http://arxiv.org/abs/2001.01328}

\bibitem{boral2023}
A.~Boral, Z.~Y. Wan, L.~{Zepeda-N{\'u}{\~n}ez}, J.~Lottes, Q.~Wang, Y.-f. Chen, J.~R. Anderson, F.~Sha, Neural {{Ideal Large Eddy Simulation}}: {{Modeling Turbulence}} with {{Neural Stochastic Differential Equations}} (Jun. 2023).
\newblock \href {http://arxiv.org/abs/2306.01174} {\path{arXiv:2306.01174}}, \href {https://doi.org/10.48550/arXiv.2306.01174} {\path{doi:10.48550/arXiv.2306.01174}}.

\bibitem{kohl2024}
G.~Kohl, L.-W. Chen, N.~Thuerey, Benchmarking {{Autoregressive Conditional Diffusion Models}} for {{Turbulent Flow Simulation}} (Jan. 2024).
\newblock \href {http://arxiv.org/abs/2309.01745} {\path{arXiv:2309.01745}}, \href {https://doi.org/10.48550/arXiv.2309.01745} {\path{doi:10.48550/arXiv.2309.01745}}.

\bibitem{ho2020}
J.~Ho, A.~Jain, P.~Abbeel, Denoising {{Diffusion Probabilistic Models}} (Dec. 2020).
\newblock \href {http://arxiv.org/abs/2006.11239} {\path{arXiv:2006.11239}}, \href {https://doi.org/10.48550/arXiv.2006.11239} {\path{doi:10.48550/arXiv.2006.11239}}.

\bibitem{song_score-based_2021}
Y.~Song, J.~Sohl-Dickstein, D.~P. Kingma, A.~Kumar, S.~Ermon, B.~Poole, \href{http://arxiv.org/abs/2011.13456}{Score-{Based} {Generative} {Modeling} through {Stochastic} {Differential} {Equations}}, arXiv:2011.13456 [cs] (Feb. 2021).
\newblock \href {https://doi.org/10.48550/arXiv.2011.13456} {\path{doi:10.48550/arXiv.2011.13456}}.
\newline\urlprefix\url{http://arxiv.org/abs/2011.13456}

\bibitem{albergo2023a}
M.~S. Albergo, E.~{Vanden-Eijnden}, Building {{Normalizing Flows}} with {{Stochastic Interpolants}} (Mar. 2023).
\newblock \href {http://arxiv.org/abs/2209.15571} {\path{arXiv:2209.15571}}, \href {https://doi.org/10.48550/arXiv.2209.15571} {\path{doi:10.48550/arXiv.2209.15571}}.

\bibitem{albergo2023stochastic}
M.~S. Albergo, N.~M. Boffi, E.~Vanden-Eijnden, Stochastic interpolants: A unifying framework for flows and diffusions, arXiv preprint arXiv:2303.08797 (2023).

\bibitem{chen2024probabilistic}
Y.~Chen, M.~Goldstein, M.~Hua, M.~S. Albergo, N.~M. Boffi, E.~Vanden-Eijnden, Probabilistic forecasting with stochastic interpolants and f$\backslash$" ollmer processes, arXiv preprint arXiv:2403.13724 (2024).

\bibitem{vangastelen2024}
T.~{van Gastelen}, W.~Edeling, B.~Sanderse, Energy-conserving neural network for turbulence closure modeling, Journal of Computational Physics 508 (2024) 113003.
\newblock \href {https://doi.org/10.1016/j.jcp.2024.113003} {\path{doi:10.1016/j.jcp.2024.113003}}.

\bibitem{sanderse2020}
B.~Sanderse, Non-linearly stable reduced-order models for incompressible flow with energy-conserving finite volume methods, Journal of Computational Physics 421 (2020) 109736.
\newblock \href {https://doi.org/10.1016/j.jcp.2020.109736} {\path{doi:10.1016/j.jcp.2020.109736}}.

\bibitem{agdestein2025}
S.~D. Agdestein, B.~Sanderse, Discretize first, filter next: {{Learning}} divergence-consistent closure models for large-eddy simulation, Journal of Computational Physics 522 (2025) 113577.
\newblock \href {https://doi.org/10.1016/j.jcp.2024.113577} {\path{doi:10.1016/j.jcp.2024.113577}}.

\bibitem{park2021}
J.~Park, H.~Choi, Toward neural-network-based large eddy simulation: Application to turbulent channel flow, Journal of Fluid Mechanics 914 (2021) A16.
\newblock \href {https://doi.org/10.1017/jfm.2020.931} {\path{doi:10.1017/jfm.2020.931}}.

\bibitem{kurz2021}
M.~Kurz, A.~Beck, Investigating {{Model-Data Inconsistency}} in {{Data-Informed Turbulence Closure Terms}}, 14th WCCM-ECCOMAS Congress 2020 (Mar. 2021).
\newblock \href {https://doi.org/10.23967/wccm-eccomas.2020.115} {\path{doi:10.23967/wccm-eccomas.2020.115}}.

\bibitem{rasp2020}
S.~Rasp, Coupled online learning as a way to tackle instabilities and biases in neural network parameterizations: General algorithms and {{Lorenz}} 96 case study (v1.0), Geoscientific Model Development 13~(5) (2020) 2185--2196.
\newblock \href {https://doi.org/10.5194/gmd-13-2185-2020} {\path{doi:10.5194/gmd-13-2185-2020}}.

\bibitem{sanderse2024a}
B.~Sanderse, P.~Stinis, R.~Maulik, S.~E. Ahmed, Scientific machine learning for closure models in multiscale problems: A review, Foundations of Data Science 7~(1) (2024) 298--337.
\newblock \href {https://doi.org/10.3934/fods.2024043} {\path{doi:10.3934/fods.2024043}}.

\bibitem{chen_probabilistic_2024}
Y.~Chen, M.~Goldstein, M.~Hua, M.~S. Albergo, N.~M. Boffi, E.~Vanden-Eijnden, \href{http://arxiv.org/abs/2403.13724}{Probabilistic {Forecasting} with {Stochastic} {Interpolants} and {Föllmer} {Processes}}, arXiv:2403.13724 [cs] (Aug. 2024).
\newblock \href {https://doi.org/10.48550/arXiv.2403.13724} {\path{doi:10.48550/arXiv.2403.13724}}.
\newline\urlprefix\url{http://arxiv.org/abs/2403.13724}

\bibitem{sagaut2006}
P.~Sagaut, Large Eddy Simulation for Incompressible Flows: An Introduction, 3rd Edition, Scientific Computation, Springer, Berlin ; New York, 2006.

\bibitem{pope2000}
S.~B. Pope, Turbulent Flows, Cambridge University Press, 2000.

\bibitem{ahmed2021}
S.~E. Ahmed, S.~Pawar, O.~San, A.~Rasheed, T.~Iliescu, B.~R. Noack, On closures for reduced order models - {{A}} spectrum of first-principle to machine-learned avenues, Physics of Fluids 33~(9) (2021) 091301.
\newblock \href {http://arxiv.org/abs/2106.14954} {\path{arXiv:2106.14954}}, \href {https://doi.org/10.1063/5.0061577} {\path{doi:10.1063/5.0061577}}.

\bibitem{kohl_benchmarking_2024}
G.~Kohl, L.-W. Chen, N.~Thuerey, \href{http://arxiv.org/abs/2309.01745}{Benchmarking {Autoregressive} {Conditional} {Diffusion} {Models} for {Turbulent} {Flow} {Simulation}}, arXiv:2309.01745 [cs] version: 2 (Jan. 2024).
\newblock \href {https://doi.org/10.48550/arXiv.2309.01745} {\path{doi:10.48550/arXiv.2309.01745}}.
\newline\urlprefix\url{http://arxiv.org/abs/2309.01745}

\bibitem{dong2024a}
X.~Dong, C.~Chen, J.-L. Wu, Data-{{Driven Stochastic Closure Modeling}} via {{Conditional Diffusion Model}} and {{Neural Operator}} (Aug. 2024).
\newblock \href {http://arxiv.org/abs/2408.02965} {\path{arXiv:2408.02965}}, \href {https://doi.org/10.48550/arXiv.2408.02965} {\path{doi:10.48550/arXiv.2408.02965}}.

\bibitem{molinaro2025}
R.~Molinaro, S.~Lanthaler, B.~Raoni{\'c}, T.~Rohner, V.~Armegioiu, S.~Simonis, D.~Grund, Y.~Ramic, Z.~Y. Wan, F.~Sha, S.~Mishra, L.~{Zepeda-N{\'u}{\~n}ez}, Generative {{AI}} for fast and accurate statistical computation of fluids (Feb. 2025).
\newblock \href {http://arxiv.org/abs/2409.18359} {\path{arXiv:2409.18359}}, \href {https://doi.org/10.48550/arXiv.2409.18359} {\path{doi:10.48550/arXiv.2409.18359}}.

\bibitem{albergo2023}
M.~S. Albergo, N.~M. Boffi, E.~{Vanden-Eijnden}, Stochastic {{Interpolants}}: {{A Unifying Framework}} for {{Flows}} and {{Diffusions}} (Nov. 2023).
\newblock \href {http://arxiv.org/abs/2303.08797} {\path{arXiv:2303.08797}}, \href {https://doi.org/10.48550/arXiv.2303.08797} {\path{doi:10.48550/arXiv.2303.08797}}.

\bibitem{chen2024c}
Y.~Chen, M.~Goldstein, M.~Hua, M.~S. Albergo, N.~M. Boffi, E.~{Vanden-Eijnden}, Probabilistic {{Forecasting}} with {{Stochastic Interpolants}} and {{F{\"o}llmer Processes}} (Mar. 2024).
\newblock \href {http://arxiv.org/abs/2403.13724} {\path{arXiv:2403.13724}}, \href {https://doi.org/10.48550/arXiv.2403.13724} {\path{doi:10.48550/arXiv.2403.13724}}.

\bibitem{beck_deep_2019}
A.~Beck, D.~Flad, C.-D. Munz, \href{https://www.sciencedirect.com/science/article/pii/S0021999119306151}{Deep neural networks for data-driven {LES} closure models}, Journal of Computational Physics 398 (2019) 108910.
\newblock \href {https://doi.org/10.1016/j.jcp.2019.108910} {\path{doi:10.1016/j.jcp.2019.108910}}.
\newline\urlprefix\url{https://www.sciencedirect.com/science/article/pii/S0021999119306151}

\bibitem{melchers2023}
H.~Melchers, D.~Crommelin, B.~Koren, V.~Menkovski, B.~Sanderse, Comparison of neural closure models for discretised {{PDEs}}, Computers \& Mathematics with Applications 143 (2023) 94--107.
\newblock \href {https://doi.org/10.1016/j.camwa.2023.04.030} {\path{doi:10.1016/j.camwa.2023.04.030}}.

\bibitem{brantner_volume-preserving_2024}
B.~Brantner, G.~d. Romemont, M.~Kraus, Z.~Li, \href{http://arxiv.org/abs/2312.11166}{Volume-{Preserving} {Transformers} for {Learning} {Time} {Series} {Data} with {Structure}}, arXiv:2312.11166 [math] (Nov. 2024).
\newblock \href {https://doi.org/10.48550/arXiv.2312.11166} {\path{doi:10.48550/arXiv.2312.11166}}.
\newline\urlprefix\url{http://arxiv.org/abs/2312.11166}

\bibitem{dener_training_2020}
A.~Dener, M.~A. Miller, R.~M. Churchill, T.~Munson, C.-S. Chang, \href{http://arxiv.org/abs/2009.07330}{Training neural networks under physical constraints using a stochastic augmented {Lagrangian} approach}, arXiv:2009.07330 [physics] (Sep. 2020).
\newblock \href {https://doi.org/10.48550/arXiv.2009.07330} {\path{doi:10.48550/arXiv.2009.07330}}.
\newline\urlprefix\url{http://arxiv.org/abs/2009.07330}

\bibitem{vangastelen2023}
T.~{van Gastelen}, W.~Edeling, B.~Sanderse, Energy-{{Conserving Neural Network}} for {{Turbulence Closure Modeling}} (Feb. 2023).
\newblock \href {http://arxiv.org/abs/2301.13770} {\path{arXiv:2301.13770}}, \href {https://doi.org/10.48550/arXiv.2301.13770} {\path{doi:10.48550/arXiv.2301.13770}}.

\bibitem{foias2001navier}
C.~Foias, O.~Manley, R.~Rosa, R.~Temam, Navier-Stokes equations and turbulence, Vol.~83, Cambridge University Press, 2001.

\bibitem{albergo_building_2023}
M.~S. Albergo, E.~Vanden-Eijnden, \href{http://arxiv.org/abs/2209.15571}{Building {Normalizing} {Flows} with {Stochastic} {Interpolants}}, arXiv:2209.15571 [cs] (Mar. 2023).
\newblock \href {https://doi.org/10.48550/arXiv.2209.15571} {\path{doi:10.48550/arXiv.2209.15571}}.
\newline\urlprefix\url{http://arxiv.org/abs/2209.15571}

\bibitem{benamou_computational_2000}
J.-D. Benamou, Y.~Brenier, \href{http://link.springer.com/10.1007/s002110050002}{A computational fluid mechanics solution to the {Monge}-{Kantorovich} mass transfer problem}, Numerische Mathematik 84~(3) (2000) 375--393.
\newblock \href {https://doi.org/10.1007/s002110050002} {\path{doi:10.1007/s002110050002}}.
\newline\urlprefix\url{http://link.springer.com/10.1007/s002110050002}

\bibitem{lippe2023}
P.~Lippe, B.~S. Veeling, P.~Perdikaris, R.~E. Turner, J.~Brandstetter, {{PDE-Refiner}}: {{Achieving Accurate Long Rollouts}} with {{Neural PDE Solvers}} (Oct. 2023).
\newblock \href {http://arxiv.org/abs/2308.05732} {\path{arXiv:2308.05732}}, \href {https://doi.org/10.48550/arXiv.2308.05732} {\path{doi:10.48550/arXiv.2308.05732}}.

\bibitem{ronneberger_u-net_2015}
O.~Ronneberger, P.~Fischer, T.~Brox, \href{http://arxiv.org/abs/1505.04597}{U-{Net}: {Convolutional} {Networks} for {Biomedical} {Image} {Segmentation}}, arXiv:1505.04597 [cs] (May 2015).
\newblock \href {https://doi.org/10.48550/arXiv.1505.04597} {\path{doi:10.48550/arXiv.1505.04597}}.
\newline\urlprefix\url{http://arxiv.org/abs/1505.04597}

\bibitem{liu_convnet_2022}
Z.~Liu, H.~Mao, C.-Y. Wu, C.~Feichtenhofer, T.~Darrell, S.~Xie, \href{http://arxiv.org/abs/2201.03545}{A {ConvNet} for the 2020s}, arXiv:2201.03545 [cs] (Mar. 2022).
\newblock \href {https://doi.org/10.48550/arXiv.2201.03545} {\path{doi:10.48550/arXiv.2201.03545}}.
\newline\urlprefix\url{http://arxiv.org/abs/2201.03545}

\bibitem{springenberg_striving_2015}
J.~T. Springenberg, A.~Dosovitskiy, T.~Brox, M.~Riedmiller, \href{http://arxiv.org/abs/1412.6806}{Striving for {Simplicity}: {The} {All} {Convolutional} {Net}}, arXiv:1412.6806 [cs] (Apr. 2015).
\newblock \href {https://doi.org/10.48550/arXiv.1412.6806} {\path{doi:10.48550/arXiv.1412.6806}}.
\newline\urlprefix\url{http://arxiv.org/abs/1412.6806}

\bibitem{zeiler_deconvolutional_2010}
M.~D. Zeiler, D.~Krishnan, G.~W. Taylor, R.~Fergus, \href{https://ieeexplore.ieee.org/document/5539957}{Deconvolutional networks}, in: 2010 {IEEE} {Computer} {Society} {Conference} on {Computer} {Vision} and {Pattern} {Recognition}, 2010, pp. 2528--2535, iSSN: 1063-6919.
\newblock \href {https://doi.org/10.1109/CVPR.2010.5539957} {\path{doi:10.1109/CVPR.2010.5539957}}.
\newline\urlprefix\url{https://ieeexplore.ieee.org/document/5539957}

\bibitem{peebles_scalable_2023}
W.~Peebles, S.~Xie, \href{http://arxiv.org/abs/2212.09748}{Scalable {Diffusion} {Models} with {Transformers}}, arXiv:2212.09748 [cs] (Mar. 2023).
\newblock \href {https://doi.org/10.48550/arXiv.2212.09748} {\path{doi:10.48550/arXiv.2212.09748}}.
\newline\urlprefix\url{http://arxiv.org/abs/2212.09748}

\bibitem{mucke_deep_2024}
N.~T. Mücke, S.~M. Bohté, C.~W. Oosterlee, \href{https://www.nature.com/articles/s41598-024-69901-7}{The deep latent space particle filter for real-time data assimilation with uncertainty quantification}, Scientific Reports 14~(1) (2024) 19447, publisher: Nature Publishing Group.
\newblock \href {https://doi.org/10.1038/s41598-024-69901-7} {\path{doi:10.1038/s41598-024-69901-7}}.
\newline\urlprefix\url{https://www.nature.com/articles/s41598-024-69901-7}

\bibitem{mucke_reduced_2021}
N.~T. Mücke, S.~M. Bohté, C.~W. Oosterlee, \href{https://www.sciencedirect.com/science/article/pii/S1877750321000934}{Reduced order modeling for parameterized time-dependent {PDEs} using spatially and memory aware deep learning}, Journal of Computational Science 53 (2021) 101408.
\newblock \href {https://doi.org/10.1016/j.jocs.2021.101408} {\path{doi:10.1016/j.jocs.2021.101408}}.
\newline\urlprefix\url{https://www.sciencedirect.com/science/article/pii/S1877750321000934}

\bibitem{geneva_transformers_2022}
N.~Geneva, N.~Zabaras, \href{https://www.sciencedirect.com/science/article/pii/S0893608021004500}{Transformers for modeling physical systems}, Neural Networks 146 (2022) 272--289.
\newblock \href {https://doi.org/10.1016/j.neunet.2021.11.022} {\path{doi:10.1016/j.neunet.2021.11.022}}.
\newline\urlprefix\url{https://www.sciencedirect.com/science/article/pii/S0893608021004500}

\bibitem{hendrycks_gaussian_2023}
D.~Hendrycks, K.~Gimpel, \href{http://arxiv.org/abs/1606.08415}{Gaussian {Error} {Linear} {Units} ({GELUs})}, arXiv:1606.08415 [cs] (Jun. 2023).
\newblock \href {https://doi.org/10.48550/arXiv.1606.08415} {\path{doi:10.48550/arXiv.1606.08415}}.
\newline\urlprefix\url{http://arxiv.org/abs/1606.08415}

\bibitem{loshchilov_decoupled_2019}
I.~Loshchilov, F.~Hutter, \href{http://arxiv.org/abs/1711.05101}{Decoupled {Weight} {Decay} {Regularization}}, arXiv:1711.05101 [cs] (Jan. 2019).
\newblock \href {https://doi.org/10.48550/arXiv.1711.05101} {\path{doi:10.48550/arXiv.1711.05101}}.
\newline\urlprefix\url{http://arxiv.org/abs/1711.05101}

\bibitem{loshchilov_sgdr_2017}
I.~Loshchilov, F.~Hutter, \href{http://arxiv.org/abs/1608.03983}{{SGDR}: {Stochastic} {Gradient} {Descent} with {Warm} {Restarts}}, arXiv:1608.03983 [cs] (May 2017).
\newblock \href {https://doi.org/10.48550/arXiv.1608.03983} {\path{doi:10.48550/arXiv.1608.03983}}.
\newline\urlprefix\url{http://arxiv.org/abs/1608.03983}

\bibitem{thygesen_stochastic_2023}
U.~H. Thygesen, Stochastic {Differential} {Equations} for {Science} and {Engineering}, Chapman and Hall/CRC, New York, 2023.
\newblock \href {https://doi.org/10.1201/9781003277569} {\path{doi:10.1201/9781003277569}}.

\bibitem{kohl_learning_2020}
G.~Kohl, K.~Um, N.~Thuerey, \href{http://arxiv.org/abs/2002.07863}{Learning {Similarity} {Metrics} for {Numerical} {Simulations}}, arXiv:2002.07863 [cs] (Jun. 2020).
\newblock \href {https://doi.org/10.48550/arXiv.2002.07863} {\path{doi:10.48550/arXiv.2002.07863}}.
\newline\urlprefix\url{http://arxiv.org/abs/2002.07863}

\bibitem{agdestein_incompressiblenavierstokesjl_2024}
S.~D. Agdestein, S.~Ciarella, B.~Sanderse, \href{https://github.com/agdestein/IncompressibleNavierStokes.jl}{{IncompressibleNavierStokes}.jl}, original-date: 2021-09-22T08:15:00Z (Nov. 2024).
\newline\urlprefix\url{https://github.com/agdestein/IncompressibleNavierStokes.jl}

\bibitem{coppola2019a}
G.~Coppola, F.~Capuano, L.~{de Luca}, Discrete {{Energy-Conservation Properties}} in the {{Numerical Simulation}} of the {{Navier}}--{{Stokes Equations}}, Applied Mechanics Reviews 71~(1) (2019) 010803.
\newblock \href {https://doi.org/10.1115/1.4042820} {\path{doi:10.1115/1.4042820}}.

\bibitem{verstappen2003}
R.~Verstappen, A.~Veldman, Symmetry-preserving discretization of turbulent flow, Journal of Computational Physics 187~(1) (2003) 343--368.
\newblock \href {https://doi.org/10.1016/S0021-9991(03)00126-8} {\path{doi:10.1016/S0021-9991(03)00126-8}}.

\bibitem{agdestein_discretize_2025}
S.~D. Agdestein, B.~Sanderse, \href{http://arxiv.org/abs/2403.18088}{Discretize first, filter next: learning divergence-consistent closure models for large-eddy simulation}, Journal of Computational Physics 522 (2025) 113577, arXiv:2403.18088 [math].
\newblock \href {https://doi.org/10.1016/j.jcp.2024.113577} {\path{doi:10.1016/j.jcp.2024.113577}}.
\newline\urlprefix\url{http://arxiv.org/abs/2403.18088}

\bibitem{sanderse2012}
B.~Sanderse, B.~Koren, Accuracy analysis of explicit {{Runge}}--{{Kutta}} methods applied to the incompressible {{Navier}}--{{Stokes}} equations, Journal of Computational Physics 231~(8) (2012) 3041--3063.
\newblock \href {https://doi.org/10.1016/j.jcp.2011.11.028} {\path{doi:10.1016/j.jcp.2011.11.028}}.

\bibitem{thygesen2023}
U.~H. Thygesen, Stochastic {{Differential Equations}} for {{Science}} and {{Engineering}}, 1st Edition, {Chapman and Hall/CRC}, Boca Raton, 2023.
\newblock \href {https://doi.org/10.1201/9781003277569} {\path{doi:10.1201/9781003277569}}.

\end{thebibliography}

\appendix
\section{Energy conservation and divergence-freeness for the filtered incompressible Navier-Stokes equations}\label{section:NS_energy}
\subsection{Kinetic energy conservation}
The Navier-Stokes equations, \eqref{eq:NS}-\eqref{eq:divfree} describe conservation of mass and momentum of a fluid. In the incompressible case, conservation of kinetic energy is a consequence of conservation of mass and momentum, and not a separate conservation law. Conservation of kinetic energy has been used for example to construct stable discretization schemes for turbulent flows \cite{coppola2019a,verstappen2003} and stable reduced order models \cite{sanderse2020}.


The kinetic energy is naturally defined as $E:=\frac{1}{2} \| \q \|_2^2$, where the $L_2$ norm is given by $\| \q \|_2^2 := \langle \q,\q \rangle$, which is induced by the standard inner product $\langle \q,\vt{v} \rangle := \int_{\Omega} \q \cdot \vt{v} \ \rd \Omega$. An equation for the evolution of $E$ is derived by differentiating $E$ in time and substituting the momentum equation:
\begin{align}
\frac{\rd E}{\rd t} &= \frac{\rd \frac{1}{2} \langle \q,\q \rangle}{\rd t} =-\langle C(\q,\q),\q \rangle - \langle \nabla p, \q \rangle + \langle D \q,\q \rangle + \langle \f(\q),\q \rangle,
\end{align}
where we introduced the convection and diffusion operators $C(\q,\q):=\nabla \cdot (\q \otimes \q)$ and $D \q := \frac{1}{\Rey} \nabla \cdot (\nabla \q + (\nabla \q)^T)$. The equation simplifies due to three symmetry properties. These symmetry properties will be guiding in designing an energy-consistent SDE. First, due to the fact that $C(\q,\q)$ can be written in a skew-symmetric form (using divergence-freeness), we have $\langle C(\q,\q),\q \rangle=0$ for periodic or no-slip boundary conditions. Second, the pressure gradient contribution disappears because $\langle \nabla p, \q \rangle = \langle p,\nabla \cdot \q \rangle =0$, again using divergence-freeness. Third, due to the symmetry of the diffusive operator we can write $\langle D(\q,\q),\q \rangle =- \langle \nabla \q,\nabla \q \rangle$. The kinetic energy balance then reduces to 
\begin{equation}\label{eqn:energy_continuous}
\frac{\rd E}{\rd t} = - \frac{1}{\Rey} \| \nabla \q \|^2_2 + \langle \f(\q),\q \rangle. 
\end{equation}
Consequently, in the absence of boundaries and body forces $\f$, the kinetic energy of the flow can only decrease in time, and in inviscid flow it is exactly conserved. The divergence-freeness of the flow field is key in deriving this result. In presence of body forces, like in the Kolmogorov flow from section \ref{sec:Kolmogorov}, the dissipation term $\frac{1}{\Rey} \| \nabla \q \|^2_2$ on average balances the work done by the body force  $\langle \f(\q),\q \rangle$.

\subsection{Filtering the Navier-Stokes equations}\label{sec:NS_filtered}
Upon filtering the Navier-Stokes equations with a convolutional filter, the velocity field stays divergence-free:
\begin{equation}\label{eq:div_free_filtered}
    \nabla \cdot \qbar = 0,
\end{equation}
because the filter and the divergence operator commute. For the discretized Navier-Stokes equations and a discrete filter, this is in general not true, as the discrete divergence operator and discrete filter do not commute. In \cite{agdestein2025} we developed a so-called \textit{face-averaging filter} which is such that \eqref{eq:div_free_filtered} also holds in a discrete sense (provided that $\qh$ is divergence-free):
\begin{equation}
    M_{h} \qhbar = 0,
\end{equation}
where $M_{h}$ is a matrix representing the discretized divergence operator (on the coarse grid), and $\qhbar:=\A \qh$. In this way, the discrete filter and discrete divergence operator still commute.

For the momentum equations, filtering does not commute with the nonlinear terms, and the filtered equations feature a so-called commutator error $\mathcal{C}(\q,\qbar)$:
\begin{equation}
    \dd{\qbar}{t} + \nabla \cdot (\qbar \otimes \qbar) = -\nabla \bar{p} + \frac{1}{\Rey} \nabla^2 \qbar  + \f(\qbar) + \mathcal{C}(\q,\qbar), 
\end{equation}
where $\mathcal{C}(\q,\qbar) = \nabla \cdot (\qbar \otimes \qbar) - \overline{\nabla \cdot (\q \otimes \q)} + \overline{\f(\q)} - \f(\qbar)$. As a consequence, the energy balance is affected, and instead of equation \eqref{eqn:energy_continuous} we have the following evolution for the kinetic energy $\bar{E}:=\frac{1}{2} \| \qbar \|_2^2$ of the filtered field:
\begin{equation}
    \frac{\rd \bar{E}}{\rd t} = - \frac{1}{\Rey} \| \nabla \qbar \|^2_2 + \langle \f(\qbar),\qbar \rangle + \langle  \mathcal{C}(\q,\qbar), \qbar \rangle.
\end{equation}
Note that we have omitted the explicit dependence on the grid size in the norm and the inner product. Including the grid size only adds a scaling factor and does not change the outcome of the derivations. In the absence of body forces, viscosity and boundary contributions, $\bar{E}$ is not a conserved quantity (in contrast to $E$), due to the additional term $\mathcal{C}(\q,\qbar)$ which can be both positive and negative, and accounts for exchange of energy with unresolved scales. In statistically stationary flow, the terms on the right hand side balance each other on average.

During time-stepping such as with the explicit schemes in \cite{agdestein_discretize_2025,sanderse2012}, first a tentative velocity field is computed as a solution to the momentum equations, and then a projection is performed to make this velocity field divergence-free. This projection of any non-divergence-free field $\qhbar^{*}$ can be written as
\begin{equation}\label{eq:project}
    \qhbar = \Pi \qhbar^{*},
\end{equation}
where $\Pi = I- M_{h}^{T} L_{h}^{-1} M_{h}$, and $L_{h} = M_{h} M_{h}^{T}$ is a Poisson matrix. In practice, equation \eqref{eq:project} is solved in a two-step process:
\begin{align}
    L_{h} \phi &= M_{h} \qhbar^{*}, \label{eqn:poisson}\\
    \qhbar &= \qhbar^{*} - M_{h}^{T} \phi. \label{eqn:velocity_update}
\end{align}

\section{Proof of Theorem \ref{theorem:energy_interpolant}} \label{appendix:proof_of_theorem}

Here, we provide a proof of Theorem \ref{theorem:energy_interpolant}.
\begin{proof}
To ease the notation, we omit the explicit dependence on $\xSI_0$ and $\xSI_1$ and write $\bI_{\tau}:=\bI_{\tau}(\xSI_0, \xSI_1)$ and $\bR_{\tau}:=\bR_{\tau}(\xSI_0, \xSI_1)$, where $(\xSI_0, \xSI_1)\sim \prob{\xSI_0, \xSI_1}$. 

We remind the reader that the dynamics of the interpolant is governed by the SDE:
\begin{align}
    \rd \bI_{\tau} = \bR_{\tau} \rd \tau + \gamma_\tau \rd \bW_\tau, \quad \tau \in [0,1], \quad \xsde_0 = \xSI_0,
\end{align}
where
\begin{align}
    \bR_{\tau} = \dot{\alpha}_\tau \xSI_0 + \dot{\beta}_\tau \xSI_1 + \dot{\gamma}_\tau \bW_\tau 
\end{align}
For any quantity of interest $\QOI(\bI_{\tau})$, we can define a process, $Y_\tau=\QOI(\bI_{\tau})$. The time evolution of $Y_\tau$ is given by Itô's lemma \cite{thygesen2023}:
\begin{align}
\begin{split}
        \rd Y_\tau &= \left[\dd{Q}{\tau} \rd \tau + \nabla_{\bI} Q \cdot \bR_\tau + \frac{1}{2} \gamma_\tau^2 (\nabla_{\bI} \nabla_{\bI}^{T} Q)\right]\rd\tau + \gamma_\tau \nabla_{\bI} Q \cdot \rd \bW_\tau \\
        &= \left[\nabla_{\bI} Q \cdot \bR_\tau + \frac{1}{2} \gamma_\tau^2 (\nabla_{\bI} \nabla_{\bI}^{T} Q)\right]\rd\tau + \gamma_\tau \nabla_{\bI} Q \cdot \rd \bW_\tau.
\end{split}
\end{align}
By setting the quantity of interest to be the kinetic energy, $\QOI(I_{\tau}) = \frac{1}{2}\norm{\bI_\tau}_2^2$, and using
\begin{align}
    \nabla_X \norm{X}^2_2 = 2 X, \quad \nabla_{X} \nabla_{X}^{T}\norm{X}^2_2 = 2 d,
\end{align}
for any $X\in \R^d$, we get the first result
\begin{align}
\begin{split}
        \rd Y_\tau = \left[\bI_{\tau} \cdot \bR_\tau + \frac{d}{2} \gamma_\tau^2\right]\rd\tau +  \gamma_\tau \bI_\tau \cdot \rd \bW_\tau.
\end{split}
\end{align}
For the second result, the expression for the expected energy evolution, we start by expanding the dot product, $\bI_{\tau} \cdot \bR_\tau$:
\begin{align}
\begin{split}
    \bI_{\tau} \cdot \bR_\tau = &\dot{\alpha} _{\tau}\alpha_{\tau}\norm{\xSI_0}_2^2 + \dot{\beta}_{\tau} \beta_{\tau}\norm{\xSI_1}_2^2 + \dot{\gamma}_{\tau} \gamma_{\tau} \norm{\bW_\tau}_2^2 \\
    + &(\dot{\beta}_{\tau} \alpha_{\tau} + \dot{\alpha}_{\tau} \beta_{\tau}) \langle \xSI_0, \xSI_1 \rangle  + \alpha_{\tau} \dot{\gamma}_{\tau} \langle \xSI_0, \bW_\tau \rangle + \beta_{\tau} \dot{\gamma}_{\tau} \langle \xSI_1, \bW_\tau \rangle.
\end{split}
\end{align}
Taking the expected value with respect to $\xSI_0$, $\xSI_1$, and $\bW_\tau$ gives:
\begin{align}
    \E_{(\xSI_0, \xSI_1, \bW_\tau)}\left[\bI_{\tau} \cdot \bR_\tau\right] = \E_{\xSI_0, \xSI_1}\left[\dot{\alpha} _{\tau}\alpha_{\tau}\norm{\xSI_0}_2^2 + \dot{\beta}_{\tau} \beta_{\tau}\norm{\xSI_1}_2^2 + (\dot{\beta}_{\tau} \alpha_{\tau} + \dot{\alpha}_{\tau} \beta_{\tau}) \langle \xSI_0, \xSI_1 \rangle \right] + \dot{\gamma}_{\tau} \gamma_{\tau} \tau d.
\end{align}
Here, we used that $\xSI_0$ and $\xSI_1$ are independent with respect to $\bW_\tau$ and that $\E[\norm{\bW_\tau}_2^2] = \tau d$. Lastly, by taking the expected value of $\rd Y_\tau$ and using that $\E\left[\bI_\tau \cdot \rd \bW_\tau\right] = 0$, we get:
\begin{align} \label{eq:final_proof_equations}
\begin{split}
        \E_{(\xSI_0, \xSI_1, \bW)}\left[\rd Y_\tau\right] &= \left(\E_{(\xSI_0, \xSI_1)}\left[\dot{\alpha} _{\tau}\alpha_{\tau}\norm{\xSI_0}_2^2 + \dot{\beta}_{\tau} \beta_{\tau}\norm{\xSI_1}_2^2 + (\dot{\beta}_{\tau} \alpha_{\tau} + \dot{\alpha}_{\tau} \beta_{\tau}) \langle \xSI_0, \xSI_1 \rangle \right]+ \dot{\gamma}_{\tau} \gamma_{\tau} \tau d + \frac{d}{2} \gamma_\tau^2\right)\rd\tau.
\end{split}
\end{align}
It does not make a difference whether we include the last two terms in the expected value in Eq. \eqref{eq:final_proof_equations}, as they are independent from $\xSI_0$ and $\xSI_1$. Hence, by including the last two terms in the expected value, we get:
\begin{align}
    \E_{(\xSI_0, \xSI_1, \bW)}\left[\rd Y_\tau\right] &= \E_{(\xSI_0, \xSI_1)}\left[ H_\tau(\xSI_0, \xSI_1; \alpha_\tau, \beta_\tau, \gamma_\tau) \right] \rd\tau,
\end{align}
with
\begin{align}
    H_\tau(\xSI_0, \xSI_1; \alpha_\tau, \beta_\tau, \gamma_\tau)=\dot{\alpha} _{\tau}\alpha_{\tau}\norm{\xSI_0}_2^2 + \dot{\beta}_{\tau} \beta_{\tau}\norm{\xSI_1}_2^2 + (\dot{\beta}_{\tau} \alpha_{\tau} + \dot{\alpha}_{\tau} \beta_{\tau}) \langle \xSI_0, \xSI_1 \rangle + \dot{\gamma}_{\tau} \gamma_{\tau} \tau d + \frac{d}{2} \gamma_\tau^2,
\end{align}
which was what we wanted to show. 
\end{proof}

\section{Metrics and quantities of interest}\label{section:appendix_metrics_and_qoi}
\paragraph{Mean squared error.} The mean squared error (MSE) is computed as:
\begin{align}
    \text{MSE}(\qhbar, \qhred) = \frac{1}{N_T N_x} \sum_{n=1}^{N_T} \norm{\qhbar^n -  \qhred^n}_2^2,
\end{align}
where $N_T$ is the number of physical time steps and $\norm{\cdot}_2^2$ is the squared $l^2$-norm. We compute the MSE between each generated trajectory and a filtered DNS trajectory simulated with the same initial condition and compute the average over all computed MSEs. Note that the MSE is not a useful metric for long-term predictions when dealing with chaotic systems.

\paragraph{Kinetic energy.} The kinetic energy for at a time step $n$ is computed by:
\begin{align}
    E(\qhbar^n) = \frac{1}{2h^2}\norm{\qhbar^n}_2^2,
\end{align}
where $h$ is the equidistant spatial grid size in $x$- and $y$- direction. We are especially interested in assessing whether the generative models generate data that follows the same kinetic energy distribution as the filtered DNS trajectories.   

\paragraph{Rate of change.} The rate of change (RoC) at time step $n$ is computed as in \cite{kohl_benchmarking_2024} by:
\begin{align}
    RoC(\qhbar^n) = \norm{(\qhbar^{n} - \qhbar^{n-1})/\Delta t}_1,
\end{align}
where $\Delta t$ is the physical step size. The RoC measures whether a trajectory follows the expected evolution. For example, if a trajectory explodes, the RoC grows substantially and if the RoC goes to zero the trajectory goes to a steady state. This metric is particularly useful for long roll-outs where the predictions are not expected to follow the true state exactly, but are expected to follow the general evolution. This is especially relevant when dealing with chaotic trajectories. 

\paragraph{Wasserstein-1 distance.} To assess the quality of the approximations of the energy distributions, we compute the Wasserstein-1 (W-1) distance. This is computed by:
\begin{align}
    W(p, q) = \inf_{\gamma \in \Gamma(p,q)} \E_{(x,y)\sim \gamma(x,y)}\left[\norm{x-y}_1 \right]
\end{align}
where $\Gamma(p,q)$ is the set of all couplings between the distributions $p$ and $q$ and $\norm{\cdot}_1$ is the $l^1$ norm. We compute the Wasserstein metric between the total kinetic energy of the filtered DNS simulations and the generated simulations. For a trajectory, the total kinetic energy is computed at each time-step. Then, the collection of energies from each time step is used to make the empirical distribution.

\paragraph{LSiM.} LSiM is a similarity metric designed for numerical simulations. The metric is computed by encoding the data with a trained neural network and then comparing the latent representations. For details, see \cite{kohl_learning_2020}. We compute the LSiM between each trajectory in the ensemble of generated states and the filtered DNS state at each time step. Then, the mean and standard deviation over all time steps and trajectories are reported. For a state with multiple fields, such as velocity in x-direction and y-direction, we consider each channel separately and report the mean. For a field, we convert the values into an RGB representation, as the LSiM is designed to handle RGB images. This is done following the same procedure as in \cite{kohl_learning_2020}.

\paragraph{Pearson correlation. } We compute the Pearson correlation coefficient between the generated state and the filtered DNS state at each time step. In the beginning the correlation is approximately 1, but will deteriorate with time. We report the time it takes for the correlation to drop below 0.8.

\section{Additional results - Kolmogorov flow}\label{section:appendix_additional_results}

\begin{figure*}[t!]
    \centering
    \begin{subfigure}[b]{0.75\textwidth}
        \centering
        \includegraphics[width=1.0\linewidth]{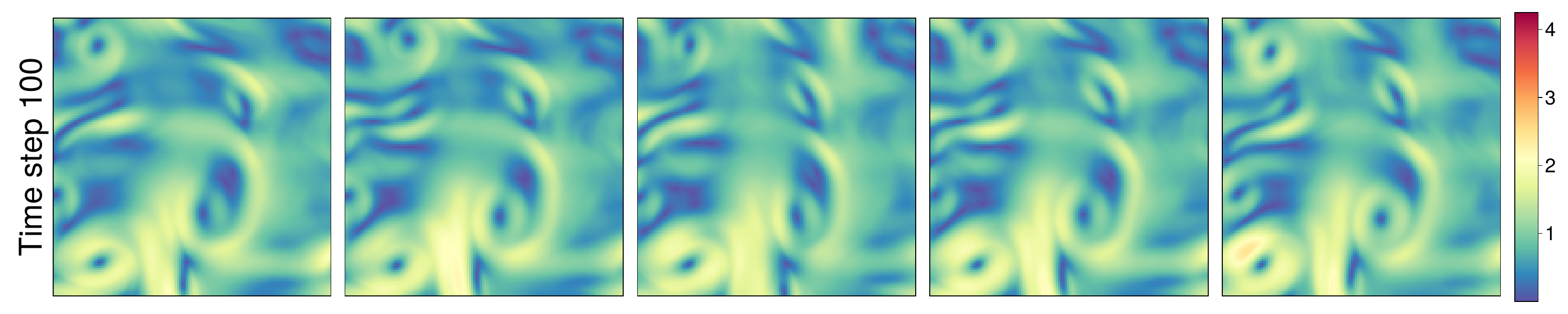}
    \end{subfigure}
    \begin{subfigure}[b]{0.75\textwidth}
        \centering
        \includegraphics[width=1.0\linewidth]{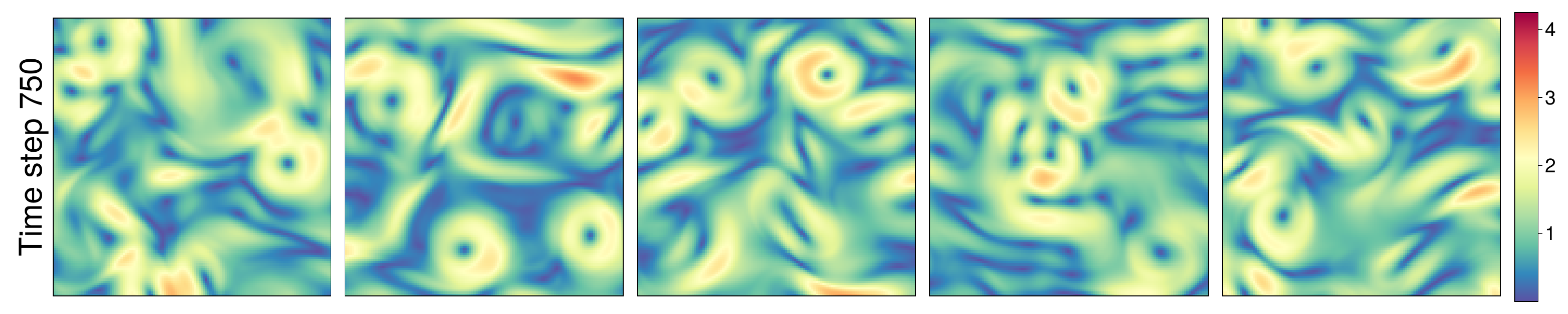}
        \caption{ACDM with 50 pseudo-steps.}
    \end{subfigure}
    \begin{subfigure}[b]{0.75\textwidth}
        \centering
        \includegraphics[width=1.0\linewidth]{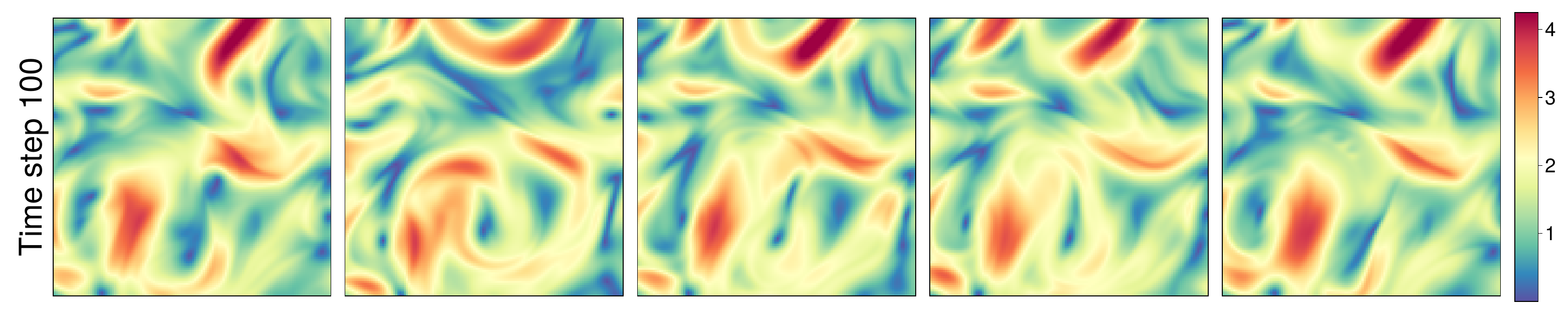}
    \end{subfigure}
    \begin{subfigure}[b]{0.75\textwidth}
        \centering
        \includegraphics[width=1.0\linewidth]{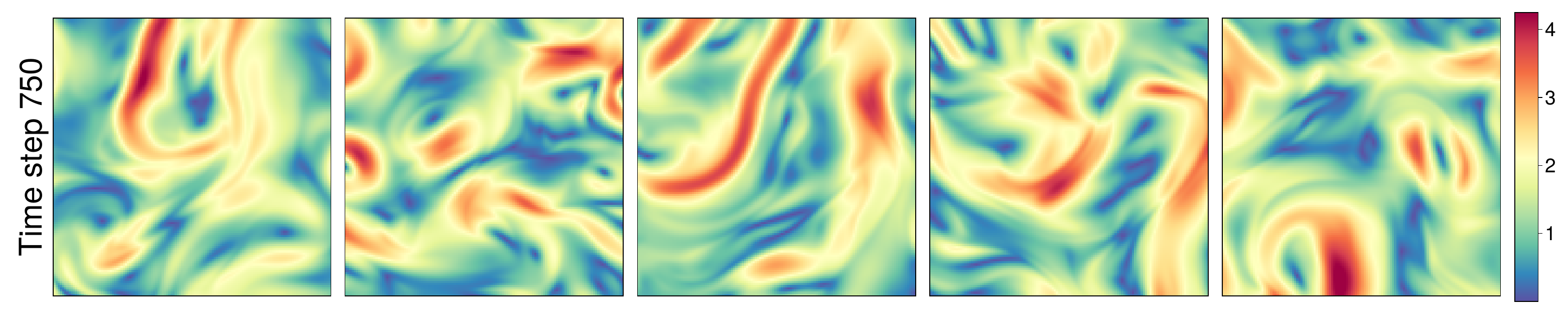}
        \caption{Refiner with 8 pseudo-steps.}
    \end{subfigure}
    \begin{subfigure}[b]{0.75\textwidth}
        \centering
        \includegraphics[width=1.0\linewidth]{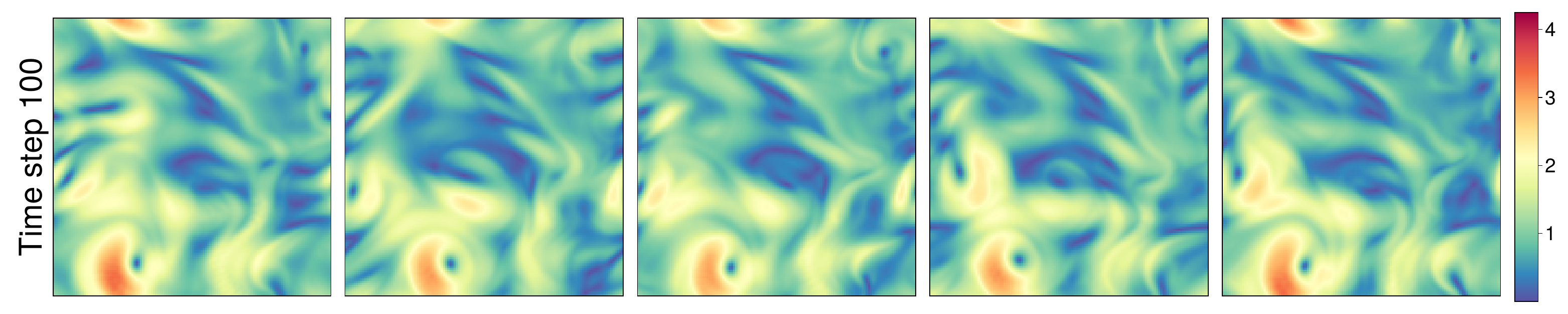}
    \end{subfigure}
    \begin{subfigure}[b]{0.75\textwidth}
        \centering
        \includegraphics[width=1.0\linewidth]{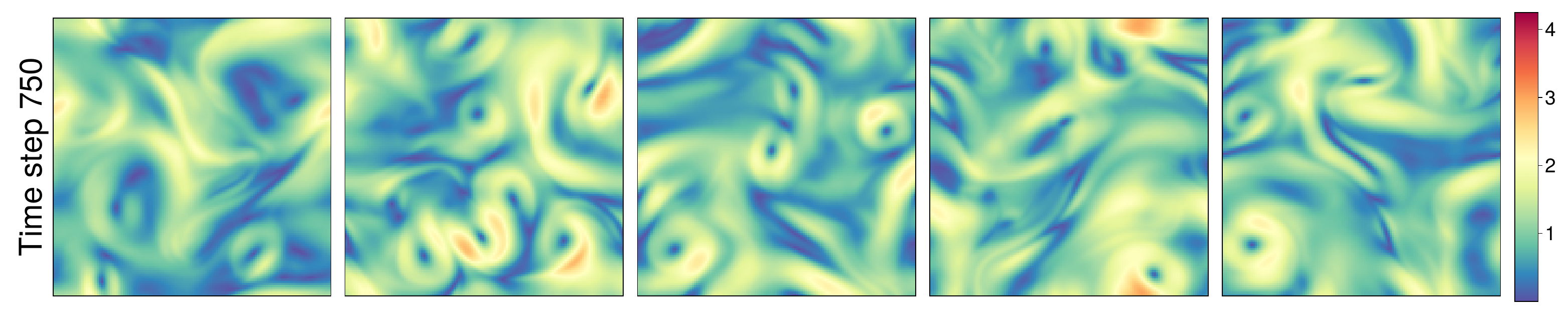}
        \caption{Stochastic inerpolant with 50 pseudo-steps.}
    \end{subfigure}
    \begin{subfigure}[b]{0.75\textwidth}
        \centering
        \includegraphics[width=1.0\linewidth]{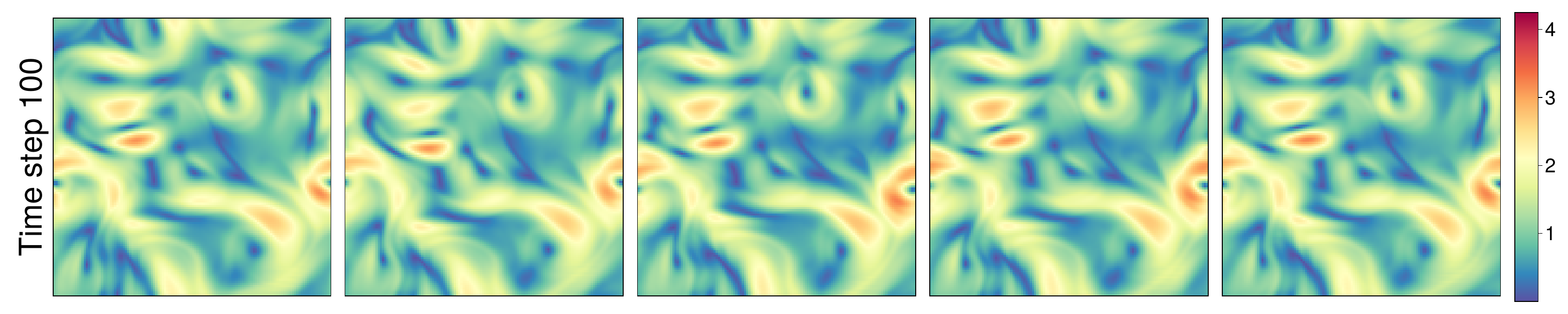}
    \end{subfigure}
    \begin{subfigure}[b]{0.75\textwidth}
        \centering
        \includegraphics[width=1.0\linewidth]{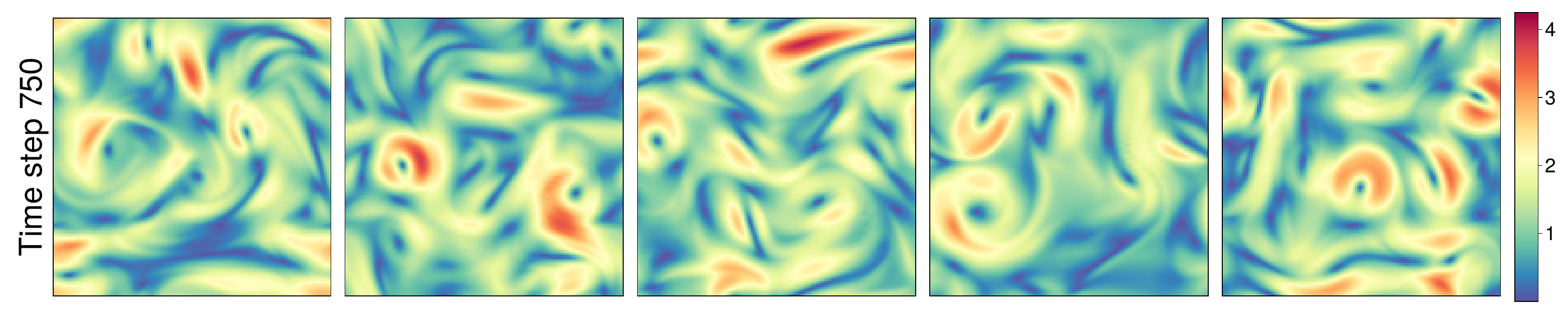}
        \caption{Optimized stochastic interpolant with divergence project and 50 pseudo-steps.}
    \end{subfigure}
    \caption{Five velocity magnitude realizations at time step 100 and 750 for various models.}
\end{figure*}

\begin{figure*}[t!]
    \centering
    \begin{subfigure}[b]{0.45\textwidth}
        \centering
        \includegraphics[width=1.0\linewidth]{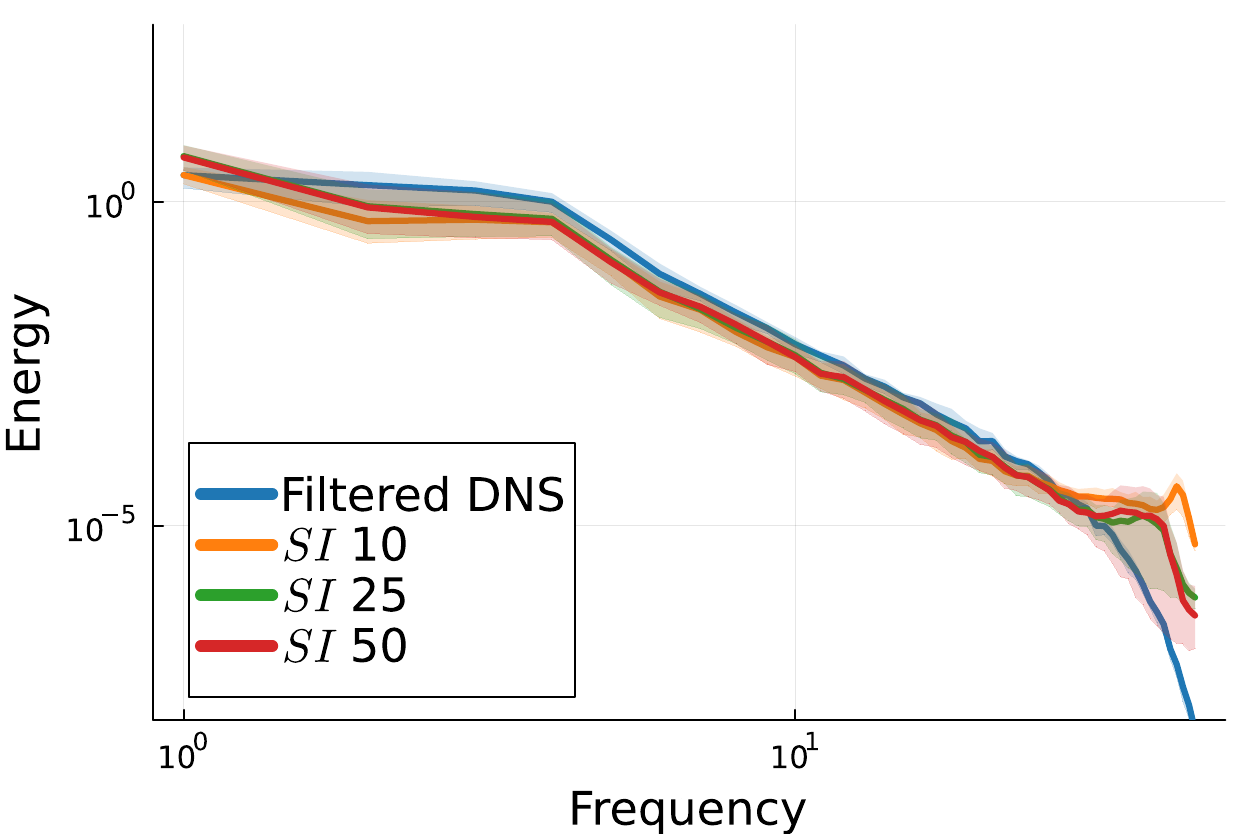}
        \caption{Stochastic interpolant, n=200}
    \end{subfigure}
    \hfill
    \begin{subfigure}[b]{0.45\textwidth}
        \centering
        \includegraphics[width=1.0\linewidth]{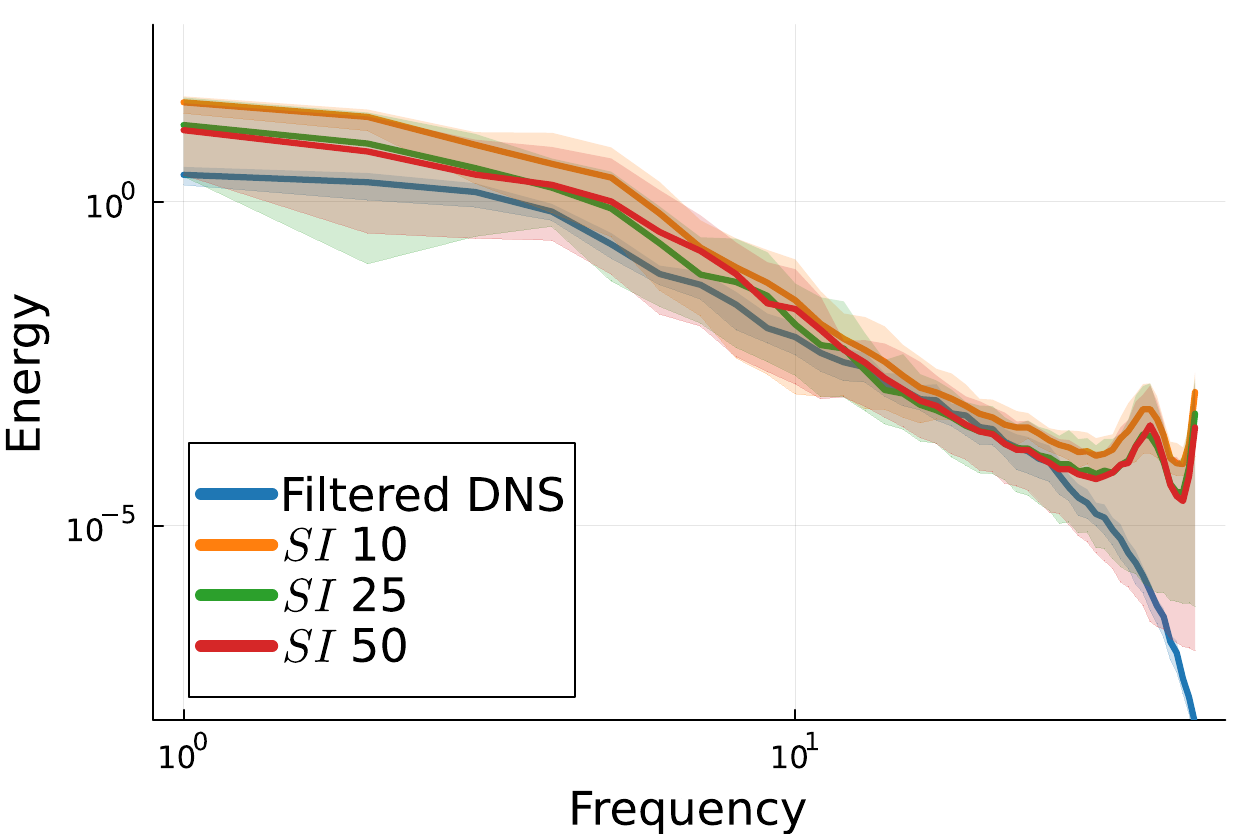}
        \caption{Stochastic interpolant, n=750}
    \end{subfigure}
    \begin{subfigure}[b]{0.45\textwidth}
        \centering
        \includegraphics[width=1.0\linewidth]{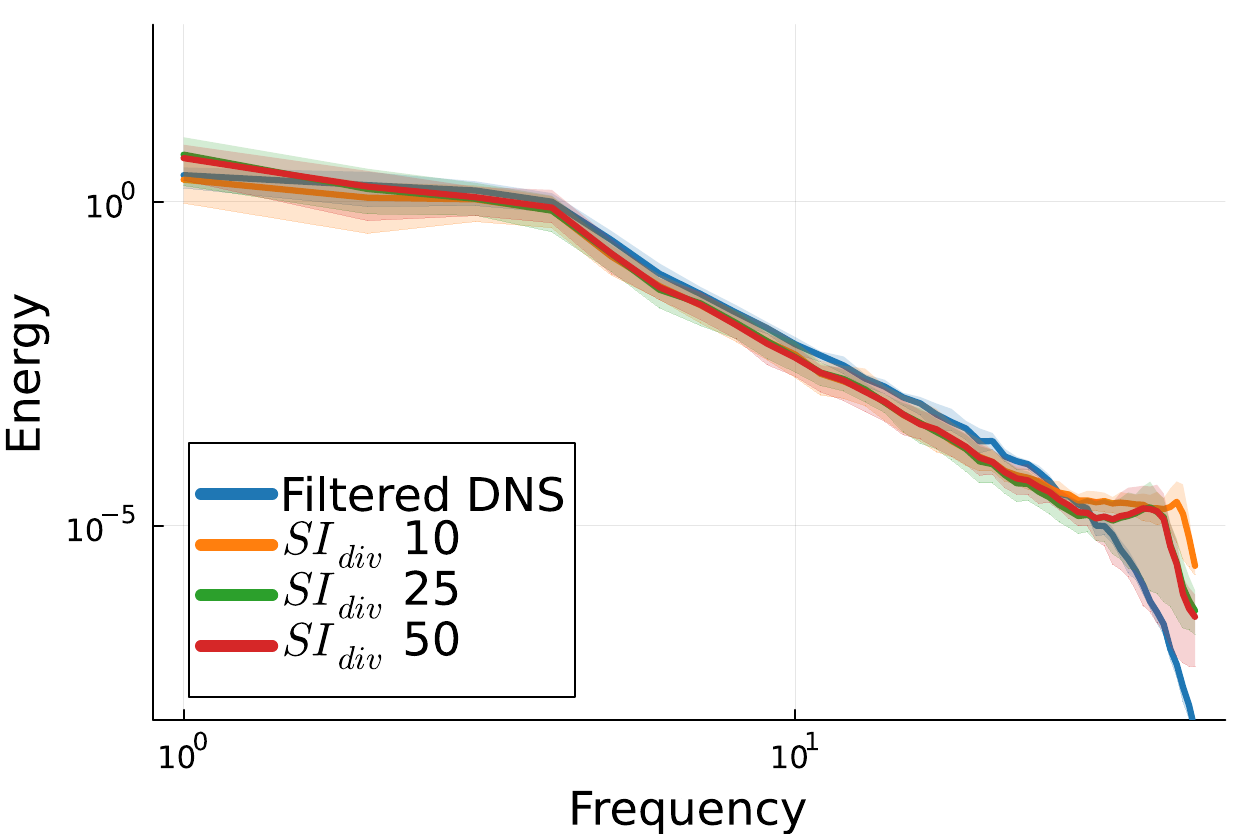}
        \caption{Stochastic interpolant with divergence-free projection, n=200}
    \end{subfigure}
        \hfill
    \begin{subfigure}[b]{0.45\textwidth}
        \centering
        \includegraphics[width=1.0\linewidth]{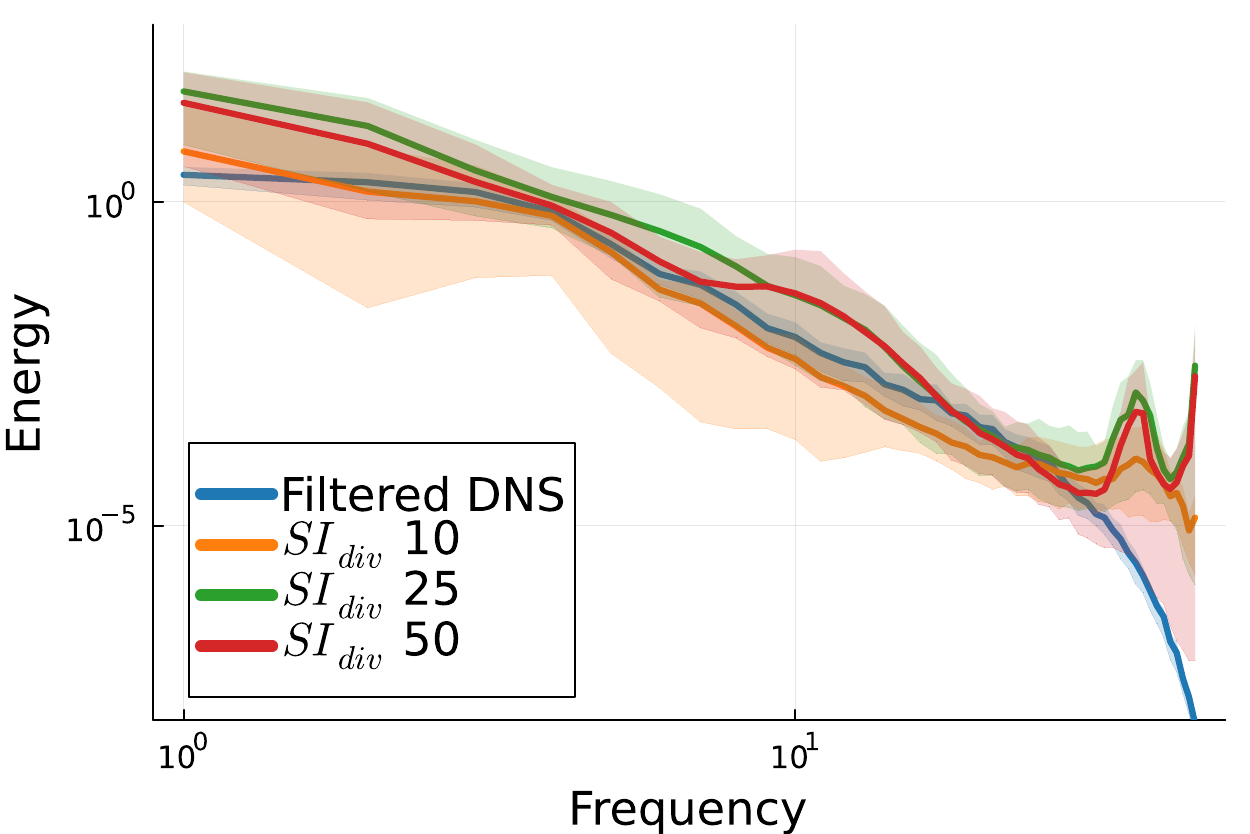}
        \caption{Stochastic interpolant with divergence-free projection, n=750}
    \end{subfigure}
    \begin{subfigure}[b]{0.45\textwidth}
        \centering
        \includegraphics[width=1.0\linewidth]{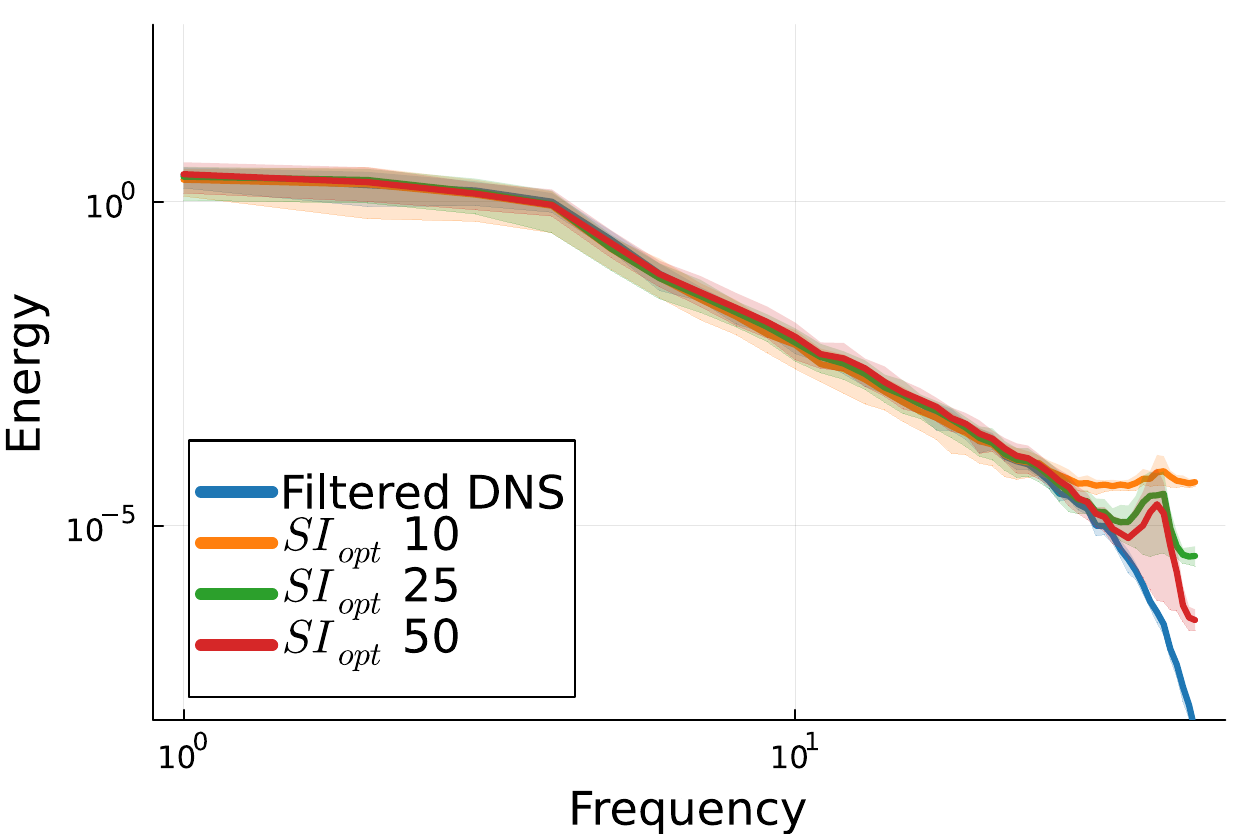}
        \caption{Optimized stochastic interpolant, n=200}
    \end{subfigure}
        \hfill
    \begin{subfigure}[b]{0.45\textwidth}
        \centering
        \includegraphics[width=1.0\linewidth]{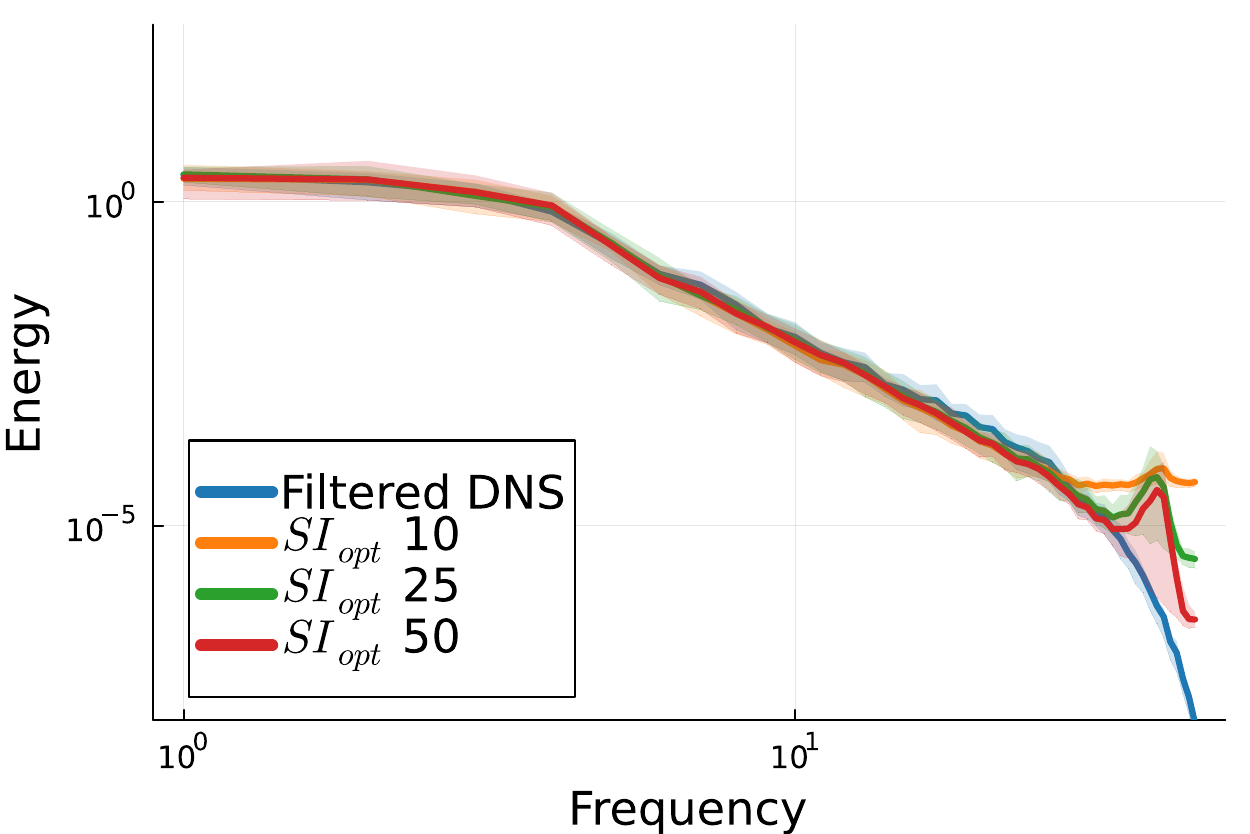}
        \caption{Optimized stochastic interpolant, n=750}
    \end{subfigure}
    \begin{subfigure}[b]{0.45\textwidth}
        \centering
        \includegraphics[width=1.0\linewidth]{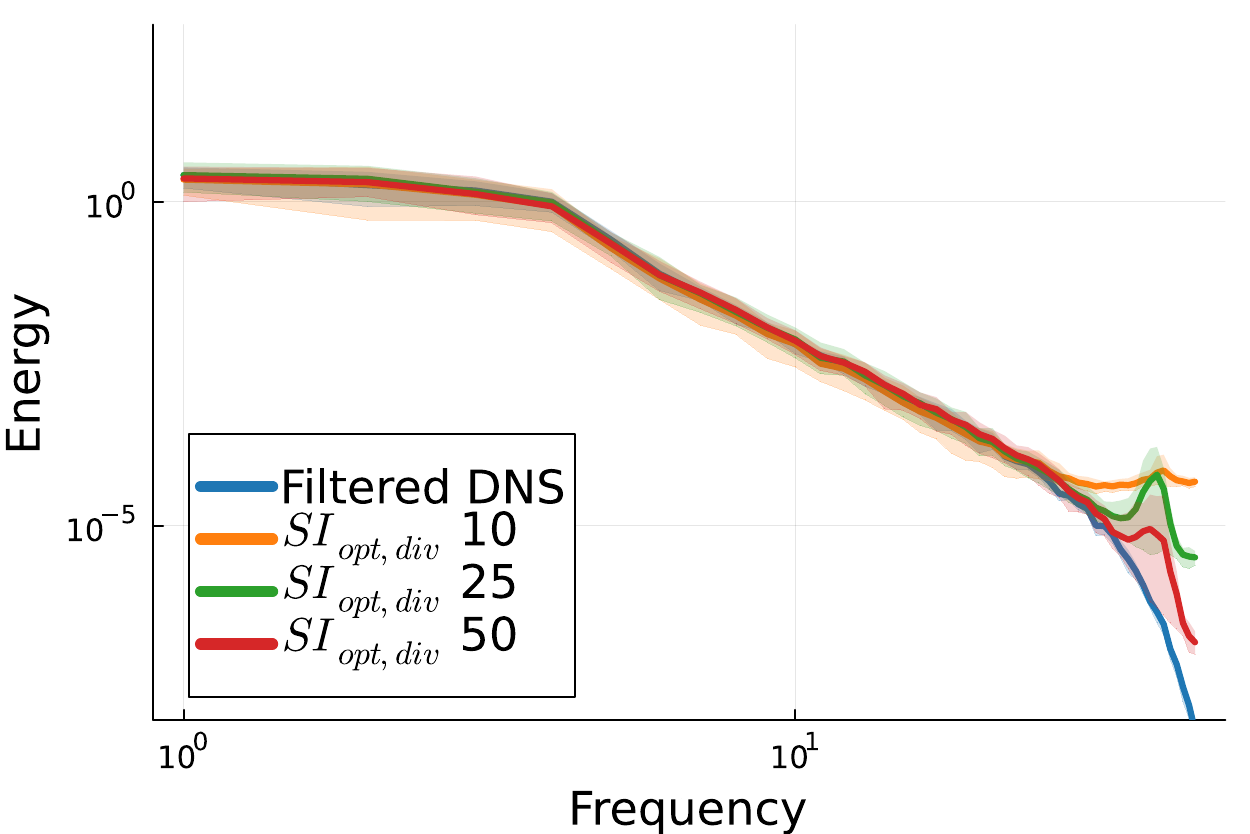}
        \caption{Optimized stochastic interpolant with divergence-free projection, n=200}
    \end{subfigure}
        \hfill
    \begin{subfigure}[b]{0.45\textwidth}
        \centering
        \includegraphics[width=1.0\linewidth]{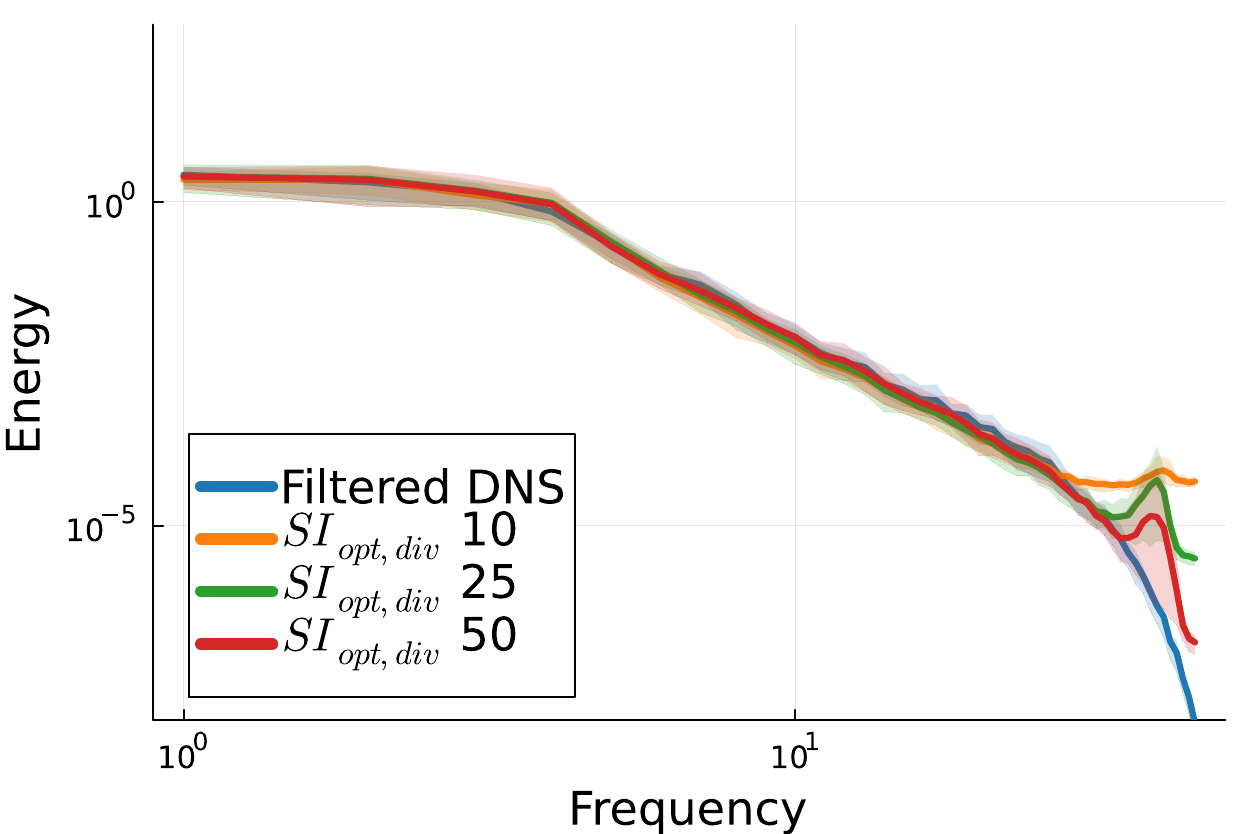}
        \caption{Optimized stochastic interpolant with divergence-free projection, n=750}
    \end{subfigure}
    \caption{Energy spectra for the stochastic interpolants.}
    \label{fig:energy_spectra_all_SI_models}
\end{figure*}

\begin{figure*}[t!]
    \centering
    \begin{subfigure}[b]{0.45\textwidth}
        \centering
        \includegraphics[width=1.0\linewidth]{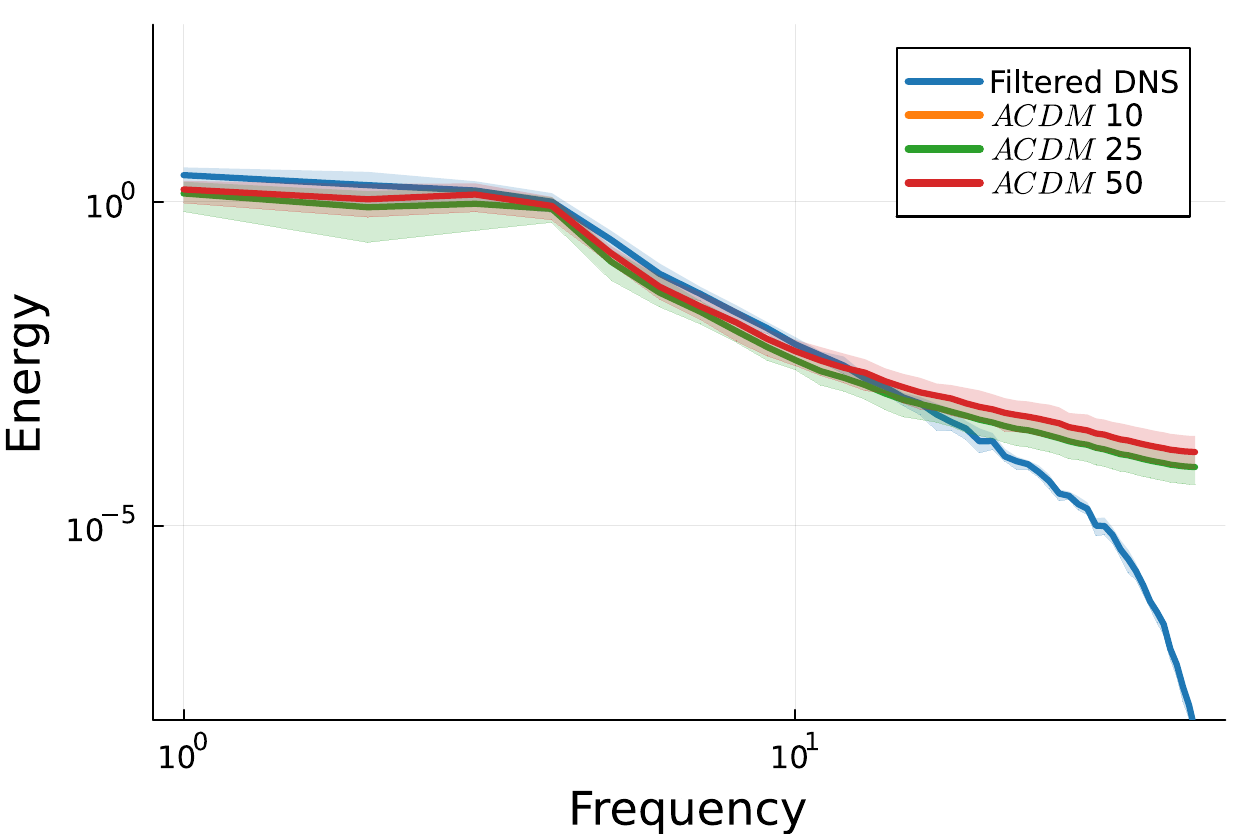}
        \caption{ACDM, n=200}
    \end{subfigure}
    \hfill
    \begin{subfigure}[b]{0.45\textwidth}
        \centering
        \includegraphics[width=1.0\linewidth]{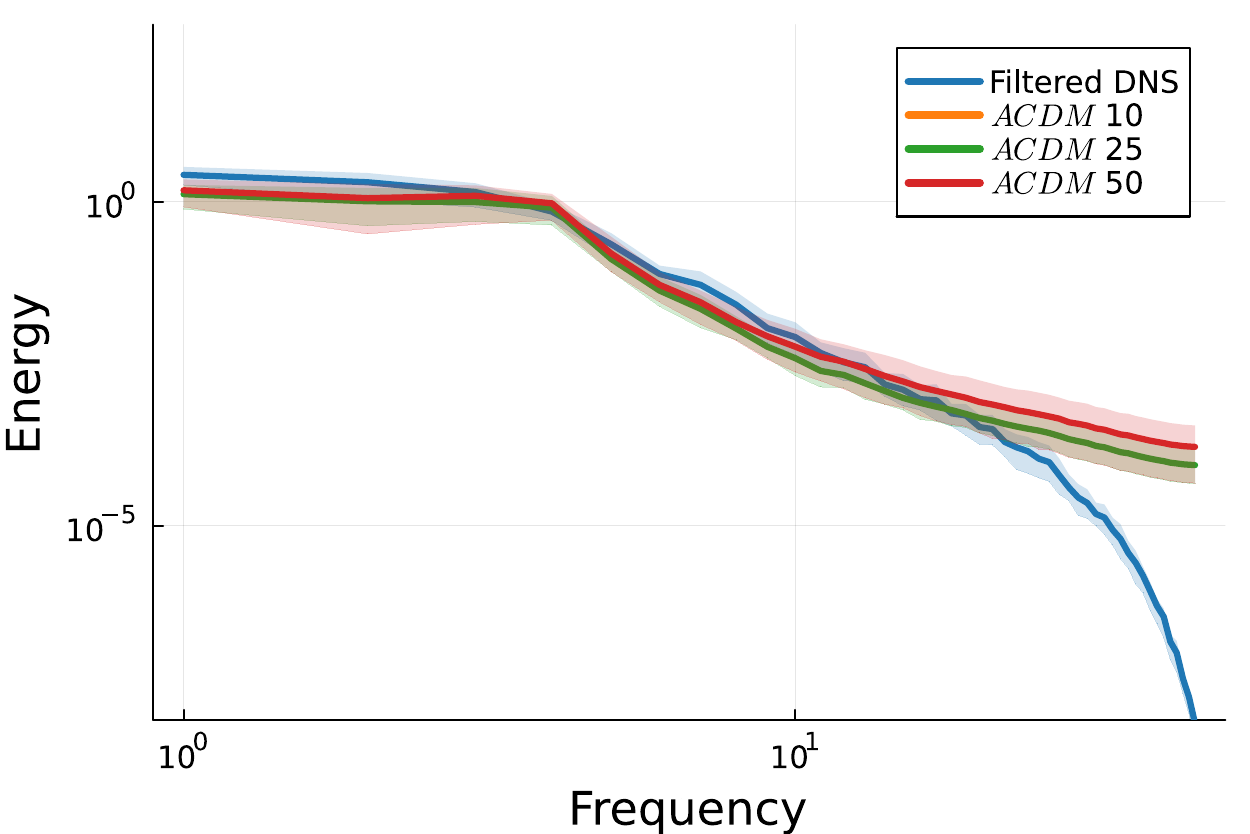}
        \caption{ACDM, n=750}
    \end{subfigure}
    \begin{subfigure}[b]{0.45\textwidth}
        \centering
        \includegraphics[width=1.0\linewidth]{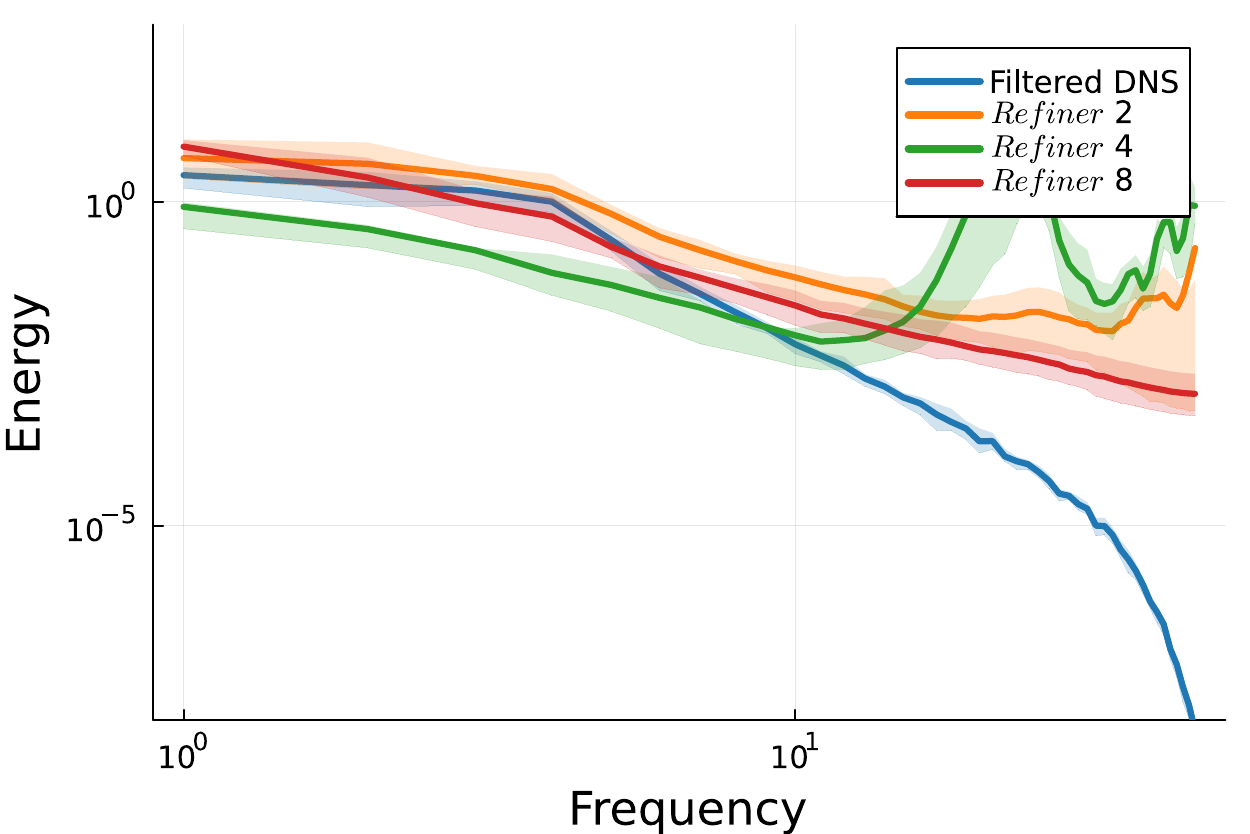}
        \caption{Refiner, n=200}
    \end{subfigure}
    \hfill
    \begin{subfigure}[b]{0.45\textwidth}
        \centering
        \includegraphics[width=1.0\linewidth]{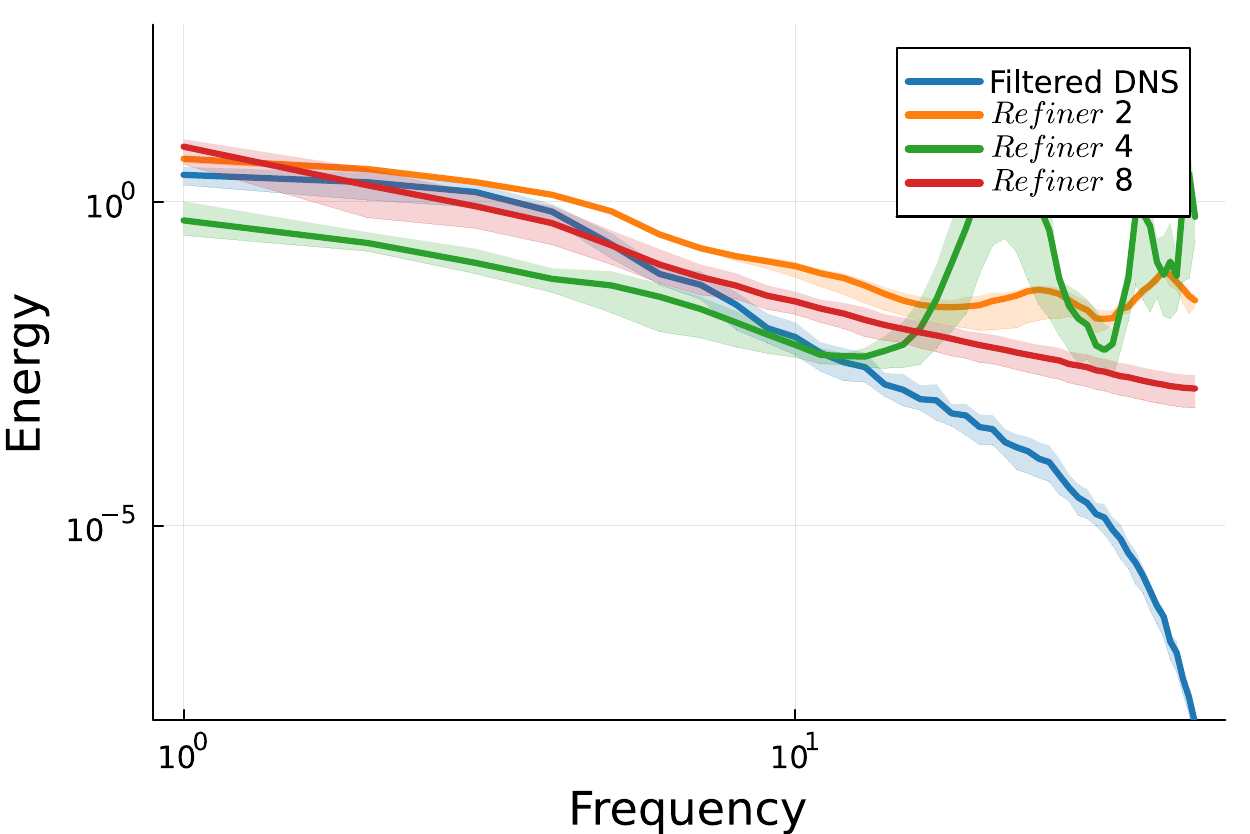}
        \caption{Refiner, n=750}
    \end{subfigure}
    \caption{Energy spectra for the ACDM and PDE-Refiner.}
\end{figure*}

\begin{figure*}[t!]
    \centering
    \begin{subfigure}[b]{0.45\textwidth}
        \centering
        \includegraphics[width=1.0\linewidth]{figures/results/kolmogorov/energy_density_SI_not_optimized_.pdf}
        \caption{Stochastic interpolant}
    \end{subfigure}
    \hfill
    \begin{subfigure}[b]{0.45\textwidth}
        \centering
        \includegraphics[width=1.0\linewidth]{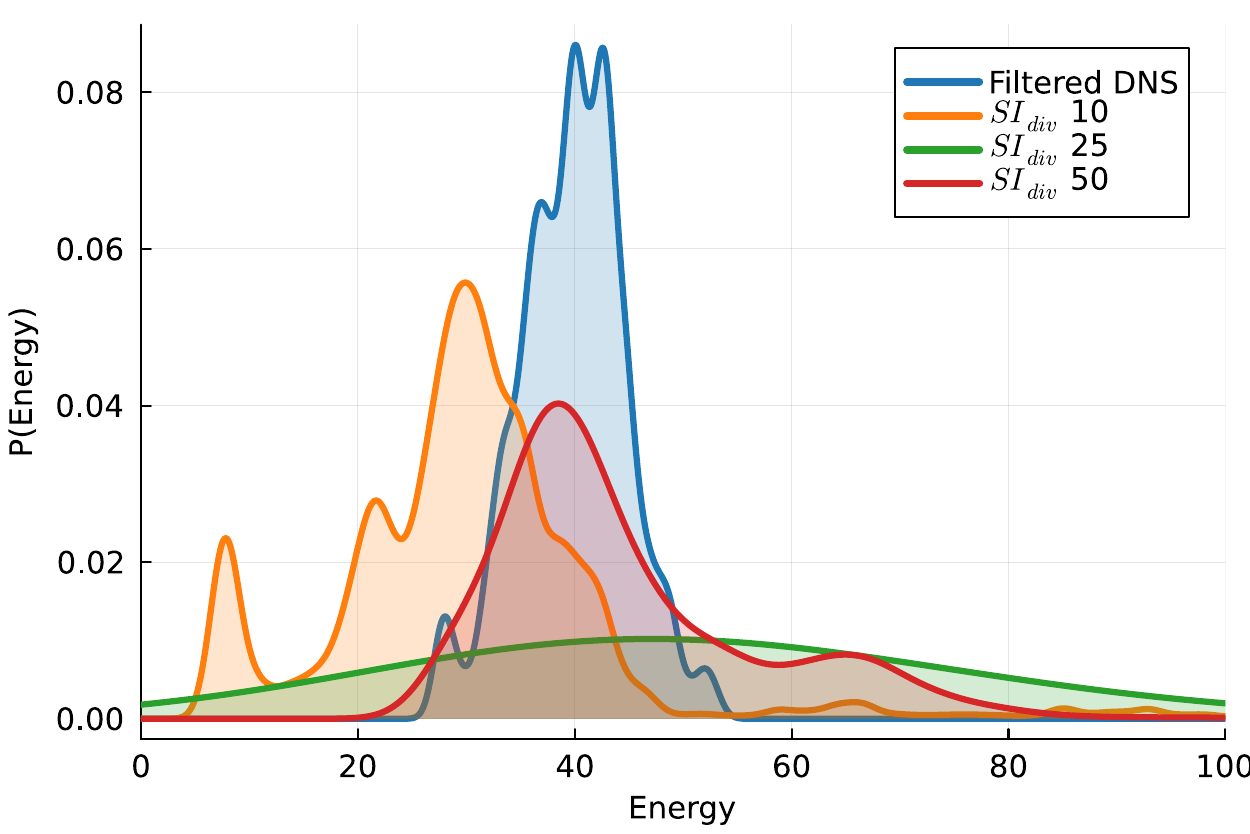}
        \caption{Stochastic interpolant with divergence-free projection}
    \end{subfigure}
    \begin{subfigure}[b]{0.45\textwidth}
        \centering
        \includegraphics[width=1.0\linewidth]{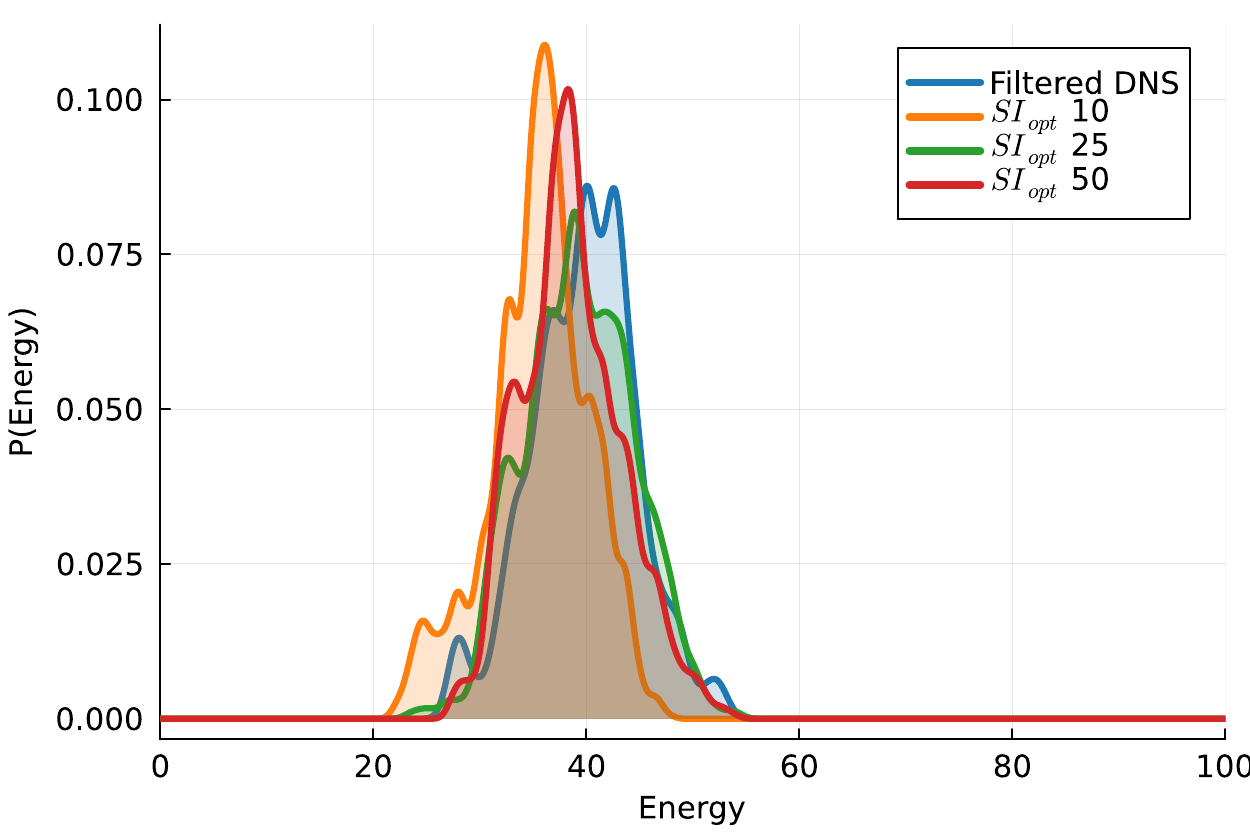}
        \caption{Optimized stochastic interpolant}
    \end{subfigure}
    \hfill
    \begin{subfigure}[b]{0.45\textwidth}
        \centering
        \includegraphics[width=1.0\linewidth]{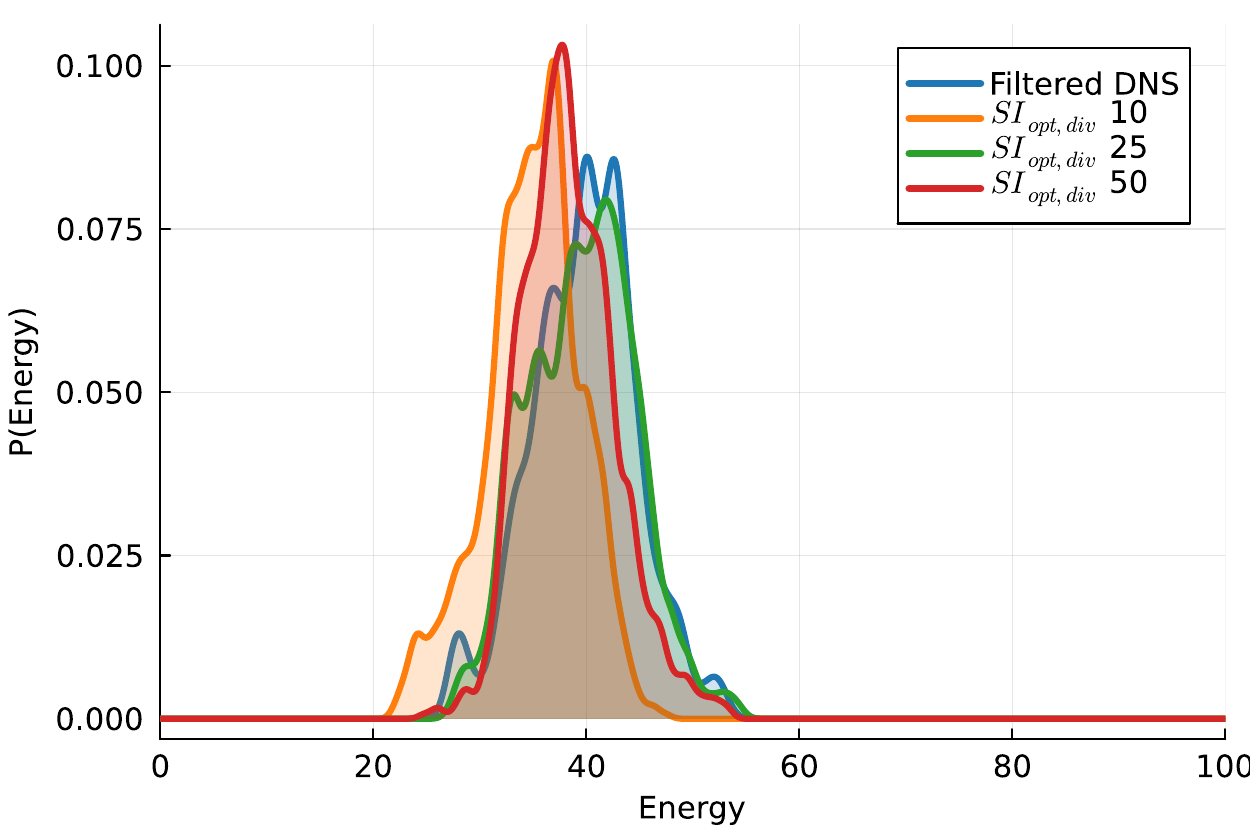}
        \caption{Optimized stochastic interpolant with divergence-free projection}
    \end{subfigure}    
    \begin{subfigure}[b]{0.45\textwidth}
        \centering
        \includegraphics[width=1.0\linewidth]{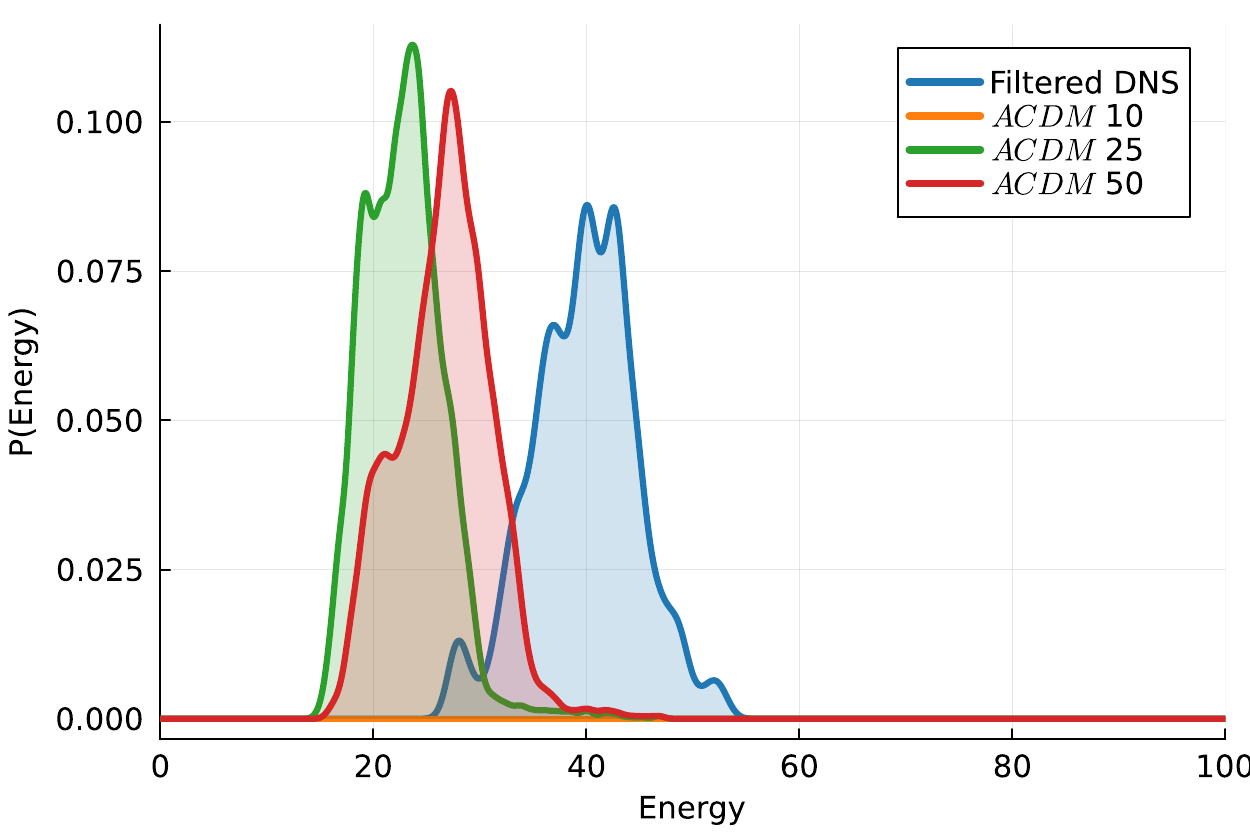}
        \caption{ACDM}
    \end{subfigure}
    \hfill
    \begin{subfigure}[b]{0.45\textwidth}
        \centering
        \includegraphics[width=1.0\linewidth]{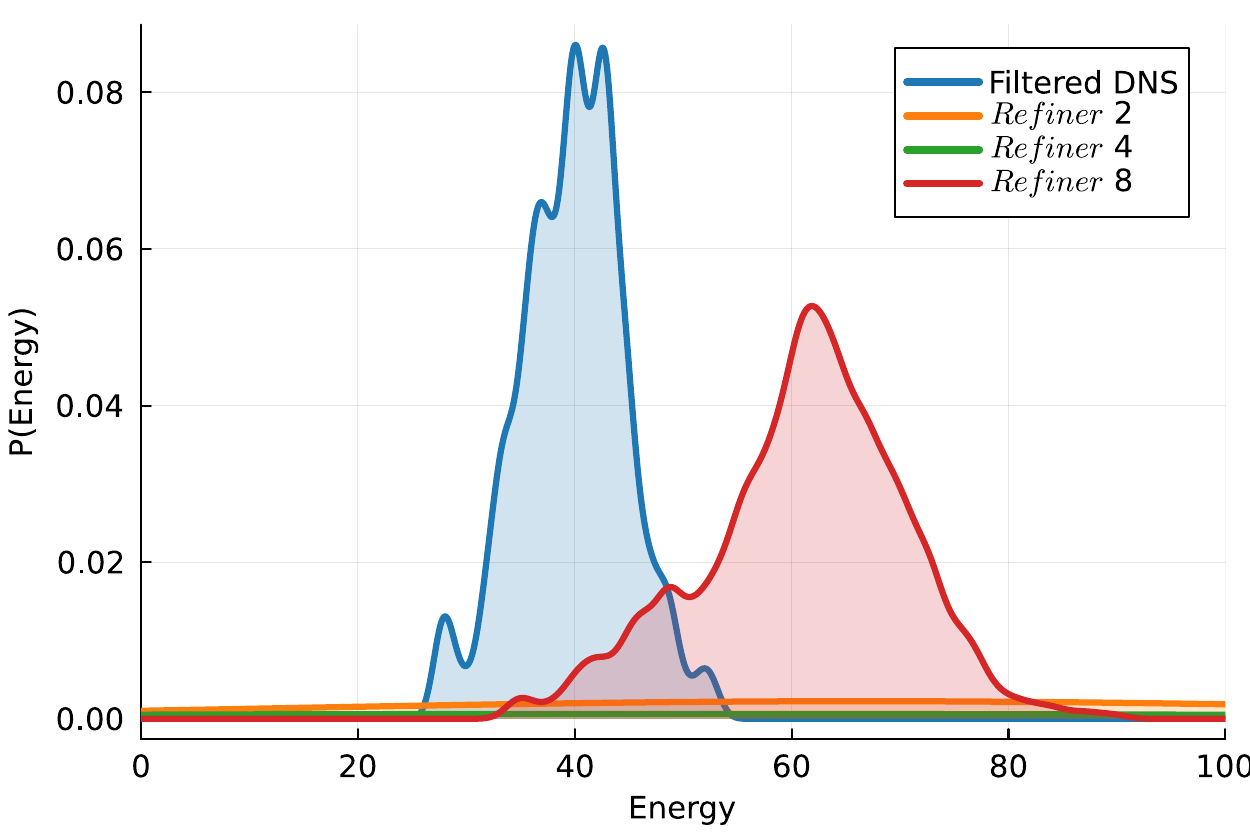}
        \caption{PDE-Refiner}
    \end{subfigure}
    \caption{Probability density function of the energy. Note that the distributions for the ACDM with 10 steps and the PDE-refiner are not in the figures, since they are centered far away from the true distribution.}
\end{figure*}

\begin{figure*}[t!]
    \centering
    \begin{subfigure}[b]{0.45\textwidth}
        \centering
        \includegraphics[width=1.0\linewidth]{figures/results/kolmogorov/total_energy_SI_not_optimized_.pdf}
        \caption{Stochastic interpolant}
    \end{subfigure}
    \hfill    
    \begin{subfigure}[b]{0.45\textwidth}
        \centering
        \includegraphics[width=1.0\linewidth]{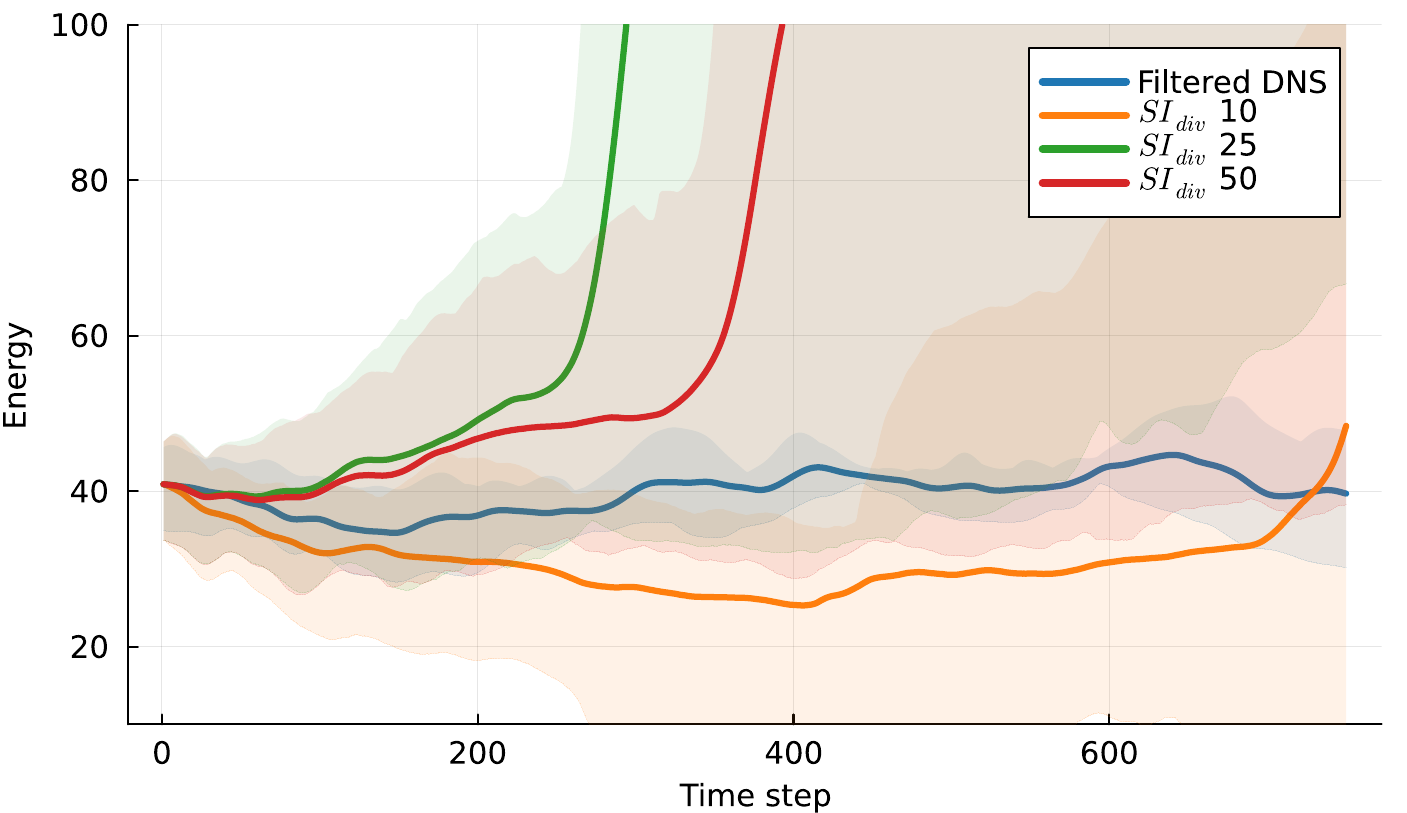}
        \caption{Stochastic interpolant with divergence-free projection}
    \end{subfigure}
    \begin{subfigure}[b]{0.45\textwidth}
        \centering
        \includegraphics[width=1.0\linewidth]{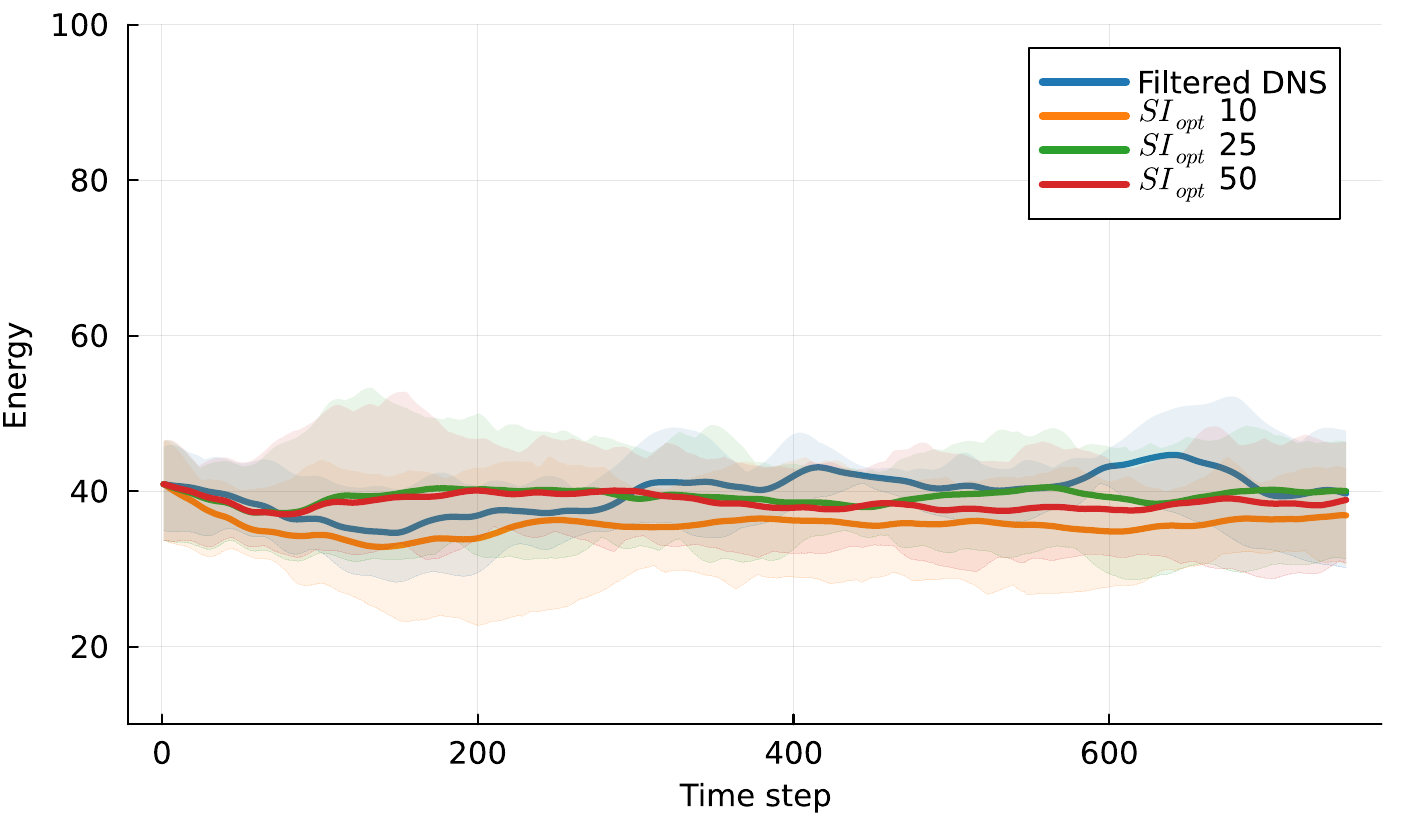}
        \caption{Optimized stochastic interpolant}
    \end{subfigure}
    \hfill    
    \begin{subfigure}[b]{0.45\textwidth}
        \centering
        \includegraphics[width=1.0\linewidth]{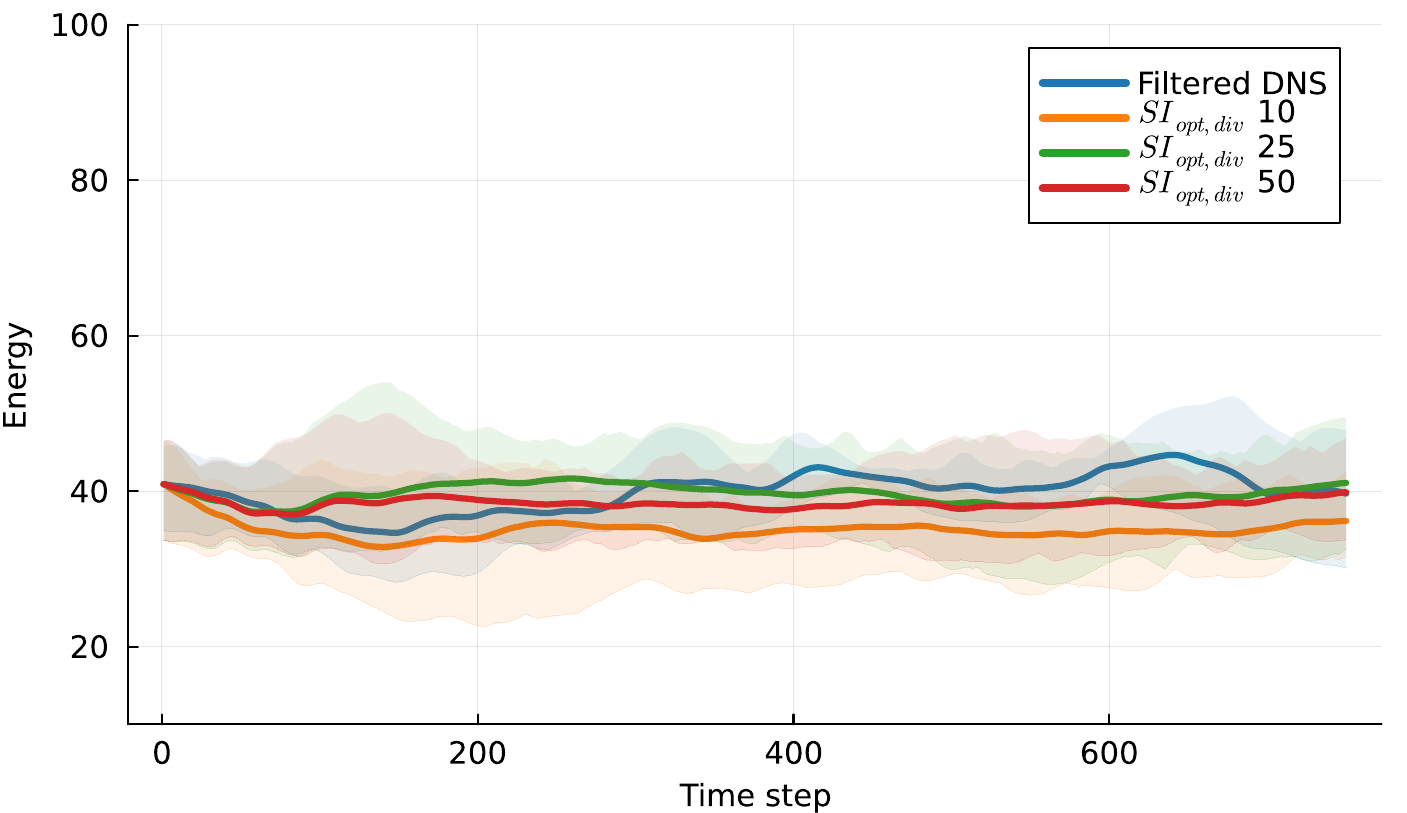}
        \caption{Optimized stochastic interpolant with divergence-free projection}
    \end{subfigure}
    \begin{subfigure}[b]{0.45\textwidth}
        \centering
        \includegraphics[width=1.0\linewidth]{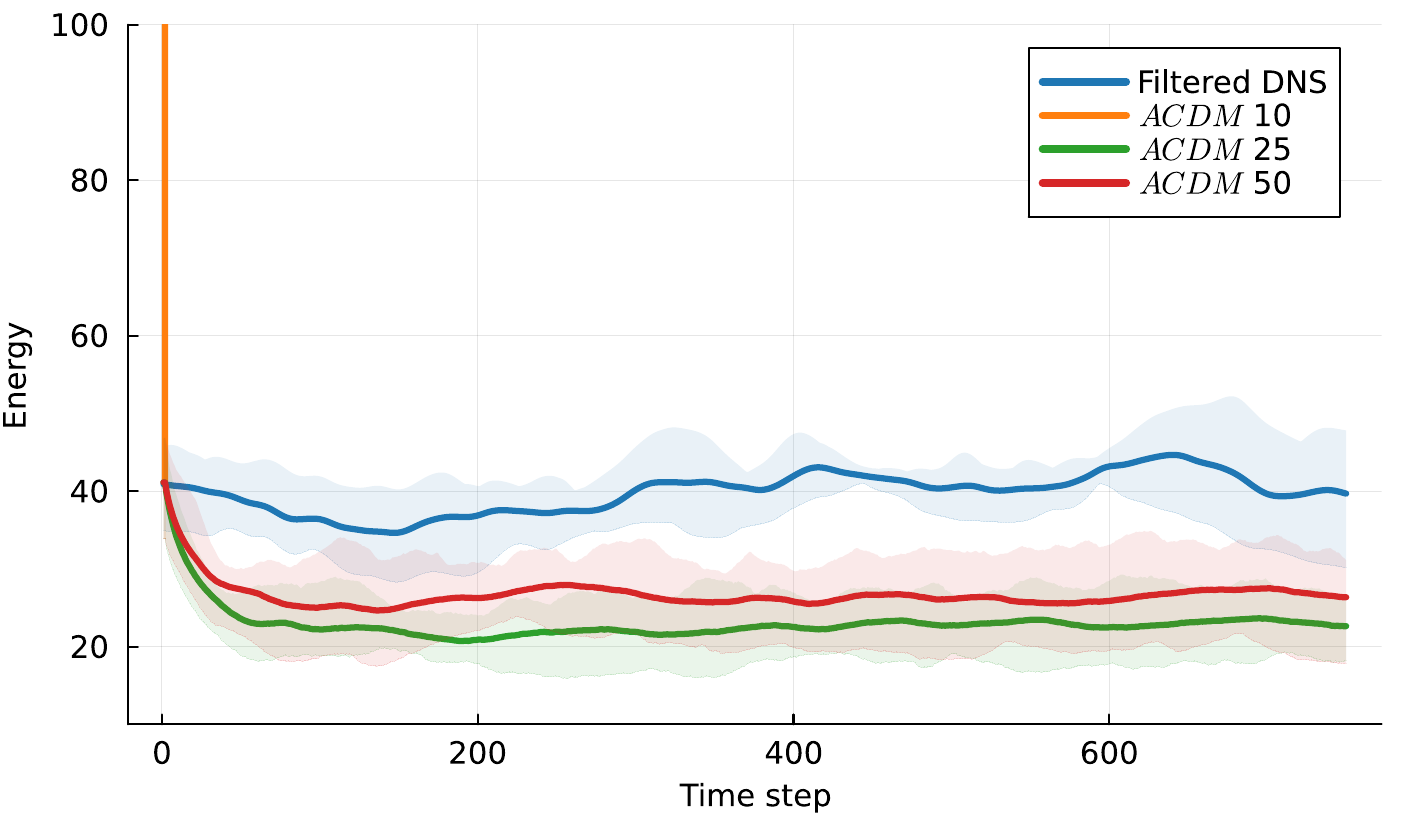}
        \caption{ACDM}
    \end{subfigure}
    \hfill    
    \begin{subfigure}[b]{0.45\textwidth}
        \centering
        \includegraphics[width=1.0\linewidth]{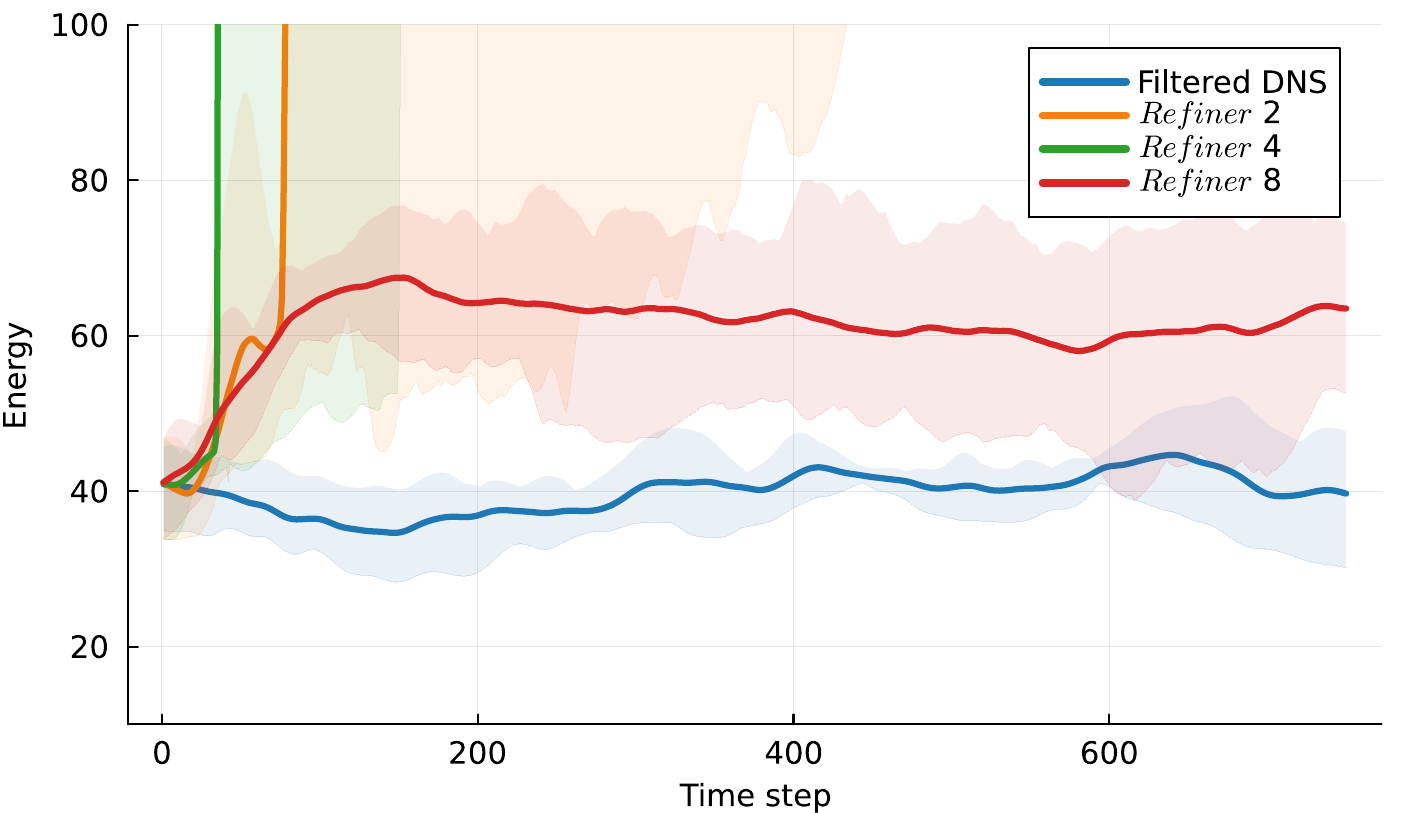}
        \caption{PDE-refiner}
    \end{subfigure}
    \caption{Energy evolution.}
\end{figure*}

\begin{figure*}[t!]
    \centering
    \begin{subfigure}[b]{0.45\textwidth}
        \centering
        \includegraphics[width=1.0\linewidth]{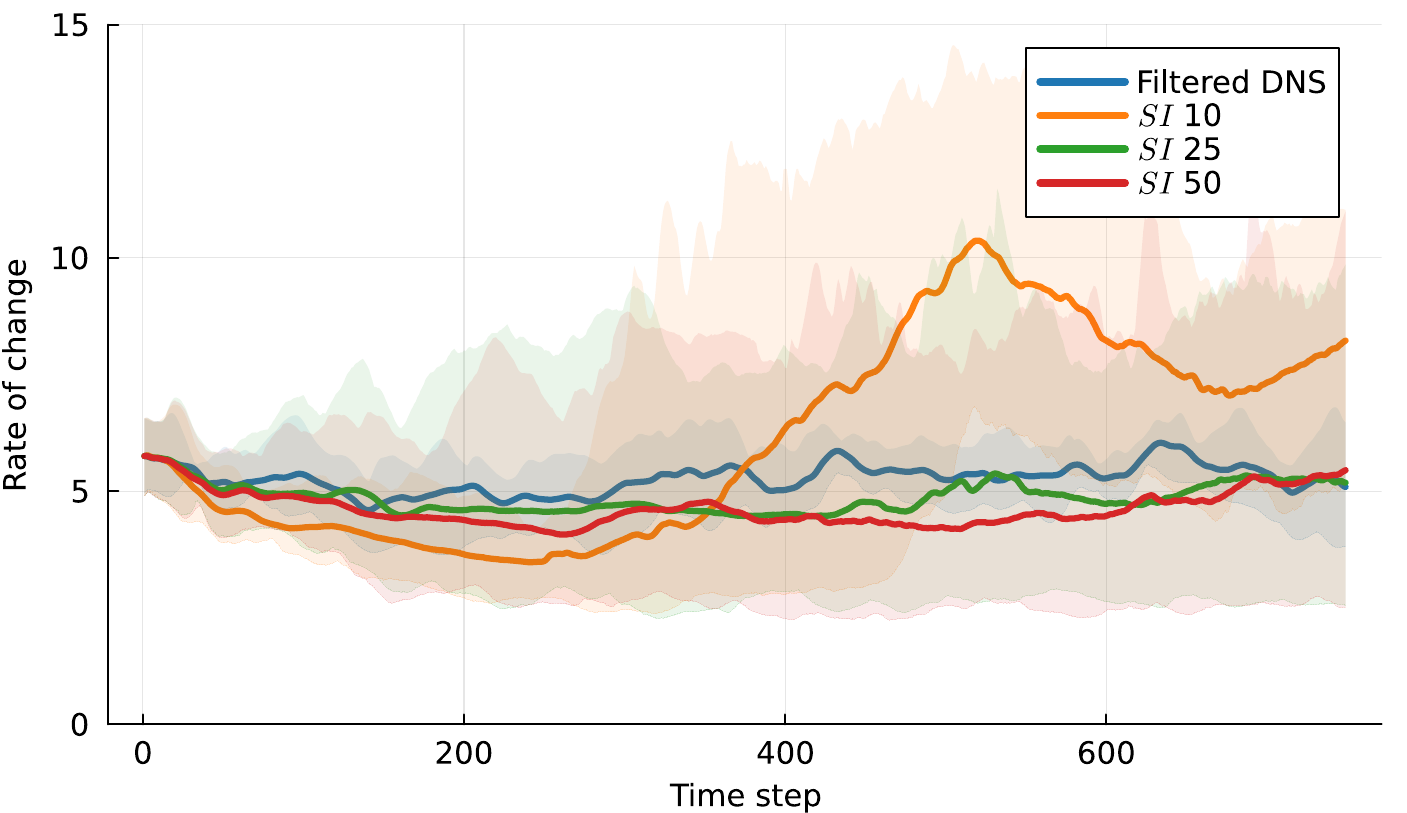}
        \caption{Stochastic interpolant}
    \end{subfigure}
    \hfill    
    \begin{subfigure}[b]{0.45\textwidth}
        \centering
        \includegraphics[width=1.0\linewidth]{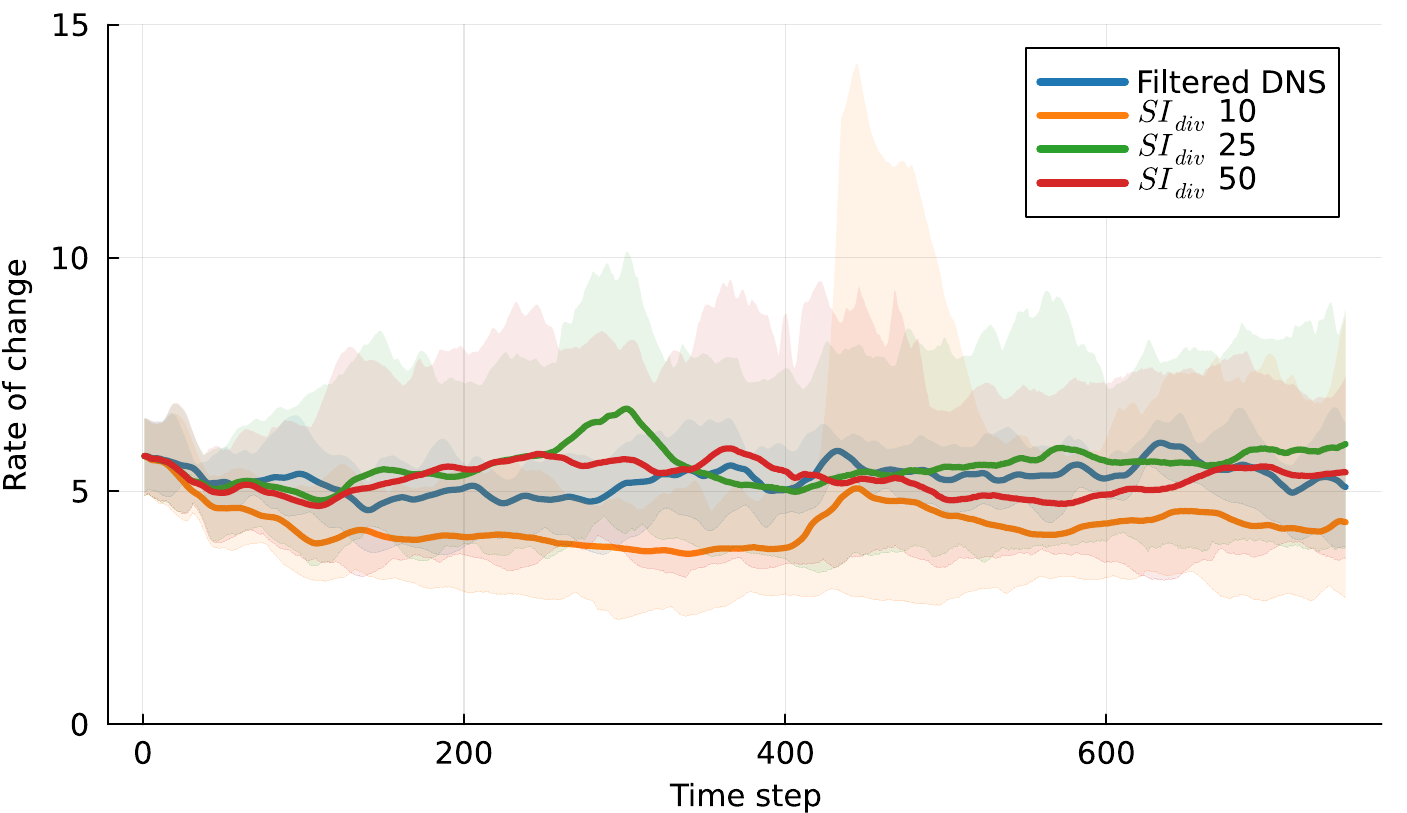}
        \caption{Stochastic interpolant with divergence-free projection}
    \end{subfigure}
    \begin{subfigure}[b]{0.45\textwidth}
        \centering
        \includegraphics[width=1.0\linewidth]{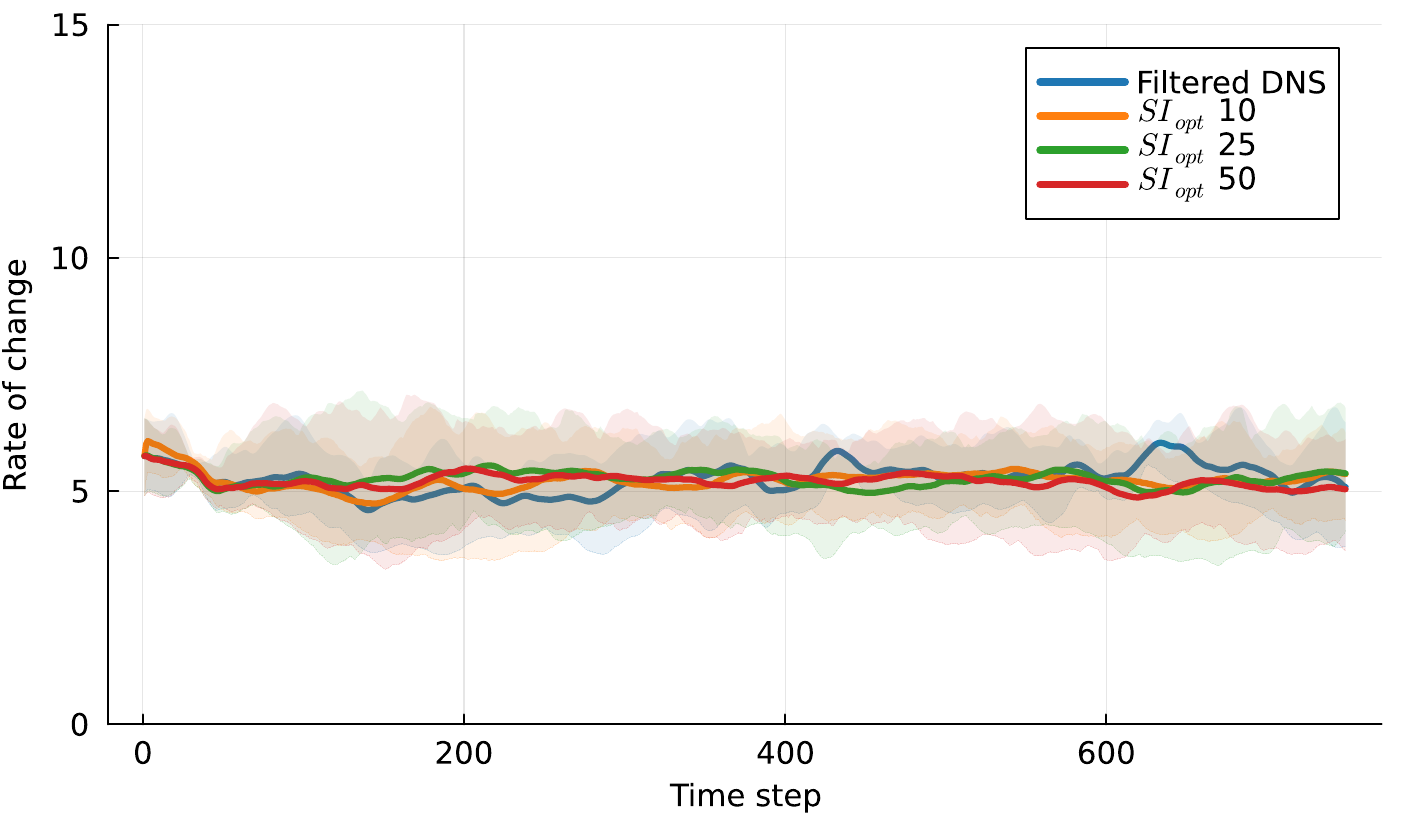}
        \caption{Optimized stochastic interpolant}
    \end{subfigure}
    \hfill    
    \begin{subfigure}[b]{0.45\textwidth}
        \centering
        \includegraphics[width=1.0\linewidth]{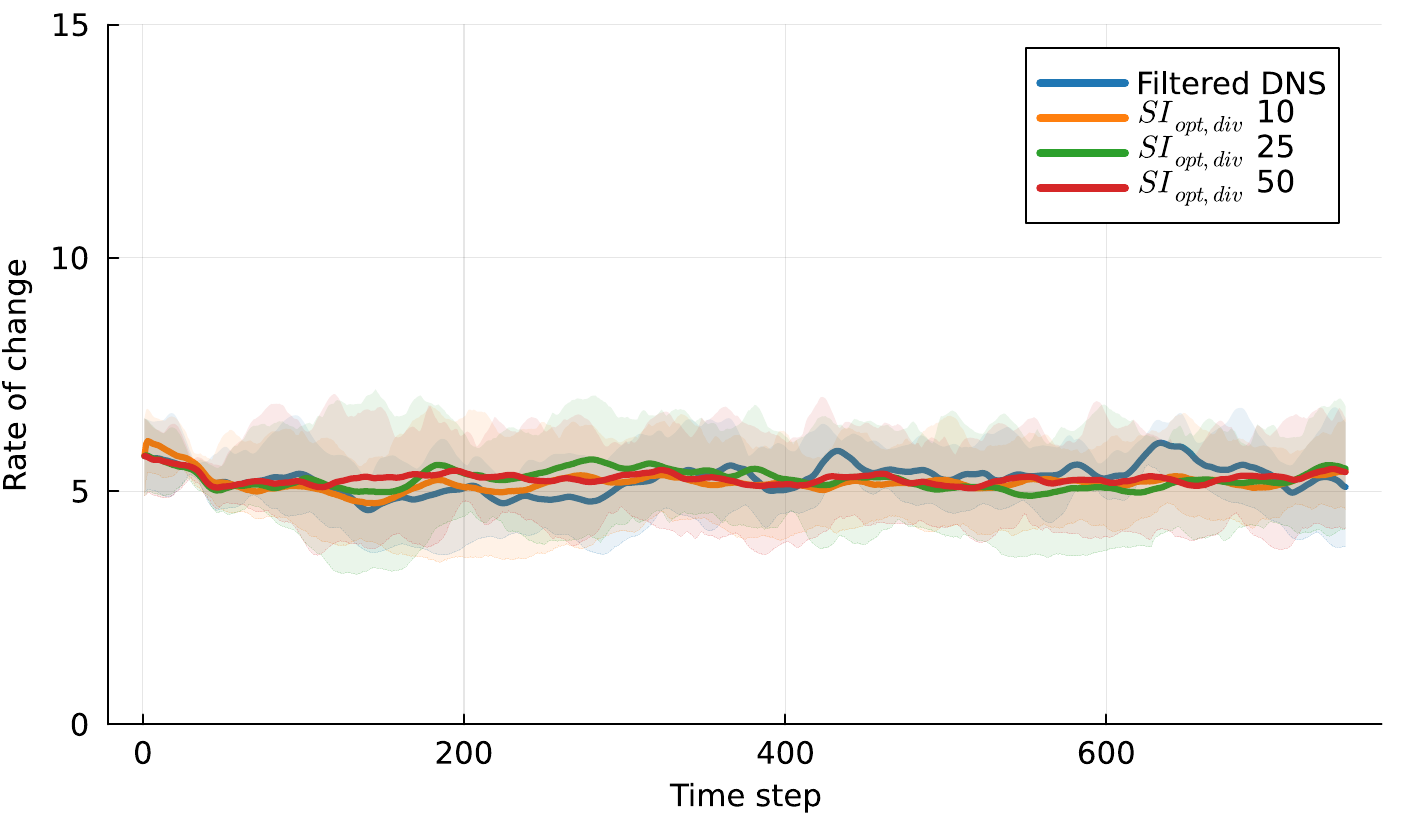}
        \caption{Optimized stochastic interpolant with divergence-free projection}
    \end{subfigure}
    \begin{subfigure}[b]{0.45\textwidth}
        \centering
        \includegraphics[width=1.0\linewidth]{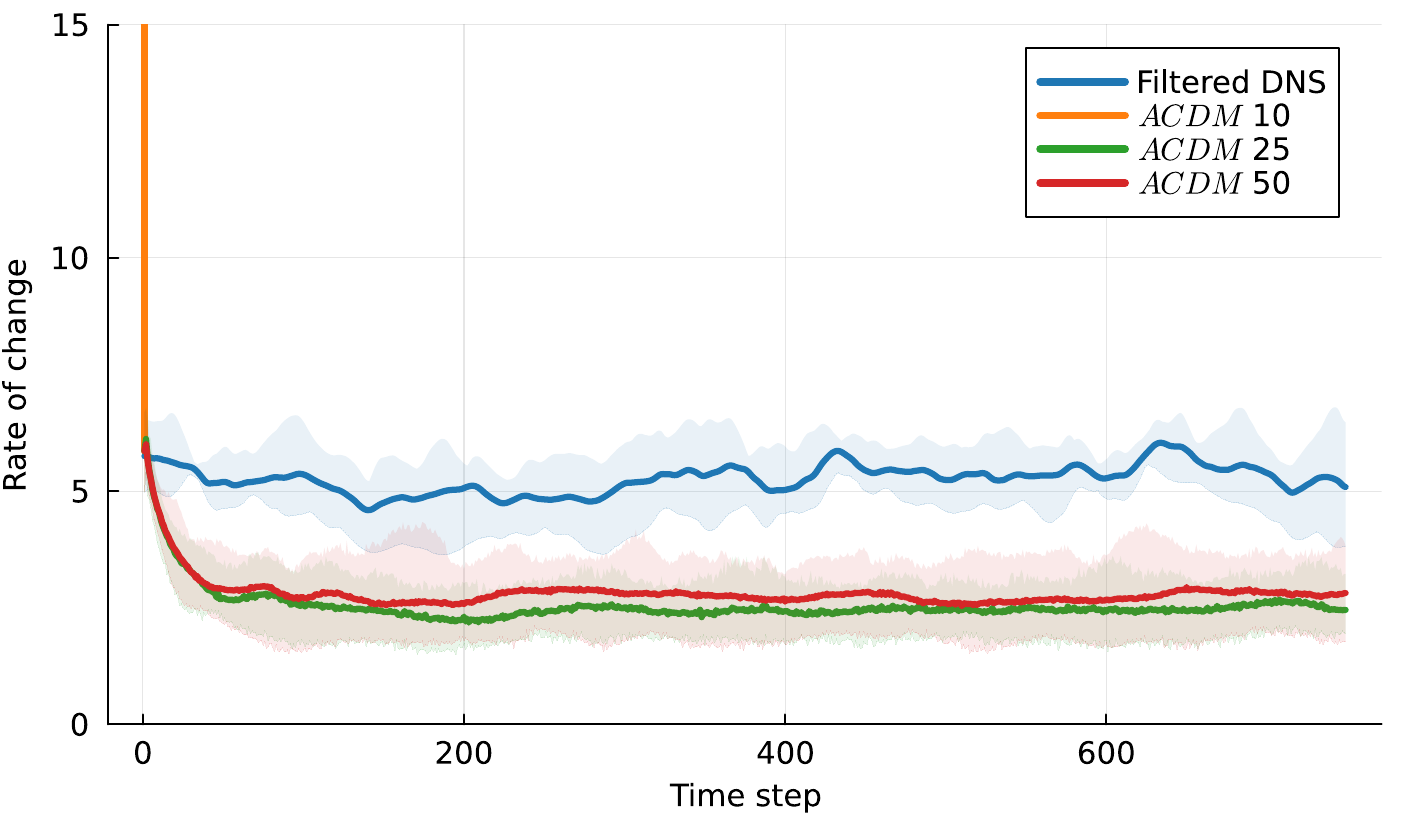}
        \caption{ACDM}
    \end{subfigure}
    \hfill    
    \begin{subfigure}[b]{0.45\textwidth}
        \centering
        \includegraphics[width=1.0\linewidth]{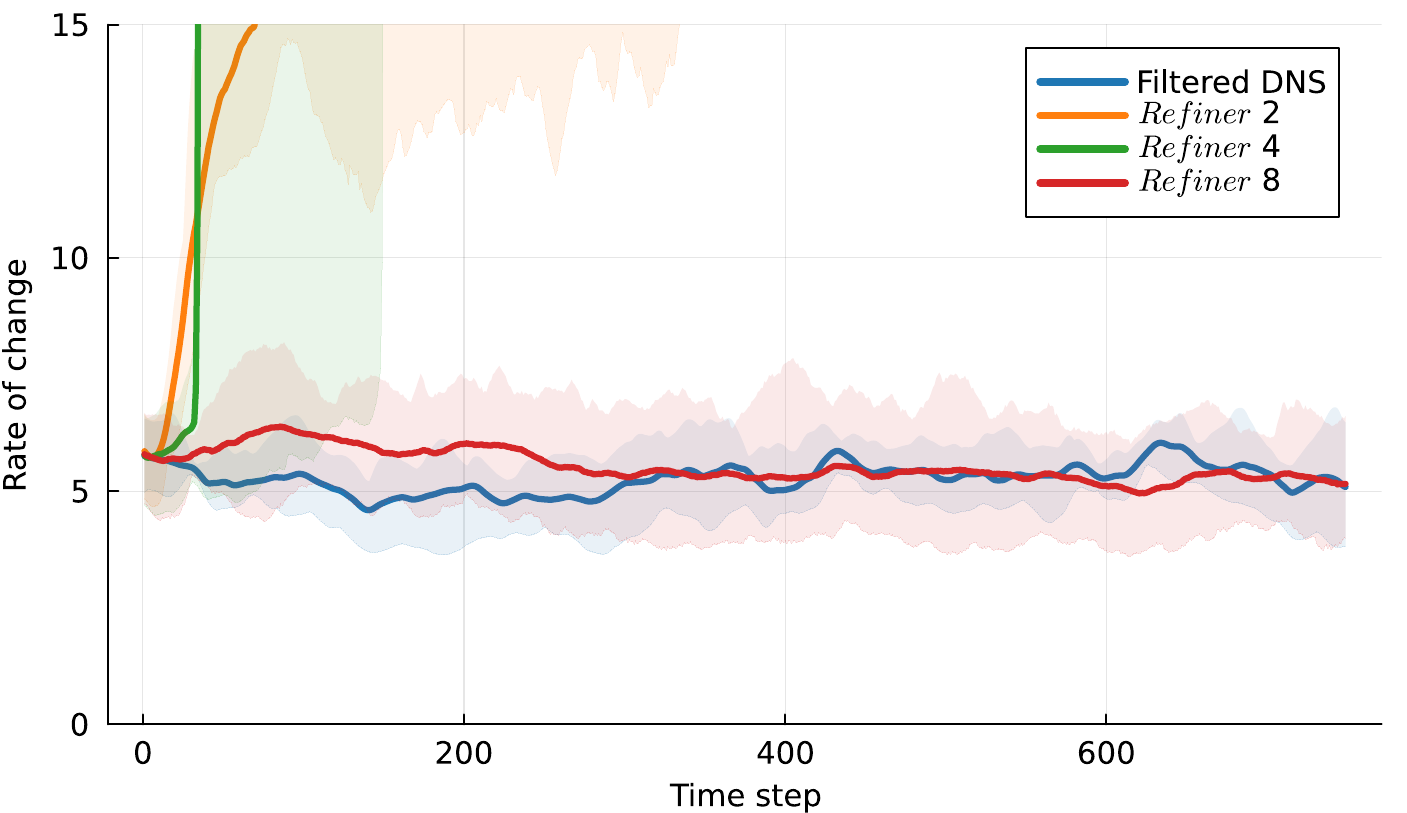}
        \caption{PDE-refiner}
    \end{subfigure}
    \caption{Rate of change.}
\end{figure*}

\end{document}